\documentclass{aa}

\usepackage[varg]{txfonts}
\usepackage{url}
\usepackage[breaklinks=true]{hyperref}
\usepackage{twoopt}
\usepackage{natbib}
\bibpunct{(}{)}{;}{a}{}{,} 
\usepackage{ctable}
\usepackage{multirow}
\usepackage{xcolor}	
\usepackage{enumitem}
\usepackage{xcolor}
\usepackage{footmisc}

\makeatletter
\newcommandtwoopt{\citeads}[3][][]{\href{http://adsabs.harvard.edu/abs/#3}%
{\def\hyper@linkstart##1##2{}%
\let\hyper@linkend\@empty\citealp[#1][#2]{#3}}}
\newcommandtwoopt{\citepads}[3][][]{\href{http://adsabs.harvard.edu/abs/#3}%
{\def\hyper@linkstart##1##2{}%
\let\hyper@linkend\@empty\citep[#1][#2]{#3}}}
\newcommandtwoopt{\citetads}[3][][]{\href{http://adsabs.harvard.edu/abs/#3}%
{\def\hyper@linkstart##1##2{}%
\let\hyper@linkend\@empty\citet[#1][#2]{#3}}}
\newcommandtwoopt{\citeyearads}[3][][]%
{\href{http://adsabs.harvard.edu/abs/#3}
{\def\hyper@linkstart##1##2{}%
\let\hyper@linkend\@empty\citeyear[#1][#2]{#3}}}
\makeatother

\newcommand{\adeg}[1]{{#1}$^{\circ}$}
\newcommand{\amin}[1]{{#1}$^\prime$}
\newcommand{\asec}[1]{{#1}$^{\prime\prime}$}
\newcommand{\thour}[1]{{#1}$^{\mathrm{h}}$}
\newcommand{\tmin}[1]{{#1}$^{\mathrm{m}}$}

\newcommand{\mjybeam}[1]{{#1}\,mJy\,beam$^{-1}$}

\newcommand{\sbeam}[2]{\asec{#1}$\times\,$\asec{#2}}
\newcommand{\pbeam}[2]{\adeg{#1}$\times\,$\adeg{#2}}
\newcommand{\temp}[2]{$T_\mathrm{#1}^\mathrm{#2}$}

\newcommand{\klambda}[1]{$\mathrm{#1}\,\mathrm{k}\lambda$}

\newcommand{\changeone}[1]{#1}
\newcommand{\changetwo}[1]{#1}

\begin{document}

\title{The GMRT 150~MHz All-sky Radio Survey}
\subtitle{First Alternative Data Release TGSS ADR1}
\titlerunning{TGSS Alternative Data Release I}

\author{
H.~T.~Intema\inst{1,2}
\and P.~Jagannathan\inst{2,3}
\and K.~P.~Mooley\inst{4,5}
\and D.~A.~Frail\inst{2}
}
\authorrunning{H.~T.~Intema et al.} 

\institute{
Leiden Observatory, Leiden University, Niels Bohrweg 2, NL-2333CA, Leiden, The Netherlands \\ 
\email{intema@strw.leidenuniv.nl}
\and
National Radio Astronomy Observatory, 1003 Lopezville Road, Socorro, NM 87801-0387, USA 
\and Department of Astronomy, University of Cape Town, Private Bag X3, Rondebosch 7701, Republic of South Africa
\and Centre for Astrophysical Surveys, University of Oxford,  Keble Road, Oxford, OX1 3RH, United Kingdom.
\and Hintze Research Fellow
}
\date{Received ... / Accepted ...}

\abstract{
We present the first full release of a survey of the 150~MHz radio sky, observed with the Giant Metrewave Radio Telescope between April 2010 and March 2012 as part of the TGSS project. Aimed at producing a reliable compact source survey, our automated data reduction pipeline efficiently processed more than 2000~hours of observations with minimal human interaction. Through application of innovative techniques such as image-based flagging, direction-dependent calibration of ionospheric phase errors, correcting for systematic offsets in antenna pointing, and improving the primary beam model, we created good quality images for over 95~percent of the 5336~pointings. Our data release covers 36,900~$\deg^2$ (or $3.6\,\pi$~steradians) of the sky between \adeg{-53} and \adeg{+90} DEC, which is 90~percent of the total sky. The majority of pointing images have a background RMS noise below \mjybeam{5} with an approximate resolution of \sbeam{25}{25} (or \changeone{\sbeam{25}{25}$/\cos{(\mathrm{DEC}-19^{\circ})}$} for pointings south of \adeg{19} DEC). We have produced a catalog of \changeone{0.62}~Million radio sources derived from an initial, high reliability source extraction at the 7~sigma level. \changetwo{For the bulk of the survey, the} measured overall astrometric accuracy is better than \asec{2} in RA and DEC, while the flux density accuracy is estimated at $\sim 10$~percent. \changetwo{Within the scope of the TGSS ADR project, the} source catalog, as well as 5336~mosaic images (\pbeam{5}{5}) and an image cutout service, are made publicly available online as a service to the astronomical community. Next to enabling a wide range of different scientific investigations, we anticipate that these survey products provide a solid reference for various new low-frequency radio aperture array telescopes (LOFAR, LWA, MWA, SKA-low), and can play an important role in characterizing the EoR foreground. \changetwo{The TGSS ADR project aims at continuously improving the quality of the survey data products. Near-future improvements include replacement of bright source snapshot images with archival targeted observations, using new observations to fill the holes in sky coverage and replace very poor quality observational data, and an improved flux calibration strategy for less severely affected observational data.}}

\maketitle 


\section{Introduction}
\label{sec:intro}

Radio continuum surveys have long played a major role in advancing observational cosmology and galactic astronomy
\citep{1999PNAS...96.4756C}. 
In the early years of radio astronomy, continuum all-sky surveys were made at decameter and meter wavelengths, often with low angular resolution and modest sensitivity. As receiver and antenna array technology improved, the drive for increased resolution and sensitivity pushed surveys into the decimeter and centimeter wavelength ranges
\citep[see][]{1977IAUS...74.....J}. 
One of the key lessons to be learned from past successful centimeter surveys is that future radio surveys must have sufficient angular resolution to enable arcsecond source identification at other wavelengths 
\citep{2015ApJ...801...26H}, 
while retaining sensitivity to low surface brightness emission
\citep{1998AJ....115.1693C}.

In recent years there has been a resurgence of low frequency surveys, brought about by new and upgraded facilities that are exploiting the latest in aperture arrays, broadband feeds with low noise receivers, and high speed data transmission and digital processing 
\citep{2013arXiv1307.0386G}. 
Notable new meter-wavelength continuum surveys include the Low Frequency Array (LOFAR) Multifrequency Snapshot Survey 
\citep[MSSS;][]{2015A&A...582A.123H} 
at 74 MHz and 151 MHz, and the Murchison Widefield Array (MWA) GaLactic and Extragalactic All-sky MWA survey 
\citep[GLEAM;][]{2015PASA...32...25W,2016MNRAS.tmp.1444H} 
at five frequencies between 72 and 231 MHz. Both of MSSS and GLEAM are achieving RMS sensitivities of 10's of milliJanskys but angular resolution remains modest at a few arcminutes. Continuum science at decameter and meter wavelengths shares some similarities with that at centimeter wavelengths; active galactic nuclei still dominate the source counts. However, this wavelength regime is also the domain of high energy astrophysics as traced by electrons and magnetic fields. AGN are dominated by their lobes not their cores, pulsars and supernova are brighter
\citep{2015arXiv151101767B}, 
magnetically active stars emit coherent flares and new types of transients appear
\citep[e.g][]{2014ApJ...788L..26O}. 
Propagation phenomena in the magneto-ionized medium of our galaxy are stronger 
\citep{2015ApJ...808..156S}, 
galaxy clusters show extended relics and halos
\citep{2012A&ARv..20...54F}, 
and both AGN and radio galaxies show evidence of earlier phases of activity through steep spectrum, extended emission
\citep{2015MNRAS.447.2468H,2016A&A...585A..29B}. 

Despite these science opportunities, there are special challenges with the processing of high-resolution, low-frequency radio observations.  Strong and widespread radio interference must be dealt with in an automatic way, direction-dependent effects (DDEs) like ionospheric phase delay must be accounted for, and  occasional instrumental instabilities all conspire to limit the dynamic range of the final images.  It is not uncommon to see compromises being made to produce a timely release of the images, and then seeing the original data being re-processed as algorithms improve.  The 8C catalog at 38 MHz was updated to reflect improvements in the calibration and source finding algorithms 
\citep{1990MNRAS.244..233R,1995MNRAS.274..447H}. 
The VLA Low Frequency Sky Survey (VLSS) at 74 MHz was re-processed after a number of software improvements were made, including a better characterization of the antenna primary beam. The result was  25~percent reduction in the RMS noise on average  and a 35~percent increase in the number of radio sources
\citep{2007AJ....134.1245C,2014MNRAS.440..327L}. 
A re-processing of the LOFAR MSSS is also planned to include the data from outer stations allowing for a large increase in the angular resolution.

It is in this spirit that we have undertaken an independent re-processing of the all-sky 150~MHz continuum survey from Giant Metrewave Radio Telescope  
\citep[GMRT;][]{1991ASPC...19..376S}. 
It is the first continuum survey at meter wavelengths with angular resolution comparable to existing centimeter surveys. Following an initial pilot project, the full survey was carried out between 2010 and early 2012. The PI-driven project was named the TIFR GMRT Sky Survey (TGSS). The survey covers the full sky visible from the observatory, a declination range of -55 to +90 degrees.  As of this writing, there have been five data releases, with only about 10~percent of the survey images having been processed and released through the TGSS project website\footnote{\url{http://tgss.ncra.tifr.res.in}}.
A small number of publications have been published using data from this initial release including
\citet{2011ApJ...736L...8B}, 
\citet{2012MNRAS.423.1053G}, 
\citet{2014A&A...562A.108S}, 
and
\citet{2014MNRAS.443.2824K}. 

In the interim, since these data were taken, there have been significant improvements in low-frequency calibration and imaging algorithms. 
\changetwo{Especially calibration of direction-dependent effect (DDEs) is seen as an essential step in converting wide-field observations into high-quality images, as is witnessed by a rich number of publications on the topic over the last decade
\citep[e.g.,][]{2004SPIE.5489..817N,2005ASPC..345..337C,2008ISTSP...2..707M,2009A&A...501.1185I,2010ISPM...27...30W,2011A&A...527A.107S,2015PASA...32...29A,2016arXiv160105422V}. 
One of the main DDEs is ionospheric dispersive delay, which causes smearing of sources (ionospheric `seeing') as well as relatively large residual sidelobe structure across the pointing image when processed using only direction-independent self-calibration
\citep[e.g.,][]{2009A&A...501.1185I} 
The pipeline used for producing the original TGSS data releases 1 through 5 does not include direction-dependent calibration.}

\changetwo{The TGSS is currently the highest resolution full-sky radio survey released below 200~MHz, making it even more sensitive to the the DDEs of the ionosphere.} Given the enormous science potential of such a survey, and its value as a reference catalog for aperture array telescopes like the MWA
\citep{2013PASA...30....7T} 
and LOFAR
\citep{2013A&A...556A...2V}, 
we re-processed the \changeone{TGSS} survey observations starting with the raw uncalibrated visibility data, all of which are publicly available from the GMRT archive. 
\changetwo{Our re-processing effort is motivated by having available a robust and fast pipeline that includes direction-dependent ionospheric calibration.}
In Section~\ref{sec:obs} we briefly describe the GMRT and the survey observations. The data processing pipeline is described in Section~\ref{sec:dpp}. This section includes the strategies that we employed to solved special imaging and calibration challenge, including direction-dependent gain variations. We characterize the properties of the survey in Section~\ref{sec:sp}, and in Section~\ref{sec:dp} we describe the standard  survey image mosaics and catalogs. In Section~\ref{sec:concl} we conclude the paper with a discussion of future work that could improve the survey.

\changetwo{Throughout the article, we use the sign convention $S \propto \nu^{\alpha}$, where $S$ is the flux density, $\nu$ the observing frequency, and $\alpha$ the spectral index. Throughout the article, the image background RMS noise is determined by fitting a Gaussian to the pixel value histogram over a given aperture, and rejecting positive outliers (i.e., true source flux).}


\section{Observations}
\label{sec:obs}

The GMRT lies 80~km north of Pune, India and consists of 30 45-m diameter stationary parabolic antennas, with 14~antennas arranged in a compact configuration and the outer antennas in a "Y-shaped" configuration. This gives the GMRT a minimum and maximum baseline of 100~m and 25~km, respectively. Each antenna is outfitted with six high-performance feeds that cover the frequency range from 150 to 1500~MHz. At 150~MHz the typical total system temperature (including ground and sky) is 615~K and the single antenna gain is 0.33~K/Jy. The total field of view to half power is \amin{186}, and the synthesized beam at the zenith is \asec{20}. The distribution of antenna baselines is such that the array at 150~MHz is only sensitive to extended emission no larger than \amin{68}. The superb angular resolution of the GMRT nicely complements the surface brightness sensitivity of MWA and the inner core of LOFAR. The entire array is currently undergoing a major wide-bandwidth upgrade. The GMRT web page has up-to-date information about system performance of the array\footnote{\url{http://gmrt.ncra.tifr.res.in}}.

\ctable[botcap,center,
caption = {\changetwo{TGSS ADR1} survey properties.},
label = tab:tgss_properties
]{l c}{
}{
\FL Frequency & 147.5~MHz
\NN Bandwidth & 16.7~MHz
\NN \changetwo{Number of pointings} & 5336
\NN \changetwo{Integration per pointing} & 15~min
\NN Total survey time & 2000~hrs
\NN Sky coverage & 36,900~$\deg^2$
\NN RMS noise (median) & \mjybeam{3.5}
\NN Resolution (DEC>\adeg{19}) & \sbeam{25}{25}
\NN Resolution (DEC<\adeg{19}) & \changeone{\sbeam{25}{25}$/\cos{(\mathrm{DEC}-19^{\circ})}$}
\NN Number of sources & \changeone{623,604}
\LL}

The 150~MHz continuum survey was undertaken as a PI-lead effort. After a pilot study in 2009 under project code 16\_279, the TGSS was fully observed in four consecutive GMRT cycles (semesters) under project codes 18\_031, 19\_043, 20\_083 and 21\_057, using over 2000~hours of observe time spread over about 200~observing sessions. The mean epoch of the TGSS is January 18, 2011 (MJD=55579.0). Summarizing the observational parameters as given on the TGSS project website, the survey consists of 5336~individual pointings on an approximate hexagonal grid following the FIRST survey strategy
\citep{1995ApJ...450..559B}. 
Data was recorded in full polarization (RR,LL,RL,LR) every 2 seconds, in 256 frequency channels across 16.7~MHz of bandwidth (140--156 MHz). Each pointing was observed for about 15 minutes, split over 3--5 scans spaced out in time to improve UV-coverage. Typically, 20--40 pointings were grouped together into single night-time observing sessions, bracketed and interleaved by primary (flux and bandpass) calibrator scans on 3C\,48, 3C\,147, 3C\,286 and 3C\,468.1. Interleaved secondary (phase) calibrator scans on a variety of sources were also included, but are typically too faint to be of much use at this frequency.

We obtained all archival TGSS data in the native LTA format by use of the NCRA Archive and Proposal System (NAPS). The modest total raw data size of about 1.8~TB takes reflects the standard time averaging for archiving down to 16~seconds. A fraction of the pointings have been observed multiple times during separate observing sessions, likely because of problems encountered in the original data processing. Some of these problems are likely related to challenging ionospheric conditions, therefore rather than filtering these observations out upfront, we have blindly passed all available data through our robust processing pipeline, and analyzed the final results to identify truly bad data.


\section{Data Processing Pipeline}
\label{sec:dpp}


\changetwo{All archival TGSS raw data was re-processed with a fully automated pipeline based on the SPAM package
\citep{2009A&A...501.1185I,2009PhDT........24I,2014arXiv1402.4889I}, 
which includes direction-dependent calibration, modeling and imaging for correcting mainly ionospheric dispersive delay. In the following sections, we provide a description of the SPAM pipeline. For more details regarding the algorithms and computing choices we refer the reader to Appendix~\ref{sec:spam_pipeline}.}

\changetwo{The pipeline consists of two parts: a \emph{pre-processing} part that converts the raw data (LTA format) from individual observing sessions into pre-calibrated visibility data sets for all observed pointings (UVFITS format), and a \emph{main pipeline} part that converts pre-calibrated visibility data per pointing into stokes I continuum images (FITS format).}


\subsection{Pre-processing}
\label{sec:dpp_pp}

\changetwo{
The purpose of the pre-processing step is to -- for each observing session -- obtain good-quality instrumental calibrations from the best available scan on one of the primary calibrators, and transfer these calibrations to the data of the observed TGSS pointings. We prefer this simple approach over combining calibration results from multiple scans and calibrator sources, mainly because ionospheric phase effects can vary strongly over time and space, which makes it very difficult (if not impossible) to sensibly interpolate calibration phases between calibrators and calibrator scans. Also, determining the necessary corrections for \temp{sys}{} variations across the sky (see below) are much simpler when using a single calibrator source.}

\changetwo{For deriving instrumental calibrations from the primary calibrator scans, we adopted simple point source models with fluxes and spectral indices following the low-frequency flux models from
\citet{2012MNRAS.423L..30S}. 
Since the source 3C\,468.1 is not included in this work, we only used 3C\,48, 3C\,147, and 3C\,286 for primary calibration. For each scan on each calibrator, after initial flagging (excision) of visibilities affected by excessive RFI, we determined time-variable complex gain solutions and time-constant bandpass solutions per antenna and per polarization. Flagging, gain calibration and bandpass calibration were repeated several times, applying increasingly strict flagging of RFI to obtain improved calibration results. After applying the bandpass solutions and time-averaged gain solutions to the calibrator data, we also determined and applied the average phase offset between the polarizations, followed by frequency channel averaging and conversion to stokes I to reduce the data size and speed up processing. The final frequency (and time) resolution leads to some image plane smearing, and is further discussed in Section~\ref{sec:sp_sss}. The resulting phase solutions from a final gain calibration are filtered to separate instrumental and ionospheric phase contributions
\citep[see][]{2009A&A...501.1185I}, 
after which only the instrumental phase solutions were applied to the calibrator data. }

\changetwo{The statistics of the final (normalized) gain amplitudes were compared for selecting the best calibrator scan. \changetwo{For each calibrator scan a weight factor was calculated proportional to the number of active antennas and the inverse variance of the gain amplitudes, and the best calibrator scan is the one with the highest weight.} Next, all \changetwo{pointings} were processed in a very similar way to the calibrator scan, by applying the same calibration tables and averaging the data. During this process, only very basic flagging of excessive RFI was performed. More elaborate RFI excision on the \changetwo{pointing} data was done during the main pipeline processing (see Section~\ref{sec:dpp_mp}).} 

\changetwo{
Before exporting the pre-processed visibilities per pointing to the main pipeline,
the data was gain corrected for system temperature variations across the sky
\citep[e.g., see][]{2009MNRAS.398..853S,2015MNRAS.451...59M}. 
For each pointing and related calibrator, we measured the sky temperatures in the all-sky 408 MHz map of
\citet{1982A&AS...47....1H} 
within the aperture of the GMRT 150~MHz primary beam, and scaled them to the observing frequency assuming \temp{sky}{}$\propto \nu^{\,\beta}$ with $\beta = -2.5 \pm 0.1$
\citep[e.g.,][]{1999A&AS..137....7R}. 
To this we added the ground pick-up and receiver temperatures as given in the GMRT status document\footnote{\url{http://www.ncra.tifr.res.in/ncra/gmrt/gtac/}} to determine the system temperatures. To correct the \changetwo{pointing} visibilities, they were scaled by the system temperature ratio of the \changetwo{pointing} over the calibrator source. The uncertainty of this method is difficult to assess, but depends strongly on the sky temperature difference between the flux calibrator and \changetwo{pointing}. For instance, we expect TGSS flux densities in extremely `hot' regions of the sky (galactic center, but also Cas~A, Cyg~A, Vir~A, etc.) to have relatively large uncertainties. Flux density accuracy is discussed in more detail in Section~\ref{sec:sp_7c}.}


\subsection{Main Pipeline}
\label{sec:dpp_mp}


\changetwo{
The purpose of the main pipeline is to convert the pre-calibrated visibility data of each pointings into a final image, which includes several steps of (self)calibration, flagging, and wide-field imaging. It is an extension of the data reduction recipe described in
\citet{2009A&A...501.1185I}, 
and
\citet{2014arXiv1402.4889I}. 
The computing choices are described in more detail in Appendix~\ref{sec:spam_pipeline}. The main pipeline consists of two parts, a direction-independent (self)calibration part and a direction-dependent (ionospheric) calibration part. Both are described in the next two subsections.}



\subsubsection{Direction-Independent Calibration}
\label{sec:dpp_mp_dic}

The flow diagram of the direction-independent calibration part of the main pipeline is depicted in Figure~\ref{fig:pipeline_main_dic}. At the start of the pipeline, the visibility data are analyzed to derive calibration, imaging, and other processing parameters (e.g., calibration solution interval, imaging pixel size, facet size and field size, flagging clip levels, etc.). A point source model of the local sky at 150~MHz around the pointing center was constructed using information from several large-area, low-frequency radio catalogs, namely NVSS at 1400~MHz
\citep{1998AJ....115.1693C},  
WENSS at 327~MHz
\citep{1997A&AS..124..259R}, 
WISH at 352~MHz
\citep{2002A&A...394...59D}, 
VLSSr at 74~MHz
\citep{2007AJ....134.1245C,2012RaSc...47.0K04L,2014MNRAS.440..327L},  
and SUMSS and MGPS-2 at 843~MHz
\citep{2003MNRAS.342.1117M,2007MNRAS.382..382M}. 
This model was used mainly as an astrometric reference, but also for a rough flux reference and for identifying bright sources outside the primary beam. For pointing centers above \adeg{-35} the astrometry is matched to NVSS,  while below it is matched to SUMSS or MGPS-2. Visibilities with amplitudes above 5 times the total model flux were immediately flagged.

Following the AIPS polyhedron/Cotton-Schwab wide-field imaging scheme
\citep{1984AJ.....89.1076S,1999ASPC..180..357C}, 
the circular primary beam area was tiled into $\sim 80$~small facets of equal size out to a \adeg{2.3} radius, or $\sqrt{2}$ times the primary beam half-power radius. These were complemented with $\sim 30$ additional facets at positions of bright outlier sources. \changetwo{Based on experience, we include all sources with an estimated true flux density above 3~Jy out to 4~times the primary beam half-power radius, as well as sources above 1.5~Jy  out to 2~times the radius.} Including outlier sources during imaging is important, as their sidelobes can negatively affect the quality of the primary beam area when not deconvolved. Image pixels are \asec{4--5} to ensure proper sampling of the central part of the PSF (point-spread function, or synthesized beam) with 4--5 pixels. After imaging all facets, they were combined into a single image covering the primary beam area.

The point source model was used as starting model for a (direction-independent) phase-only gain calibration on a 16 second timescale, to properly capture the time-varying effects of ionospheric phase delay averaged across the primary beam. Note that these calibration results are kept as separate tables, and are applied on the fly during imaging rather than being applied permanently.

Next, wide-field imaging was done using Briggs weighting with the \texttt{robust} parameter set to -1. Using this robust imaging weighting scheme shifted slightly towards uniform weighting suppresses the abundance of short baselines in GMRT observations, producing a near-Gaussian central PSF while suppressing the broad wings that are typical for centrally condensed arrays like GMRT. This trade-off between sensitivity, resolution and PSF shaping comes at the cost of reduced sensitivity for large-scale emission. In addition, we also exclude the visibilities within \klambda{0.2} distance of the UV-plane origin from imaging, since image reconstruction of strong emission at large angular scales (few tens of arcminutes; e.g., the galactic plane, or the lobes of Cen~A) is found to be extremely difficult even with full synthesis observations
\citep[e.g.,][]{2014MNRAS.442.2867W}. 
The resulting sensitivity to extended emission will therefore not go beyond angular scales of \amin{10}--\amin{20} rather than \amin{68} (see Section~\ref{sec:obs}). We recognize that the current choice of imaging parameters makes the survey relatively less sensitive to extended sources, which negatively impacts the measured flux density of extended sources as well as the source catalog completeness.

Imaging performs a single-scale CLEAN deconvolution down to 3 times the central background noise (as measured in the central quarter radius of the primary beam), with automated CLEAN boxes placed at positive peaks of at least (i) 5 times the local background noise (as measured in each facet), and (ii) 1.25 times the magnitude of the most negative local feature. In this initial case the imaging is done iteratively (down to 3 times the central noise, so that the noise measurements and CLEAN boxes could be updated inbetween iterations. In later imaging steps the CLEAN box updates are based on the previous image, and images are deconvolved down to 2 times the central noise.

The first image was used as an input model for an updated phase-only self-calibration. Similar to the primary calibrator, the gain phase solutions are filtered to separate ionospheric from instrumental effects, which typically improves the estimate of the instrumental contribution to the phase. The instrumental phase contributions were removed from visibilities and gain phase solutions. The outlier facets were split into sources with high and low apparent flux, \changetwo{i.e., sources that are and are not surrounded by noticable sidelobes. Based on experience, the dividing line lies at 0.2~Jy.} The CLEAN component models of the fainter group were subtracted from the visibilities (while temporarily applying the self-calibration) and removed from the active list of facets, \changetwo{with typically 5--10~outlier facets remaining}.

Before making a new (second) image, the visibilities were examined for bad data using a custom-built flagging function based on work by e.g., En\ss{}lin \& Kronberg (see their AIPS-tutorial\footnote{\url{http://www.mpa-garching.mpg.de/~ensslin/Paper/}}). This function takes residual visibilities (with CLEAN components from all facets subtracted)
and makes a residual image. This image is then Fourier-transformed back to the visibility domain, creating pseudo-visibilities which have the visibility amplitudes, visibility weights and imaging weights imprinted. Any ripple artifacts in the image background show up as localized, high-amplitude features in the pseudo-visibilities, which are then easy to identify and flag automatically in the original data (e.g., see Figure~\ref{fig:ripple_killer}). Since the density and amplitude of visibilities is naturally high in the center of the UV-plane, it is necessary to prevent overflagging of short baseline data. This is done by downscaling the pseudo-visibility amplitudes with the square-root of the density in a cell, by detecting outliers using annuli in UV-space, and by repeating this flagging operation multiple times on on a series of residual images with decreasing image cell sizes (or increasing UV-cell sizes). 

Gain amplitudes are also a very effective way of detecting antenna-based problems. The second image is used as input for an amplitude \& phase self-calibration, with a solution interval similar to the visibility time resolution. The time series of gain amplitudes are filtered per antenna per scan, rejecting significant outliers. The resulting gain amplitudes are smoothed per scan and applied to the visibility data (no gain phases are applied), which effectively flags visibility all visibility data related to the rejected gain solutions. This is followed by a phase-only self-calibration and making a new (third) image.

The final round of self-calibration starts with further excision of RFI, first by rejecting outliers in the residual visibility amplitudes,
then by another round of image-based ripple detection as described above, then by subtracting quasi-continuous RFI using the Obit task \texttt{LowFRFI()} (see Obit Development Memo Series\footnote{\url{http://www.cv.nrao.edu/~bcotton/Obit.html}} no.~16). The latter step attemps to isolate, model and subtract ground-based RFI based on the fringe rotation of the visibility phases introduced by tracking the celestial radio sky. Conservative subtraction parameters are used to prevent subtraction of true source flux (see EVLA Memo Series\footnote{\url{http://www.aoc.nrao.edu/evla/memolist.shtml}} no.~161). 

Phase-only gain calibration and imaging is preceded by a bandpass calibration on the \changetwo{pointing visibilities}, determining one amplitude and phase correction per frequency channel for the whole observation. This step is introduced to remove the time-averaged frequency-dependent ($\Delta\phi \propto 1 / \nu$) ionospheric phase difference between calibrator and \changetwo{pointing}, but also helps in removing the average, apparent spectral index difference between calibrator and \changetwo{pointing}. 

The source positions of the apparent radio sky model that serves as input for the final phase-only self-calibration are compared against the astrometric reference catalogs (NVSS for DEC above \adeg{-35}, SUMSS/MGPS-2 for DEC below \adeg{-35}), and any systematic position offset is removed by shifting the source positions. Phase-only gain calibration and wide-field imaging yields the final (fourth) self-calibration image


\subsubsection{Direction-Dependent Calibration}
\label{sec:dpp_mp_ddc}

The flow diagram of the direction-dependent calibration part of the main pipeline is depicted in Figure~\ref{fig:pipeline_main_ddc}. The gain phases and apparent sky model that result from the direction-independent calibration part of the pipeline are typically found to be sufficient to successfully start off the direction-dependent calibration part of the pipeline. Direction-dependent calibration of ionospheric phase delay is the core functionality of the SPAM package, and described in detail in
\citet{2009A&A...501.1185I} 
and
\citet{2009PhDT........24I}. 
Here we will shortly summarize this functionality and describe the additional steps.

Direction-dependent gain phases are obtained by peeling the apparently brightest sources in the wide-field image. In this process, candidate sources are characterized, tested for compactness, and ordered by peak flux. After residual visibilities are formed by subtracting the full sky model from the visibilities (while temporarily applying the phase calibration), the CLEAN model of the first (brightest) source and its immediate surroundings (out to a radius of approximately \amin{10}) is added back to form an approximate visibility data set containing only the first source. This data set is self-calibrated and imaged several times, thus updating the CLEAN model and the gain calibration phases in the direction of the source. In case of image improvement (a relative improvement of the peak-to-noise ratio) the new image and calibration information is saved, and new residual visibilities are created by subtracting the new source model; otherwise the source is discarded. This process is repeated for all candidate sources, up to a maximum of 20 (see below). Inbetween calibrations, to propagate the astrometry of the reference catalog (NVSS or SUMSS/MGPS-2), the peak of the source model is shifted to the nearest reference source position, taking into account differences in resolution and possible absence of a reference source within a reasonable search radius.

The direction-dependent gain phases of the peeled sources are dominated by ionospheric phase delay, and provide a sparse sampling of the ionospheric volume over the GMRT at the time of observing. The gain phases per time stamp are spatially fitted with a 2-layer phase screen model, which reduces the noise on the individual gain phases, and drastically reduces the number of free parameters fitted to the data through peeling. The model is used to  predict antenna-based ionospheric phase delays for arbitrary positions within the wide-field image.

The list of peeled sources is filtered to improve the quality of the ionospheric model fit, removing entries whose peak position is further than one pixel (about \asec{4}) shifted from the position in the last wide-field image (the self-calibration image). A minimum selection of 4~peeling sources is required for model fitting. \changetwo{In 2--3~percent of the cases, fewer than 4~peeling sources are found. This is commonly caused by the presence of a very bright source in the field, whose residual sidelobes dominate the background noise. For these cases, the wide-field imaging is repeated while applying the gain phases of the brightest peeling source, which typically results in better suppression of the sidelobes and subsequently in a larger yield of peeling sources.}

Per time interval, the peeling gain phases are fitted with an ionosphere model consisting of a smooth, large-scale (second order polynomial) phase screen at 300~km height over the GMRT, and subsequently a small-scale, double-layer turbulent screen model \changetwo{(based on the discrete Karhunen-Lo\`eve transform assuming Kolmogorov turbulence; see references given above)} at 250 and 350~km height.
Residual instrumental phase contributions per antenna are easily identified after the model fitting, as they are common in all calibration directions These slowly varying residual phases are smoothed in time and removed from the visibilities and the peeling gain phases before refitting the ionosphere model. For time stamps with persistent high phase residuals on all antennas after model fitting, all visibility data is flagged (typically a few time stamps per observations). 

The ionosphere model is used to generate individual gain tables for each of the small facets covering the primary beam area and the nearby bright outlier sources. Additional facets are added at the exact location of the peeled sources, which also receive individual gain tables. The gain tables contain the model ionospheric phase corrections per antenna per time stamp, as well as a delay term that captures the (approximate linear) behaviour of ionospheric phase over frequency accross the 16~MHz observing band. For the very brightest sources \changetwo{(with a flux density to noise ratio larger than 500)} we choose to use the peeling gain table directly,  which suppresses strong sidelobes better than when using the model gain table. Similar to described in Section~\ref{sec:dpp_mp_dic}, wide-field imaging is performed while temporarily applying the relevant facet-based gain tables at the appropriate time on the fly, yielding a new (fifth) image. 


\changetwo{Next,} we perform a number of additional calibration and flagging operations. The first step involves a bandpass and subsequent gain amplitude self-calibration (while pre-applying all direction-dependent calibrations) to solve for residual, direction-independent instrumental gains across frequency and time. This is followed by imaging (sixth image). During the second step, image-based ripple detection as described in Section~\ref{sec:dpp_mp_dic} is performed. In addition to generating residual images on the fly for this, we also analyze the last wide-field image (sixth image) for the presence of ripples across the primary beam area, as well as locally near bright sources. Furthermore, the residual visibility amplitudes are searched for statistical outliers, which are flagged. This is again followed by imaging (seventh image). 

\changetwo{At this point we repeat the peeling process, starting off with better residual visibility data using the improved calibration tables and sky model. The ionosphere model is updated and followed by wide-field imaging (eighth image). Two more images are created, repeating the steps in the previous paragraph and thereby refining the} direction-independent instrumental calibration and the excision of bad data. Only the final (tenth) image is exported to a FITS file, and the \changetwo{(flagged and calibrated)} visibility data used for creating this image is exported to a UVFITS file, \changetwo{together with the ionosphere model}. As mentioned in Section~\ref{sec:dpp_mp}, these files together with the log file are kept as the output products of the pipeline run.


\subsubsection{Difficult and Failed Pointings}
\label{sec:dpp_mp_dfp}

The SPAM pipeline has successfully processed $\sim 95$~percent of the TGSS observations in one run. This includes most fields in areas of high sky temperature, like the galactic plane and the lobes of Cen~A. For pointings for which the pipeline failed, analyzing the pipeline log files was in most cases sufficient to identify the problem and find a solution. Most common were `bookkeeping' problems, where observational data was mislabelled or missing from the archive. Strong preference was given to processing visibilities that were correlated using the new GMRT software back-end (GSB), but in a few cases we had to revert to hardware correlator data (recorded in parallel) due to corrupted or missing GSB data.

A common point of failure in the pipeline processing was the transition between direction-independent and direction-dependent calibration, when too few peeled sources were found for ionospheric modeling. For this we introduced an additional imaging and peeling round (Section~\ref{sec:dpp_mp_ddc}). The remaining failed pointings typically contained a very bright (50--100~Jy) source, causing the (increased) background noise to be dominated by artifacts due to dynamic range limitations. These pointings were guided through problematic pipeline steps manually. On various pointings containing an extremely bright source ($>300$~Jy; Cyg~A, Cas~A, Cen~A, etc.) any effort to obtain direction-dependent calibrations failed (because too few or no other sources were detectable). For these cases, we kept the direction-independent pointing images.

Out of 5336~pointing positions, there are 177 (or 3.3~percent) that currently have no image. Our main priority is to fill in these missing images as soon as possible (also see Section~\ref{sec:concl}), but since it involves a minor fraction of the total survey we have decided not to delay the first public data release for this reason. We found a high number of failures for pointings in the lowest two DEC rows, indicated with names RxxD00 and RxxD01, covering the DEC range between \adeg{-55} and \adeg{-53}. The intermediate pipeline images of these pointings are generally of very poor quality. This is most likely caused by extreme baseline shortening due to very low elevation projection, and strong phase effects due to a large ionospheric air mass. For this data release, we have chosen to leave all these pointings out \changetwo{(144~pointings)}. This means that the lowest DEC included in TGSS ADR1 is \adeg{-53}. 

There were in the order of 10~observing sessions for which the data quality was generally very poor. For almost all of these, make-up observations were available. The observing session on January 28, 2011, has been particularly problematic, yet no make-up observations are available. Based on the erratic behaviour of the self-calibration gain phases, bad ionospheric weather seems to be the cause. We have chosen not to include the 33~relevant pointings in this data release. The absence of the pointings with names in the ranges R25D51--R29D51, R26D52--R28D52, R27D53--R29D53, R21D54--R28D54, R27D55--R29D55, R21D56--R28D56, and R27D57--R29D57 causes incomplete coverage in a region between RA \thour{6.5}--\thour{9.5} and DEC \adeg{25}--\adeg{39}.


\subsection{Mosaicking}
\label{sec:dpp_mos}

We produced 5452~pointing images at 5159~unique locations, which is 96.7~percent of the full TGSS pointing grid. These images are combined into \pbeam{5}{5} mosaics for further processing. The properties of the pointing images are all slightly different. Each pointing has been imaged at its intrinsic resolution, largely defined by the specific UV-coverage obtained during the observation and after flagging. Generation of the mosaics requires that partially overlapping pointing images are optimally aligned in terms of astrometry, flux density and resolution. Here we describe the post-imaging checks and steps that were performed to obtain the final survey mosaics.


\subsubsection{Astrometry Corrections}
\label{sec:dpp_mos_ac}

\changetwo{A complication for wide-field low-frequency radio images is that the ionosphere introduces differential astrometric shifts within the primary beam area. We have applied direction-dependent calibration to compensate for this, but there will be residual astrometric errors. This is because we have a limited density of in-beam calibrator sources, and a limited accuracy with which we can match the calibrator sources to NVSS source positions. The latter is due to differences in resolution and surface brightness sensitivity, and due to source spectral structure.}

The source positions in \changetwo{each pointing image were checked} for \emph{a systematic astrometric offset} in comparison with the positions in the NVSS catalog (for fields above \adeg{-35} DEC) and the SUMSS / MGPS-2 catalog (for fields below \adeg{-35} DEC). We used PyBDSM 
\citep[][also see Section~\ref{sec:sp_se}]{2015ascl.soft02007M} 
to create basic Gaussian source catalogs and robustly cross-matched the positions against typically 50--100 counterparts in the reference catalog. \changetwo{Almost all} systematic astrometric offsets were \changetwo{much smaller than} \asec{25}, and the few outliers we found all relate to either to the few poorly calibrated fields (e.g., at very low DEC) or fields with severe dynamic range limitations (e.g., near Cyg~A). \changetwo{Keeping the processing consistent,} systematic offsets were removed from \changetwo{all pointing} images by adjustment of their reference sky position. 





\begin{figure}
\begin{center}
\resizebox{\hsize}{!}{
\includegraphics[angle=0]{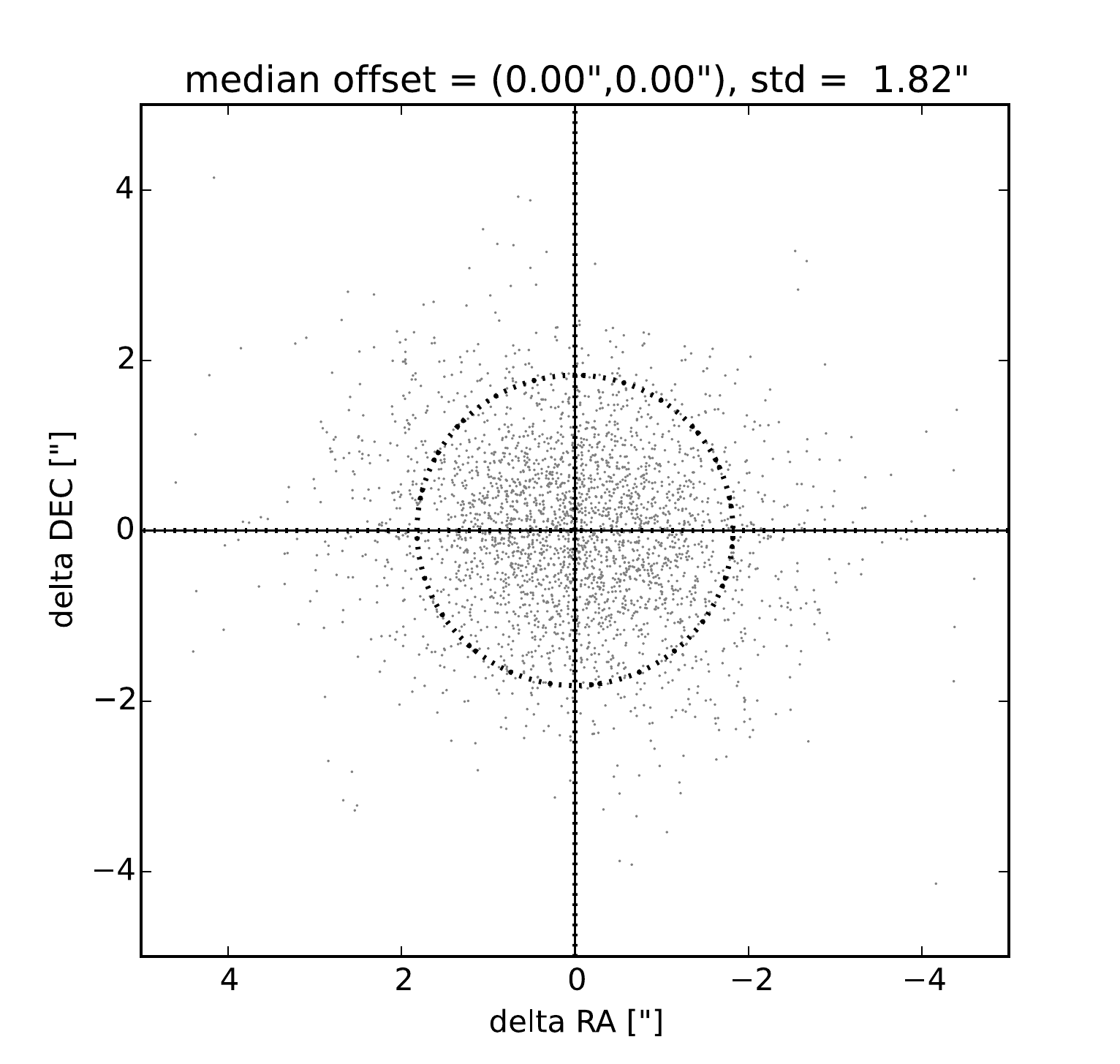}}
\caption{The measured relative position offsets of sources that appear in multiple pointing images (in overlap regions) of 360 pointings in five DEC rows close to \adeg{20}).  The dotted lines indicate the position of the median offset in RA and DEC, while the dotted circle indicates the extend of the standard deviation of the offset radii.}
\label{fig:relative_astrometry}
\end{center}
\end{figure}

To determine the relative astrometric accuracy, we compared positions of a few thousand duplicate sources in overlapping regions between neighboring pointings for a representative (high-elevation) set of pointings. Figure~\ref{fig:relative_astrometry} shows that for a representative subset of the pointings, the relative position offsets are characterized by a Gaussian distribution with a radial standard deviation of less than \asec{2}. This can be considered an upper limit for the relative astrometry, since our source sample also includes faint and/or resolved sources that have larger uncertainties in their measured positions. 

\subsubsection{Flux Density Corrections}
\label{sec:dpp_mos_fdc}


The pointing images need to be corrected for primary beam attenuation. Following the AIPS standard, we adopt a parameterized axisymmetric model of the form:
\begin{equation}
A(r,\nu) = 1 + \sum_{i=1}^{5} C_i \times 10^{-3i} (r \nu)^{2i}
\label{eq:beam_model}
\end{equation}
where $r$ is the radial (angular) distance from the pointing center in arcminutes, $\nu$ the observing frequency in GHz, and $C_i$ the model coefficients. Our initial choice was to use the model coefficients as provided in the GMRT Observer's Manual\footnote{\label{fn:gmrt_manual}\url{http://gmrt.ncra.tifr.res.in/gmrt_hpage/Users/Help/help.html}}, and listed in Table~\ref{tab:pbeam_model} (center column). As the fractional bandwidth is small ($\sim 10$~percent), we use same the central frequency (147.5~MHz) beam model for all frequency channels.

\ctable[botcap,center,
caption = {Primary beam model coefficients for the GMRT at 150~MHz, as defined in Equation~\ref{eq:beam_model}.},
label = tab:pbeam_model
]{c c c}{
}{
\FL Coefficient & Original value & Updated value
\ML $C_1$ & -4.04 & -2.460
\NN $C_2$ & +7.62 & +1.076
\NN $C_3$ & -6.88 & -0.6853
\NN $C_4$ & -2.203 & +3.573
\NN $C_5$ & 0 & -2.582
\LL}

\begin{figure}
\begin{center}
\resizebox{\hsize}{!}{
\includegraphics[angle=0]{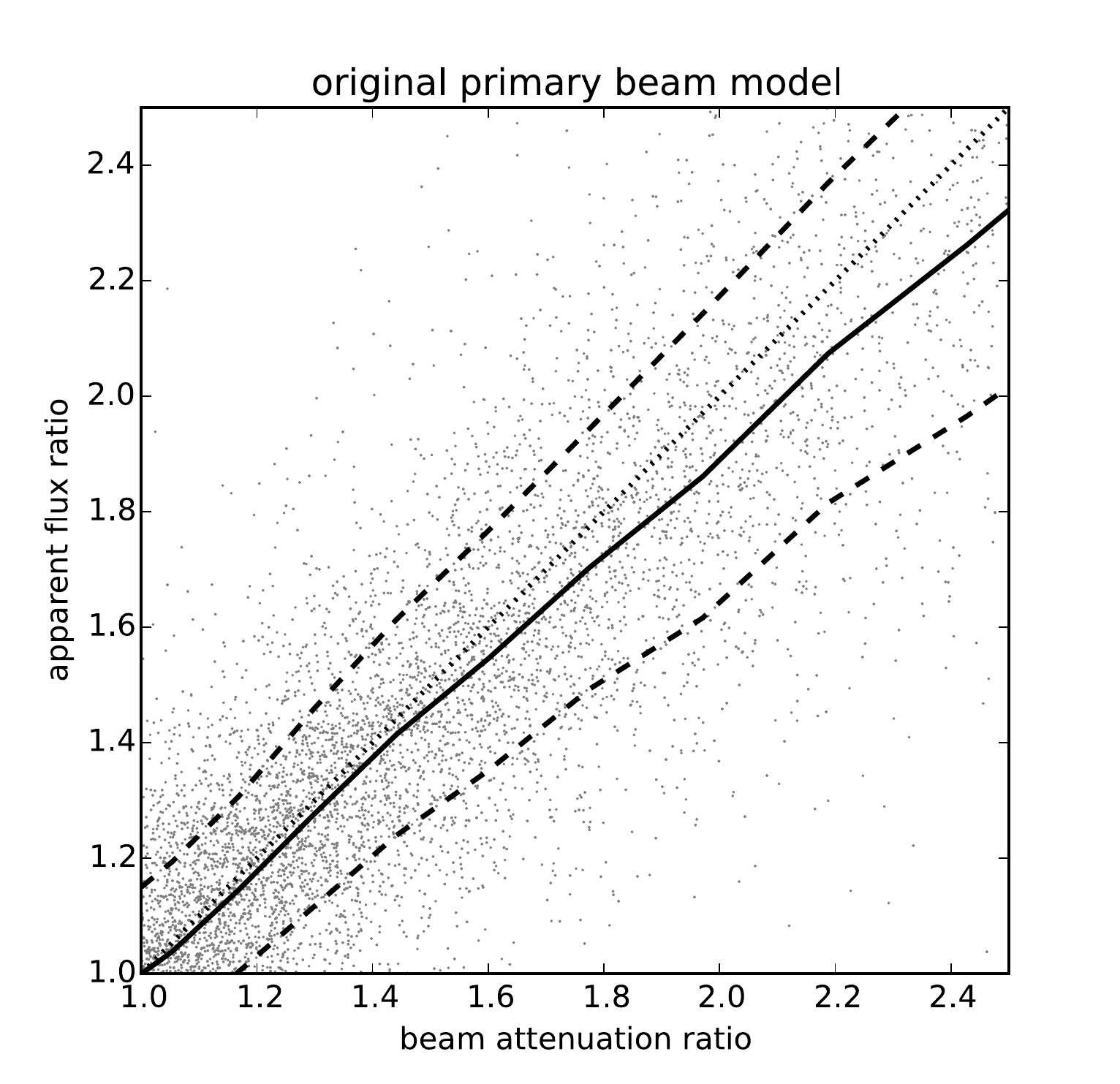}}
\resizebox{\hsize}{!}{
\includegraphics[angle=0]{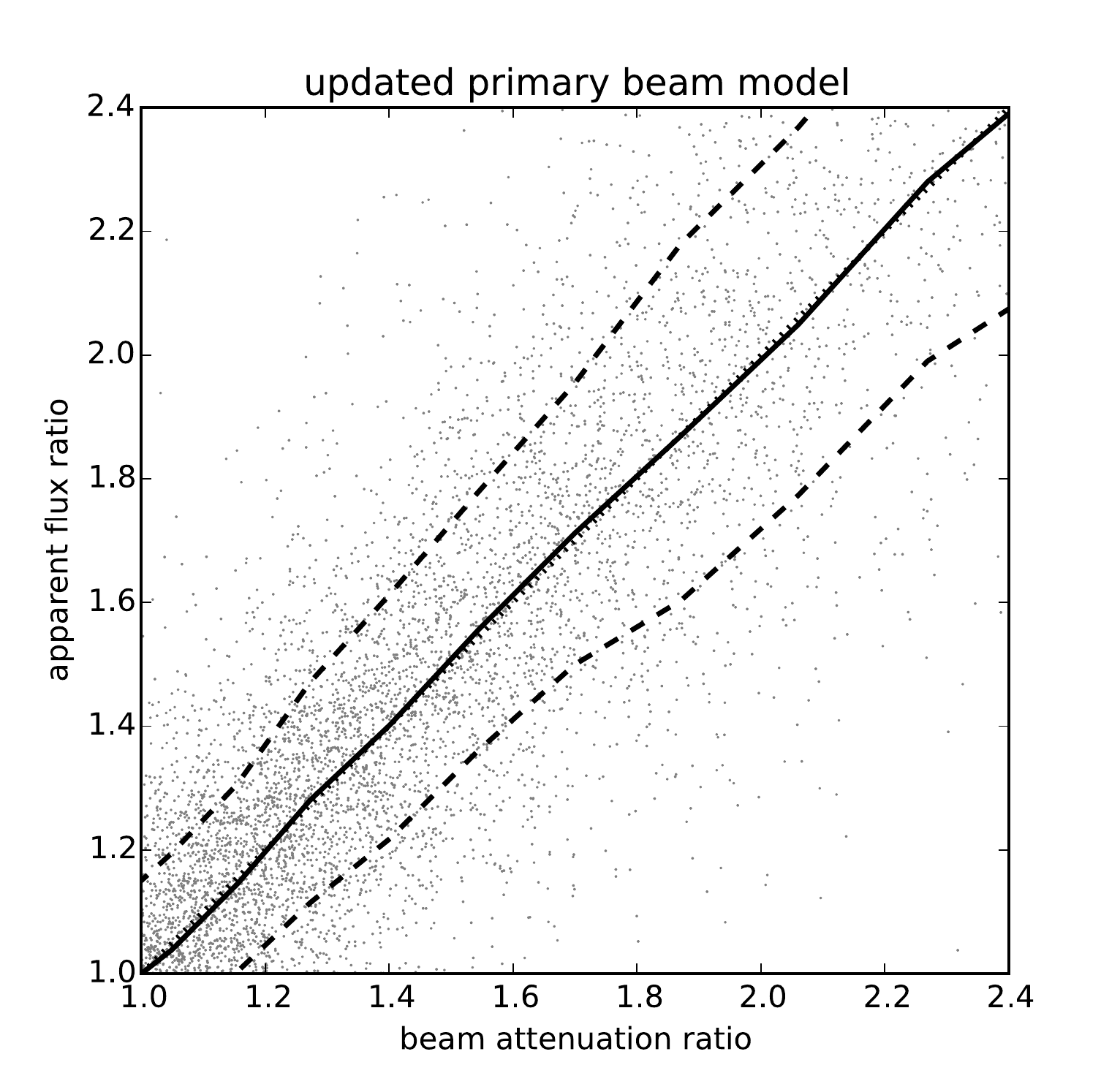}}
\caption{For a selection of bright, compact sources that are found in overlap areas between neighboring pointings in a DEC range close to the GMRT latitude, this figure plots the ratio of measured apparent flux density (gray dots) as a function of the ratio of the primary beam model attenuations. \emph{Top:} When using the original GMRT primary beam model (see Figure~\ref{fig:beam_models_flux}), there is a systematic deviation from the ideal line of unity (black dotted line), which is emphasized by plotting the median trend of binned apparent flux ratios (solid line; dashed lines represent the $\pm 1$~sigma scatter). \emph{Bottom:} When using the updated primary beam model, this systematic deviation is strongly suppressed.}
\label{fig:relative_fluxes}
\end{center}
\end{figure}

Similar to Section~\ref{sec:dpp_mos_ac}, to determine the relative flux density accuracy, we compared flux densities of a few thousand duplicate sources in overlapping regions between neighboring pointings for a representative (high-elevation) set of pointings. Figure~\ref{fig:relative_fluxes} shows the measured ratio of the apparent flux densities as compared to the expected ratio based on the primary beam model. The apparent flux ratio is systematically lower than the beam attenuation ratio above one, which indicates that the apparent flux density of sources drops less rapidly with increasing radius from the center than predicted by the primary beam model.

\begin{figure}
\begin{center}
\resizebox{\hsize}{!}{
\includegraphics[angle=0]{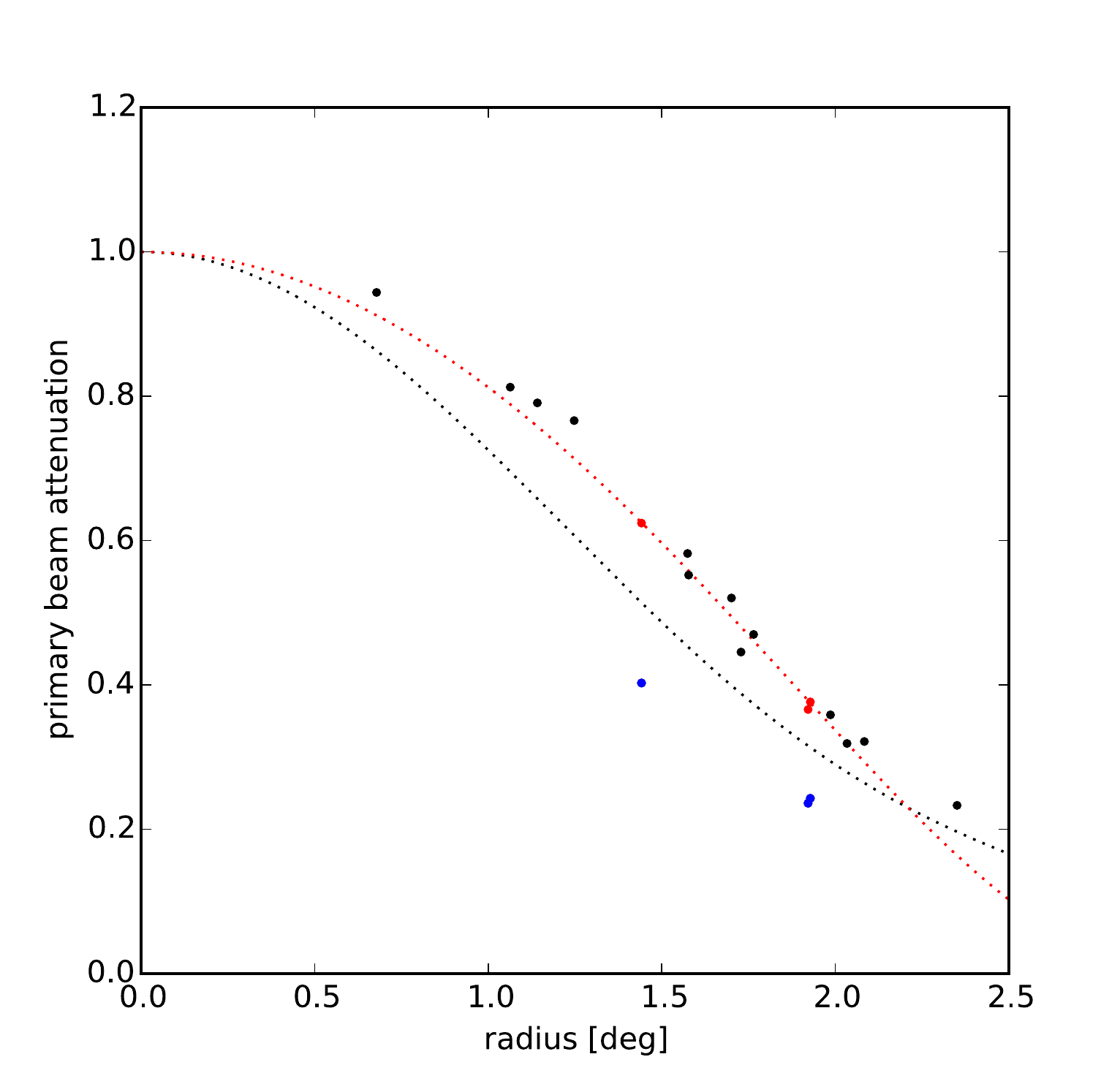}}
\caption{Radial primary beam models as a function of radius from the pointing center. The black dotted line indicates the original model as provided by the GMRT, while the red dotted line indicates the updated model used for this work. The black dots are ratios of the apparent over model flux density of all flux calibrators (except 3C\,286) described by
\citet{2012MNRAS.423L..30S}, 
plotted as a function of radial distance from the center of the pointings in which they appear (typically multiple pointings per calibrator). The updated beam model fits these data points within a few percent, except for 3C\,286 (blue dots). The latter flux densities needs to be scaled up by a factor of about 1.55 to make them line up with the beam model (red dots).}
\label{fig:beam_models_flux}
\end{center}
\end{figure}

To inspect the accuracy of the primary beam model, we measured in the relevant pointing images the apparent flux densities of primary calibrators 3C\,48, 3C\,147, 3C\,196, 3C\,286, 3C\,295, and 3C\,380, and divided these by the model flux densities as given in 
\citet{2012MNRAS.423L..30S} 
to obtain beam attenuations. Figure~\ref{fig:beam_models_flux} shows these beam attenuations as a function of radial distance from the center of their respective pointings. All measurements except those of 3C\,286 follow a clear trend. The original primary beam model is a poor match to these measurements. Refinement of the primary beam model is not uncommon while doing low-frequency surveys
\citep[e.g.,][]{2014MNRAS.440..327L}. 
Ignoring the measurements on 3C\,286 for the moment, we obtained new values for the coefficients by fitting the parameterized model in Equation~\ref{eq:beam_model} to the measurements, also listed in Table~\ref{tab:pbeam_model} (right column). Besides providing a better fit to the measured beam attenuations of the primary calibrators (see Figure~\ref{fig:beam_models_flux}), the new primary beam model also (independently!) corrects the systematic deviation in the apparent flux ratios in pointing overlap regions (Figure~\ref{fig:relative_fluxes}). For this work, we chose to continue using the updated primary beam model.

Regarding 3C\,286, we investigated possible reasons for the deviation in apparent flux. This source is a well-established flux calibrator with no evidence of significant variability\footnote{\url{https://science.nrao.edu/facilities/vla/docs/manuals/cal/flux/monitor}}. Since there is one scaling factor that makes the three apparent flux measurements on 3C\,286 compatible with the new primary beam model (1.55; see Figure~\ref{fig:beam_models_flux}), we \changetwo{suspected} a common cause, either in the data or in the data processing. We exclude non-axisymmetry in the primary beam as a possible cause, since that would also affect the measurements of the other flux calibrators. The observations of pointings containing 3C\,286 are all from one observing session on July 6, 2010. We used 3C\,48 to calibrate all pointings in that session, including the ones containing 3C\,286 (pointings R41D52, R42D53, and R41D54). The magnitude of the \temp{sys}{} correction (with an uncertainty estimate) is $0.91 \pm 0.05$, which is not large enough to explain an order 50~percent deviation. 

\changetwo{Only recently have we identified simultaneous and persisting phase delay jumps on about 10~antennas as being the cause. Although delay jumps are a known problem of the GMRT\footref{fn:gmrt_manual}, this rare phenomena is usually short-lived and occurs on up to 2~antennas at any given time. Flagging based on gain amplitudes is effective in removing the affected data. During this particular observation, there was not enough contrast between healthy and affected data, therefore the flagging failed. When combined during calibration or imaging, the partially decorrelated visibility data across frequency results in pointing images for which the flux density is systematically too low, which is what we observe for 3C\,286. Correcting for these delay jumps is largely possible through additional calibration, but this needs to be integrated in the data processing in an early stage. For ADR1, we find that for the bulk of observations the flux scale is not seriously affected (e.g., see Section~\ref{sec:sp_7c}). The delay jump corrections will be an integral part of the future TGSS second alternative data release (ADR2) together with other fixes and improvements.}

\begin{figure}
\begin{center}
\resizebox{\hsize}{!}{
\includegraphics[angle=0]{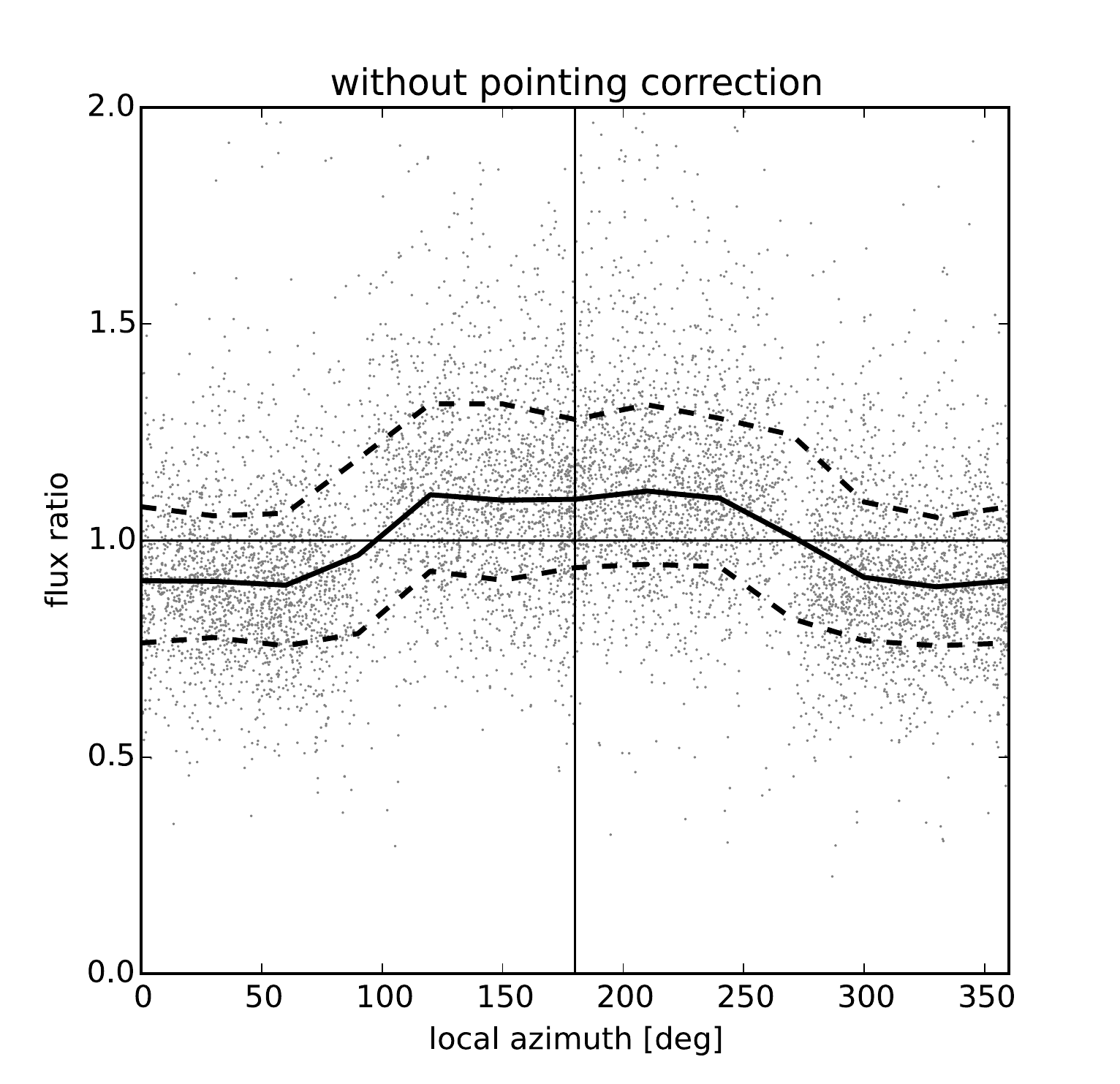}}
\resizebox{\hsize}{!}{
\includegraphics[angle=0]{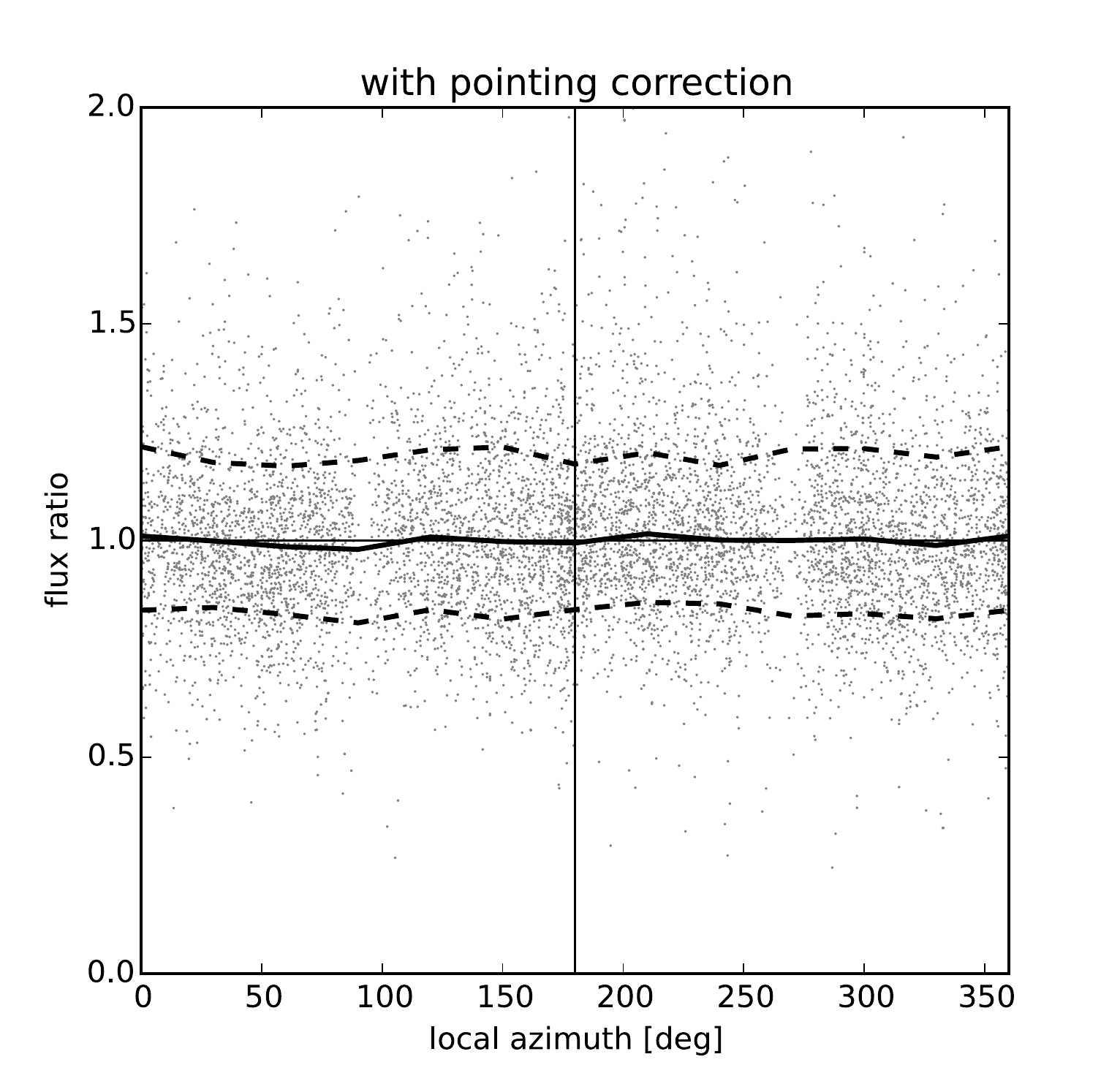}}
\caption{For a selection of bright, compact sources that are found in overlap areas between neighboring pointings in a DEC range close to \adeg{-45}, this figure plots the ratio of measured true flux densities (gray dots) as a function of local azimuth of the source position in the first pointing. \emph{Top:} With no pointing correction, there is a systematic deviation from unity (thin black line). The median trend of binned flux ratios (solid line; dashed lines represent the $\pm 1$~sigma scatter) is sinusoidal with a peak at a local azimuth of \adeg{180}, which indicates an average negative pointing offset along the DEC axis (which is fitted for). \emph{Bottom:} When using the pointing offset model (see Figure~\ref{fig:pointing_offsets}) on the same sources, this systematic deviation is strongly suppressed.}
\label{fig:pointing_fluxes}
\end{center}
\end{figure}

In comparing flux densities in overlapping fields, we noticed another systematic deviation. Suspecting pointing offsets to be the cause of this
\citep[e.g.,][]{2007MNRAS.376.1251G}, 
we compared true (primary-beam corrected) flux densities of a few thousand duplicate sources in overlapping regions between neighboring pointings for a representative set of pointings at very low-declination. Figure~\ref{fig:pointing_fluxes} (top panel) shows that the average measured flux ratio systematically deviates from unity when plotted against azimuth of the source position w.r.t. the pointing center corresponding to the flux ratio nominator. The flux ratios follow a sinusoidal trend as a function of azimuth, which is exactly what is expected for a systematic pointing offset of all antennas. Furthermore, the minimum and maximum of the average flux ratio lie close to \adeg{0} and \adeg{180}, respectively, which indicates that the antennas systematically pointed too low. The origin of this systematic pointing offset is likely the combination of gravitational sagging of the antennas, and upwards refraction of the radio sky due to the bulk ionosphere. 

The image-plane effect as seen in Figure~\ref{fig:pointing_fluxes} is an average effect over all antennas and over a range of observing hour angles. We correct for this average effect by fitting for a systematic offset of the primary beam attentuation model that minimizes the systematic deviation of the flux density ratios from unity. Figure~\ref{fig:pointing_fluxes} (bottom panel) demonstrates the effect on the flux ratios when applying a single systematic pointing offset of about \amin{5} in DEC southward of the intended pointing centers. 

\begin{figure}
\begin{center}
\resizebox{\hsize}{!}{
\includegraphics[angle=0]{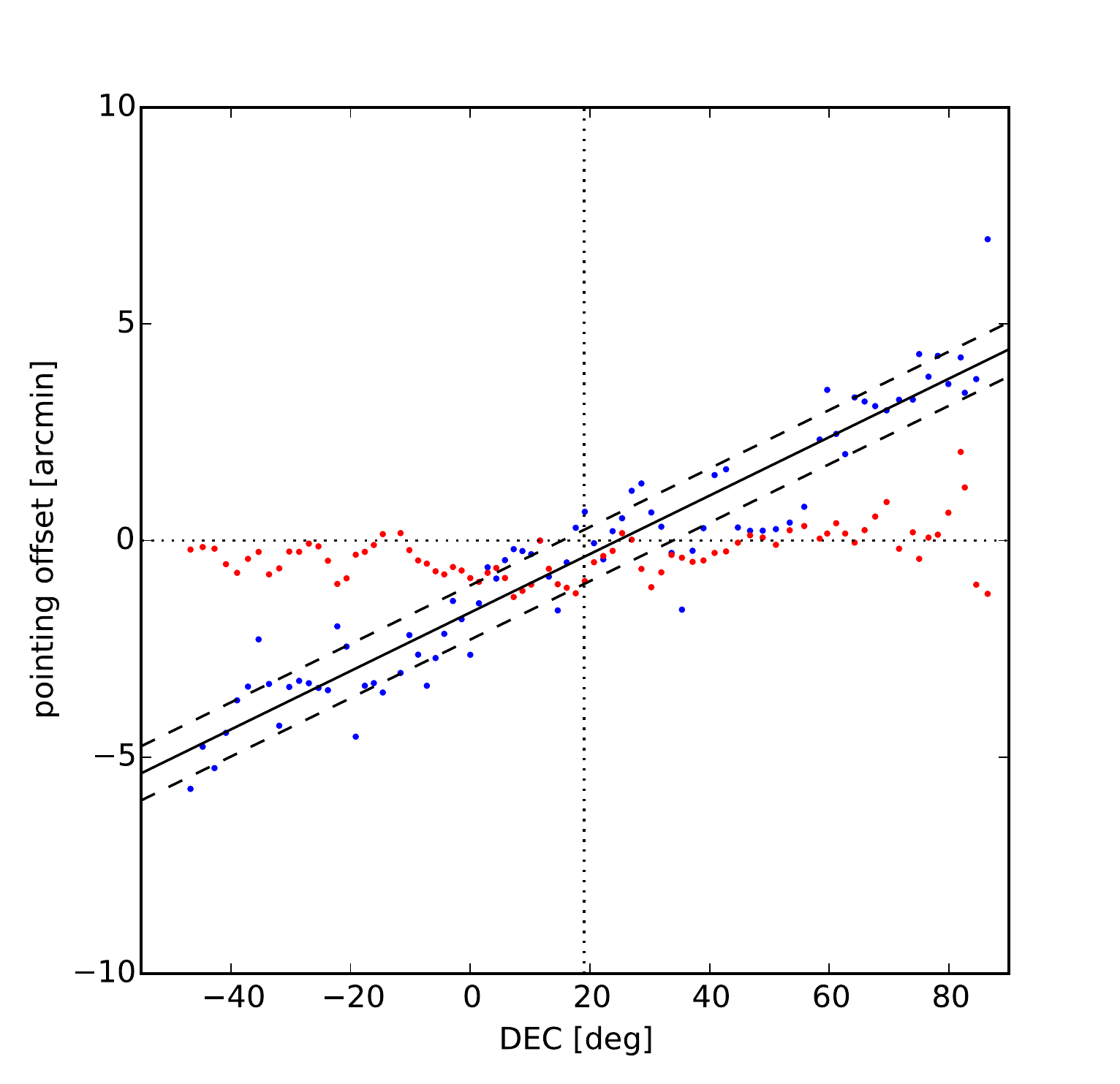}}
\caption{Estimated pointing offsets in RA (red dots) and DEC (blue dots) as a function of declination. On average, the offsets in RA are close to zero, while the offsets in DEC appear to follow a linear trend than can be fitted with a simple model (solid line, with dashed lines representing the $\pm 1$~sigma uncertainty).}
\label{fig:pointing_offsets}
\end{center}
\end{figure}

We extended this method for all pointings, comparing flux density ratios in overlapping regions along rows of fixed DEC. As a cross-check, in fitting for an offset we also allowed for a shift in RA. The results in Figure~\ref{fig:pointing_offsets} show a clear trend of measured offsets in DEC as a function of DEC row, while the measured offsets in RA are close to zero. Note that the magnitude of the pointing offset is minimal for DEC values close to the latitude of GMRT (\adeg{19}). Given the scatter in the measurements and absence of a physical model, we fitted the pointing offset in DEC with a simple linear function, given by:
\begin{equation}
\Delta\mathrm{DEC} = 4.05^{\prime\prime} \times\,(\mathrm{DEC} - 24.6^\circ),
\label{eq:pointing_model}
\end{equation}
with DEC in degrees and $\Delta$DEC in arcseconds. Figure~\ref{fig:pointing_offsets} shows this pointing model, as well as the model uncertainty estimated from the scatter in the RA offsets around zero. For this work, we incorporate this pointing model when constructing the final mosaics, by adjusting the DEC of the primary beam model center while keeping the RA the same.

\changetwo{After correcting for primary beam and pointing offsets, the scatter in the flux ratios plotted in Figures~\ref{fig:relative_fluxes} and \ref{fig:pointing_fluxes} imply an internal flux consistency of $\sim 15$~percent in the pointing overlap regions. However, this includes a large number of faint sources with relatively large uncertainties in their measured flux densities, thus can be considered an upper limit.}


\subsubsection{Building Mosaics}
\label{sec:dpp_mos_bm}

We used the pointing grid coordinates to generate mosaics that form the basis for the source extraction described in Section~\ref{sec:sp_se}. By creating \pbeam{5}{5} mosaics we also facilitate the creation of up to \pbeam{1}{1} cutouts anywhere on the covered sky without the need for additional mosaicking. Also, by limiting the mosaics to this size, the variation in orientation of the final restoring beam is still fairly small.

\begin{figure}
\begin{center}
\resizebox{\hsize}{!}{
\includegraphics[angle=0]{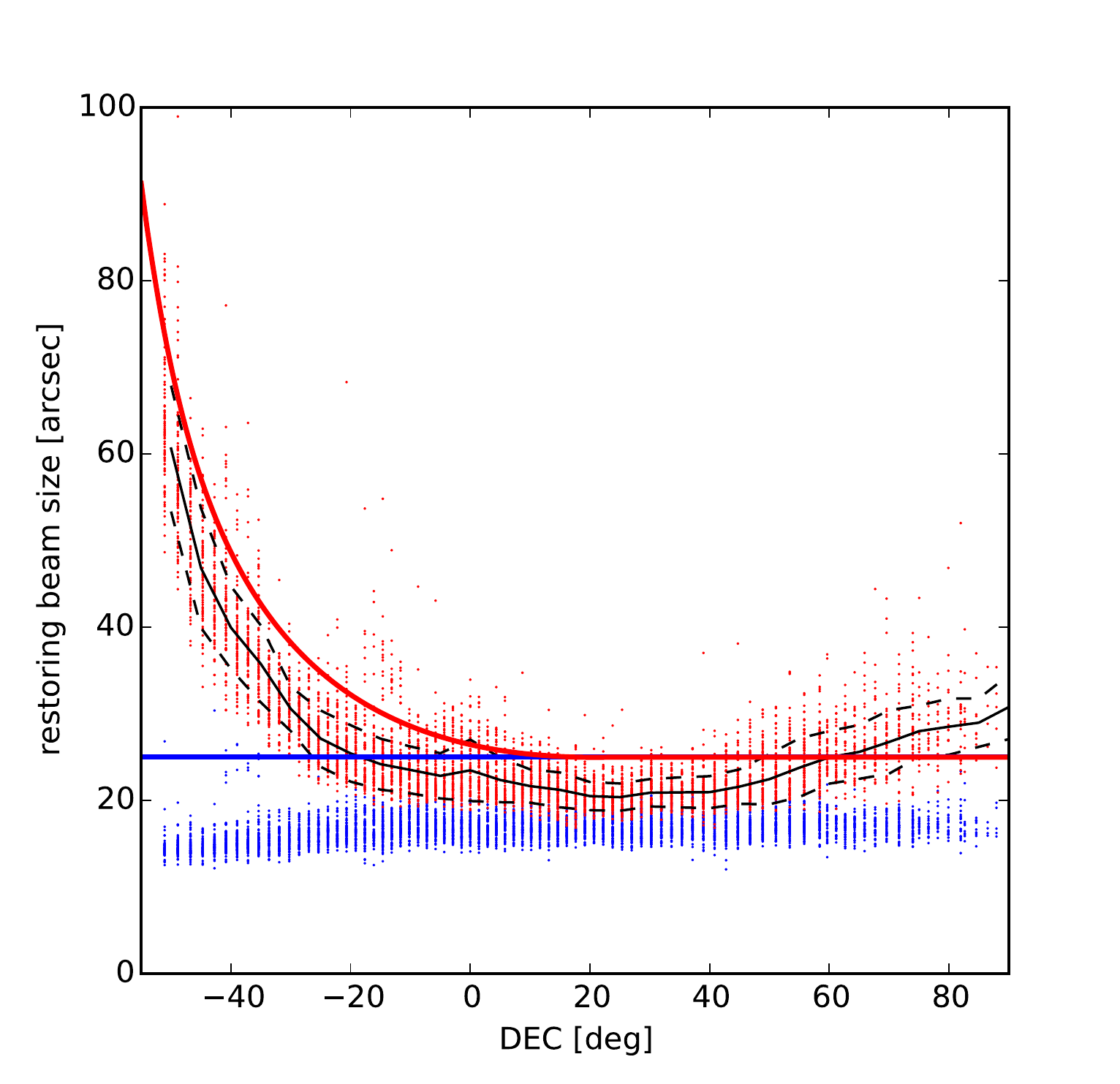}}
\caption{Restoring beam sizes as a function of declination. For all pointings, the red and blue dots are the fitted sizes of the restoring beam major and minor axes, respectively. On average, the binned major axis sizes follow an increasing trend (black solid line; dashed lines represent the $\pm 1$~sigma scatter) that is compatible with the shortening of N-S baselines due to elevation projection. The choice of restoring beam sizes for the mosaics (red and blue solid lines for major and minor axes, respectively) are a trade-off between enveloping the majority of the fitted beam sizes, maintaining high resolution, and enforcing a circular beam towards the north celestial pole.}
\label{fig:beam_sizes}
\end{center}
\end{figure}

In the pipeline processing, the CLEAN restoring beam was determined by fitting a Gaussian to the center part of the PSF, and therefore varies from pointing to pointing. For mosaicking over the full survey area it is necessary to enforce a well-behaved restoring beam. Figure~\ref{fig:beam_sizes} shows for all pointing images the fitted CLEAN beam size as a function of DEC. The smallest, most circular CLEAN beams correspond to pointings with a DEC close to the GMRT latitude. When moving away in DEC, the systematic shortening of the baselines in N-S direction due to projection is clearly visible through the increase of the CLEAN beam major axis size, while the minor axis stays roughly constant. Although not plotted, also the beam position angles preferentially line up N-S because of this.

The choice of restoring beam used for mosaicking is a trade-off between several factors, namely (i) maintaining high spatial resolution, (ii) encapsulating as many beam sizes as possible, (iii) having a circular beam near the north celestial pole where the change in beam position angle is strongest, and (iv) well-described global properties. We adopted a circular \sbeam{25}{25} mosaic beam for the sky north of the GMRT latitude (\adeg{19}\amin{05}\asec{47.5}), and a N-S elongated \changeone{\sbeam{25}{25}$/\cos{(\mathrm{DEC}-\mathrm{LAT}_\mathrm{GMRT})}$} mosaic beam south of the GMRT latitude. The mosaic beam dimensions are overplotted in Figure~\ref{fig:beam_sizes} to illustrate how it relates to the fitted CLEAN beam dimensions.


For each unique position of the observing pointing grid, a \pbeam{5}{5} mosaic is created. For each mosaic, the mosaic beam is fully defined by the mosaic center DEC. Pointing images that overlap with the mosaic area are convolved (and renormalized) so that the resulting resolution matches the size and orientation of the mosaic beam.  Relative rotation of the pointings with respect to the mosaic center is taken into account. \changetwo{Per image, the convolution operates on both the CLEANed source flux and non-CLEANed background, which retains the close resemblance between the CLEAN beam and the central peak of the dirty beam. This is important for mosaicking and extracting faint sources with a significant fraction of non-CLEANed flux density.} In cases where the CLEAN beam is not contained within the restoring beam, the original CLEAN components (kept in a table with the pointing image) are first removed from the image, then a best effort is made to convolve the residual (background) image to a resolution as close as possible to the mosaic beam, and lastly the CLEAN components are added back using the mosaic beam size.

The pointing images are regridded to a pixel size of \asec{6.2}, which samples the mosaic beam minor axis with just over 4~pixels. The images are then corrected for primary beam attentuation while applying the updated beam model and the pointing offset model described in Section~\ref{sec:dpp_mos_fdc}. The pixels in the final mosaic are the weighted average of corresponding (interpolated) pixels in the corrected pointing images, where the weight is the inverse square of the local background RMS noise (the inverse variance) which maximizes the local signal-to-noise in the mosaic. A sliding box \changetwo{(circular, 91-pixel diameter)} is used to estimate the spatial variation in local background RMS noise across each pointing image. The weighted averaging naturally suppresses the influence of poor-quality high-noise images to the quality of the final mosaic.


\subsection{Source Catalog Extraction}
\label{sec:sp_se}

We draw on the large number of catalog completeness and reliability studies 
\citep[e.g.,][]{2012MNRAS.422.1812H,2013ApJ...768..165M,2015PASA...32...37H} 
to select PyBDSM\footnote{http://www.astron.nl/citt/pybdsm/} as our source finding algorithm of choice. We experimented with different input parameters for PyBDSM, compared the results with a different algorithm (SAD in AIPS) on selected mosaic images, and manually inspected the sources in the resulting test catalogs. From these different tests, we selected the default parameters of PyBDSM and a 7-sigma detection threshold. This corresponds to a median source detection threshold of \mjybeam{24.5} across the survey region. Assuming a \asec{25} circular beam, we have 0.2~billion synthesized beams over the entire survey region, and thus a $<0.1$~percent probability of having a false detection due to interferometeric noise. The 7-sigma threshold also mitigates the chances of accepting sidelobes artifacts around bright sources in our catalog. By default, PyBDSM calculates the \changetwo{local} background noise (sigma) within a (pre-calculated) sliding box size of about 300~pixels (\amin{30}), but near very bright sources the box size was sometimes a few times larger. Since this generates a lot of spurious detections, we manually forced the box size to the median value of 291~pixels.

We generated a PyBDSM source catalogs for each of our 5336~TGSS \pbeam{5}{5} mosaics. In its default mode, PyBDSM uses a 3-sigma threshold to find islands of pixels in the image \changetwo{surrounding the peak above 7-sigma}. 
These image pixels are fitted with Gaussian components from which the image point spread function is deconvolved to provide an estimate of extracted flux. The Gaussian components are then grouped appropriately to generate a list of sources, combined source flux densities, and flags denoting whether single/multiple Gaussian components went into the reconstruction of the sources. \changetwo{Details on all these operations are given in the PyBDSM online documentation.}

Adjacent mosaic images overlap one another significantly, so the same source can be present in multiple mosaic catalogs. While the complete survey covers 37,000~deg$^2$, the area covered by 5336~mosaics of \pbeam{5}{5} is 3.6~times larger. \changeone{Before elimination of duplicate entries, there are a total of 2.24~Million sources present in the combined mosaic catalogs. Thus we expect the number of unique sources to be approximately 0.62~Million. The elimination of duplicate entries was executed as follows.}

\ctable[topcap,center,sideways,
doinside=\tiny,
caption = {Random sample of 50~entries from the TGSS ADR1 source catalog. Descriptions of the columns are given in Section~\ref{sec:sp_se}. The full table with \changeone{623,604}~entries will be available in electronic form through CDS.},
label = tab:catalog
]{c c c c c c c c c c c c c}{
}{
\FL Name & RA & $\sigma_\mathrm{RA}$ & Dec & $\sigma_\mathrm{Dec}$ & $S_{\rm total}$ & $S_{\rm peak}$ & Maj & Min & PA & RMS noise & Code & Mosaic
\NN {\changeone{(IAU)}} & [\adeg{}] & [\asec{}] & [\adeg{}] & [\asec{}] & [mJy] & [\mjybeam{}] & [\asec{}] & [\asec{}] & [\adeg{}] & [\mjybeam{}] & {\changeone{(S/M/C)}} & {}
\NN (1) & (2) & (3) & (4) & (5) & (6) & (7) & (8) & (9) & (10) & (11) & (12) & (13)
\ML
\NN 	TGSSADR J062343.5-045829	&	95.93138	&	2.00	&	-4.97454	&	2.00	&	1414.3	$\pm$	158.4	&	1001.5	$\pm$	107.8	&	35.0	$\pm$	0.3	&	27.3	$\pm$	0.2	&	3.1	$\pm$	1.6	&	7.6	&	M	&	R20D29
\NN 	TGSSADR J145035.8-155123	&	222.64928	&	2.00	&	-15.85620	&	2.00	&	603.9	$\pm$	68.4	&	463.0	$\pm$	49.6	&	36.2	$\pm$	0.3	&	25.4	$\pm$	0.2	&	164.0	$\pm$	1.0	&	3.3	&	M	&	R45D22
\NN 	TGSSADR J014739.7-143254	&	26.91535	&	2.01	&	-14.54806	&	2.01	&	266.0	$\pm$	33.5	&	252.5	$\pm$	29.3	&	29.9	$\pm$	0.5	&	26.1	$\pm$	0.4	&	11.3	$\pm$	5.3	&	4.0	&	S	&	R06D24
\NN 	TGSSADR J050022.0+362952	&	75.09184	&	2.74	&	+36.49803	&	3.07	&	62.9	$\pm$	14.0	&	35.7	$\pm$	8.9	&	35.4	$\pm$	5.5	&	31.1	$\pm$	4.4	&	176.2	$\pm$	52.1	&	4.9	&	S	&	R16D57
\NN 	TGSSADR J201741.0-071411	&	304.42103	&	2.01	&	-7.23632	&	2.01	&	227.3	$\pm$	27.0	&	172.1	$\pm$	19.9	&	31.5	$\pm$	0.5	&	29.6	$\pm$	0.5	&	68.1	$\pm$	10.5	&	2.6	&	S	&	R62D29
\NN 	TGSSADR J191927.9+415131	&	289.86608	&	2.45	&	+41.85888	&	2.38	&	40.1	$\pm$	10.0	&	32.5	$\pm$	6.9	&	29.3	$\pm$	3.5	&	26.4	$\pm$	2.9	&	124.0	$\pm$	47.6	&	3.5	&	S	&	R59D59
\NN 	TGSSADR J125746.9-174352	&	194.44556	&	2.25	&	-17.73096	&	2.27	&	73.4	$\pm$	14.0	&	54.7	$\pm$	9.7	&	34.5	$\pm$	2.8	&	29.8	$\pm$	2.1	&	40.5	$\pm$	22.2	&	4.0	&	S	&	R40D21
\NN 	TGSSADR J161104.4+093841	&	242.76830	&	2.10	&	+9.64499	&	2.08	&	58.2	$\pm$	10.4	&	51.5	$\pm$	7.9	&	28.1	$\pm$	1.6	&	25.4	$\pm$	1.3	&	117.9	$\pm$	22.6	&	2.7	&	C	&	R49D40
\NN 	TGSSADR J013729.6+771013	&	24.37342	&	2.10	&	+77.17042	&	2.08	&	140.5	$\pm$	19.5	&	84.8	$\pm$	12.1	&	35.8	$\pm$	1.7	&	28.9	$\pm$	1.1	&	48.3	$\pm$	3.7	&	3.4	&	S	&	R02D79
\NN 	TGSSADR J102049.0+002704	&	155.20424	&	2.42	&	+0.45133	&	2.55	&	66.4	$\pm$	10.4	&	31.5	$\pm$	5.8	&	44.7	$\pm$	4.3	&	31.0	$\pm$	2.4	&	36.9	$\pm$	11.8	&	2.5	&	C	&	R32D33
\NN 	TGSSADR J171746.8-182418	&	259.44519	&	2.08	&	-18.40478	&	2.24	&	76.0	$\pm$	14.0	&	65.3	$\pm$	10.4	&	35.6	$\pm$	2.4	&	26.6	$\pm$	1.4	&	4.0	$\pm$	9.6	&	3.8	&	S	&	R53D21
\NN 	TGSSADR J131530.1+603646	&	198.87524	&	2.94	&	+60.61294	&	2.58	&	26.5	$\pm$	6.2	&	17.6	$\pm$	4.1	&	36.9	$\pm$	5.7	&	25.6	$\pm$	2.9	&	62.8	$\pm$	21.3	&	2.2	&	S	&	R21D69
\NN 	TGSSADR J063835.5-402256	&	99.64808	&	2.08	&	-40.38195	&	2.88	&	34.8	$\pm$	8.8	&	36.9	$\pm$	6.5	&	46.4	$\pm$	4.9	&	24.1	$\pm$	1.3	&	0.2	$\pm$	4.6	&	3.1	&	S	&	R21D07
\NN 	TGSSADR J111740.8+052858	&	169.42006	&	2.00	&	+5.48298	&	2.00	&	657.1	$\pm$	72.9	&	491.1	$\pm$	52.1	&	30.9	$\pm$	0.2	&	25.8	$\pm$	0.2	&	79.2	$\pm$	1.7	&	3.0	&	M	&	R35D37
\NN 	TGSSADR J082017.3-294824	&	125.07190	&	2.38	&	-29.80641	&	3.41	&	33.5	$\pm$	10.3	&	30.4	$\pm$	7.1	&	40.4	$\pm$	6.5	&	27.2	$\pm$	3.0	&	175.6	$\pm$	15.1	&	4.1	&	S	&	R26D13
\NN 	TGSSADR J010219.2+001736	&	15.57999	&	2.45	&	+0.29341	&	2.58	&	27.4	$\pm$	9.2	&	27.1	$\pm$	6.4	&	27.0	$\pm$	3.9	&	24.7	$\pm$	3.3	&	16.0	$\pm$	65.8	&	3.7	&	S	&	R04D33
\NN 	TGSSADR J122002.9-165739	&	185.01197	&	2.05	&	-16.96079	&	2.06	&	112.0	$\pm$	16.5	&	90.8	$\pm$	12.3	&	32.1	$\pm$	1.2	&	28.9	$\pm$	1.0	&	152.7	$\pm$	15.1	&	3.1	&	S	&	R38D21
\NN 	TGSSADR J004303.2-440503	&	10.76330	&	2.01	&	-44.08395	&	2.11	&	218.1	$\pm$	28.2	&	193.2	$\pm$	22.8	&	59.7	$\pm$	1.6	&	26.8	$\pm$	0.3	&	0.3	$\pm$	0.7	&	3.9	&	S	&	R03D05
\NN 	TGSSADR J212325.7+275643	&	320.85706	&	2.51	&	+27.94552	&	2.44	&	35.6	$\pm$	8.7	&	27.4	$\pm$	6.0	&	30.6	$\pm$	3.8	&	26.6	$\pm$	3.0	&	125.2	$\pm$	36.4	&	3.1	&	S	&	R65D51

\LL
}

\changeone{For each mosaic catalog, all sources belonging to one island were temporarily merged into a single catalog entry, with the position being the (flux-weighted) barycenter of the sources. This yielded a total of 2.11~Million entries. Next, per mosaic catalog, the position of each entry was checked to see whether it was closest to its own mosaic center. If the radial distance to any other mosaic center was smaller, the entry was removed. This reduced the total number of entries to 0.59~Million. As a check, each mosaic catalog was matched against all other catalogs from overlapping mosaics using a \asec{10} search radius, which yielded a total of just 7~additional double entries. This number remained the same when increasing the search radius to \asec{50}, showing that these are well-isolated cases. These double entries occured on the equidistance line of two overlapping pointings, where noise and pixelation effects shifted the two measured positions of the same source slightly towards their own respective mosaic centers. Similarly, we expect a roughly equal number ($\lesssim 10$) of cases where both sources were discarded. \changetwo{Attempts to find and add back these lost sources led to the re-introduction of many more redundant source entries, and were therefore abandoned.} For the double entries, we discarded one of the two by directly comparing their radial distance to their mosaic center and keeping the smallest. After this, all remaining islands were expanded into sources again, and the mosaic source catalogs were merged into a single catalog containing 0.63~Million sources, close to what was expected (see above).}




We further eliminated entries in the catalog having rather extreme values for the fitted parameters.
Based on the manual inspection of false detections and the histogram distributions of parameter values in the catalog, we clipped the parameters in our source catalog in order to retain only sources with the following properties:
\begin{enumerate}
\item $S_\mathrm{p} / \sigma_\mathrm{L} > 7$ 
\item $S_\mathrm{i} / S_\mathrm{p} > 0.5$
\item $B_\mathrm{min} > 0.7\,\times\,$\asec{25},
\end{enumerate}
with $S_\mathrm{i}$ and $S_\mathrm{p}$ being the integrated (total) and peak flux, respectively, $\sigma_\mathrm{L}$ the local RMS noise, and $B_\mathrm{min}$ the source size minor axis. Thus we eliminated \changeone{about 9,500~sources}
in the catalog and are left with a total of \changeone{623,604}~sources in the final TGSS ADR1 catalog. Sample entries from the catalog are shown in Table~\ref{tab:catalog}, and the properties of the catalog are explored in Section~\ref{sec:sp}. The column descriptions for the ADR1 catalog are as follows:
\begin{enumerate}[label={(\arabic*)}]
 \item ID denotes the source name according to IAU convention: TGSSADR Jhhmmss.s+ddmmss.
 \item The right ascension (RA) of the source in decimal degrees.
 \item The 1-sigma uncertainty in the right ascension ($\sigma_\mathrm{RA}$) in arcseconds.
 \item The declination (Dec) of the source in decimal degrees.
 \item The 1-sigma uncertainty in the declination ($\sigma_\mathrm{Dec}$) in arcseconds.
 \item $S_{\rm total}$ denotes the total (integrated) flux density and the associated uncertainty in units of mJy.
 \item $S_{\rm peak}$ denotes the peak flux and the associated uncertainty in units of \mjybeam{}.
 \item Major axis (Maj) of the Gaussian fit to the source and the uncertainty in arcseconds.
 \item Minor axis (Min) of the Gaussian fit to the source and the uncertainty in arcseconds.
 \item The position angle (PA) of the Gaussian fit to the source and the uncertainty in degrees. \changeone{The uncertainty was clipped at \adeg{$\pm 90$}, which indicates an undetermined PA.}
 \item The local RMS noise computed in a \changeone{$40\times40$}~pixel square centered on the source.
 \item A code that defines the multiplicity of the source structure in terms of Gaussian components: ‘S’ refers to an \changeone{isolated} single-Gaussian source, 'C' refers to a single-Gaussian source \changeone{partly overlapping with other sources}, and ‘M’ is a source fit by multiple Gaussians.
 \item The name of the \pbeam{5}{5} image mosaic from which the the source entry was extracted. \changeone{A list of mosaic names and their center RA,DEC are provided through the TGSS ADR project website (see Section~\ref{sec:dp_im}).}
\end{enumerate}
The quoted uncertainties are 1-sigma uncertainties resulting from the fitting process, following
\citet{1997PASP..109..166C}. 
A systematic contribution of \asec{2} was added in quadrature to the uncertainties in RA and DEC (see Section~\ref{sec:sp_gaa}). A systematic scaling error of 10~percent was added to the uncertainties in total flux density and peak flux (see Section~\ref{sec:sp_7c}). \changeone{For more details on the derived source properties, we refer the reader to the PyBDSM documentation\footnote{\url{http://www.astron.nl/citt/pybdsm/}}}.
 

\section{Survey Properties}
\label{sec:sp}



\subsection{Noise Properties}
\label{sec:sp_noise}

\begin{figure}
\begin{center}
\resizebox{\hsize}{!}{
\includegraphics[angle=0]{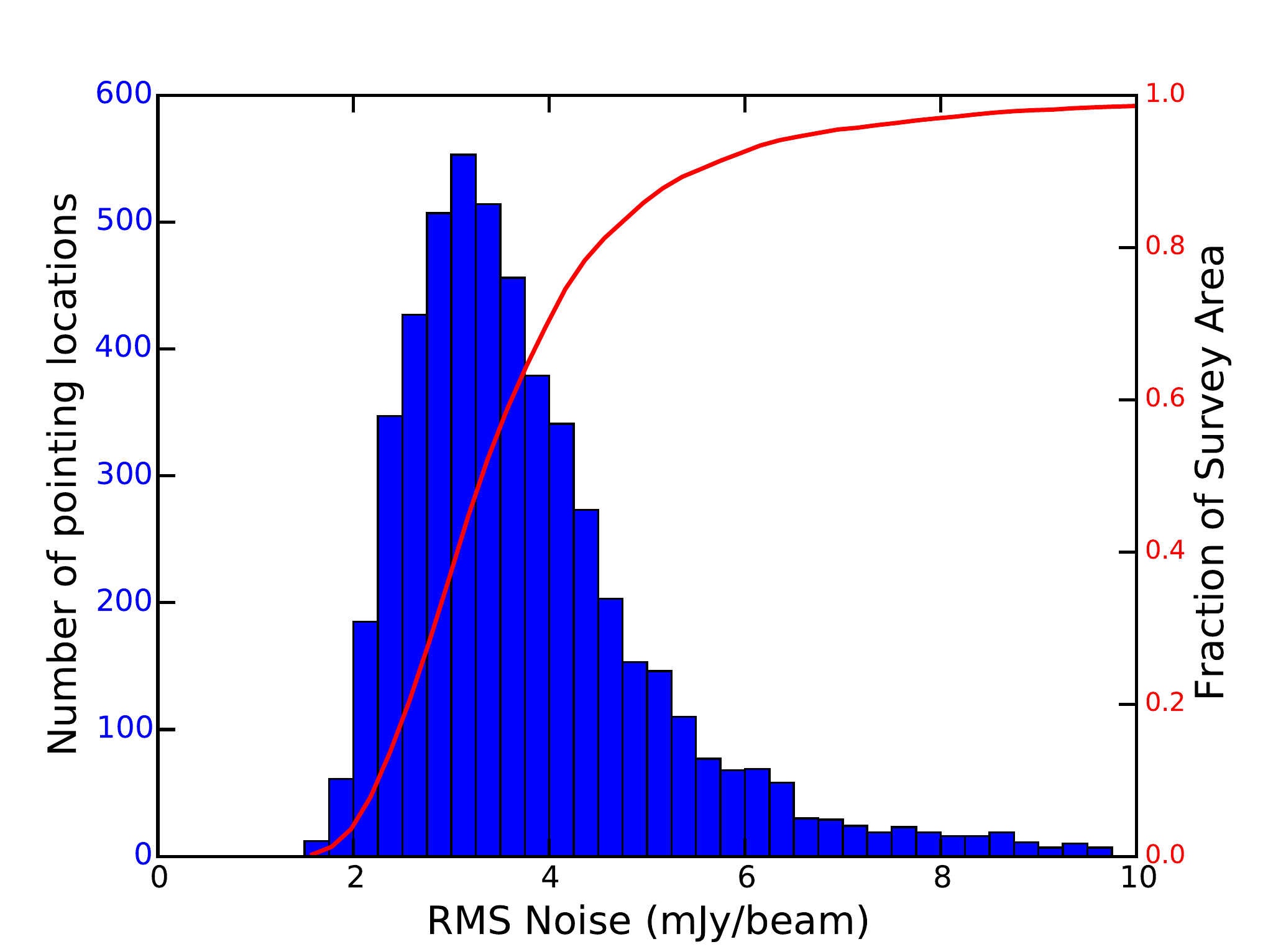}}
\caption{Histogram (blue) and cumulative (red) distribution of the RMS noise in mosaic images. The median RMS noise is \mjybeam{3.5}, while 80~percent of all measurements lie below \mjybeam{5}. The tail of high RMS values is produced by pointings toward the Galactic plane and known bright radio sources.}
\label{fig:rms_histo}
\end{center}
\end{figure}

\begin{figure*}[!ht]
\begin{center}
\resizebox{0.85\hsize}{!}{
\includegraphics[angle=0]{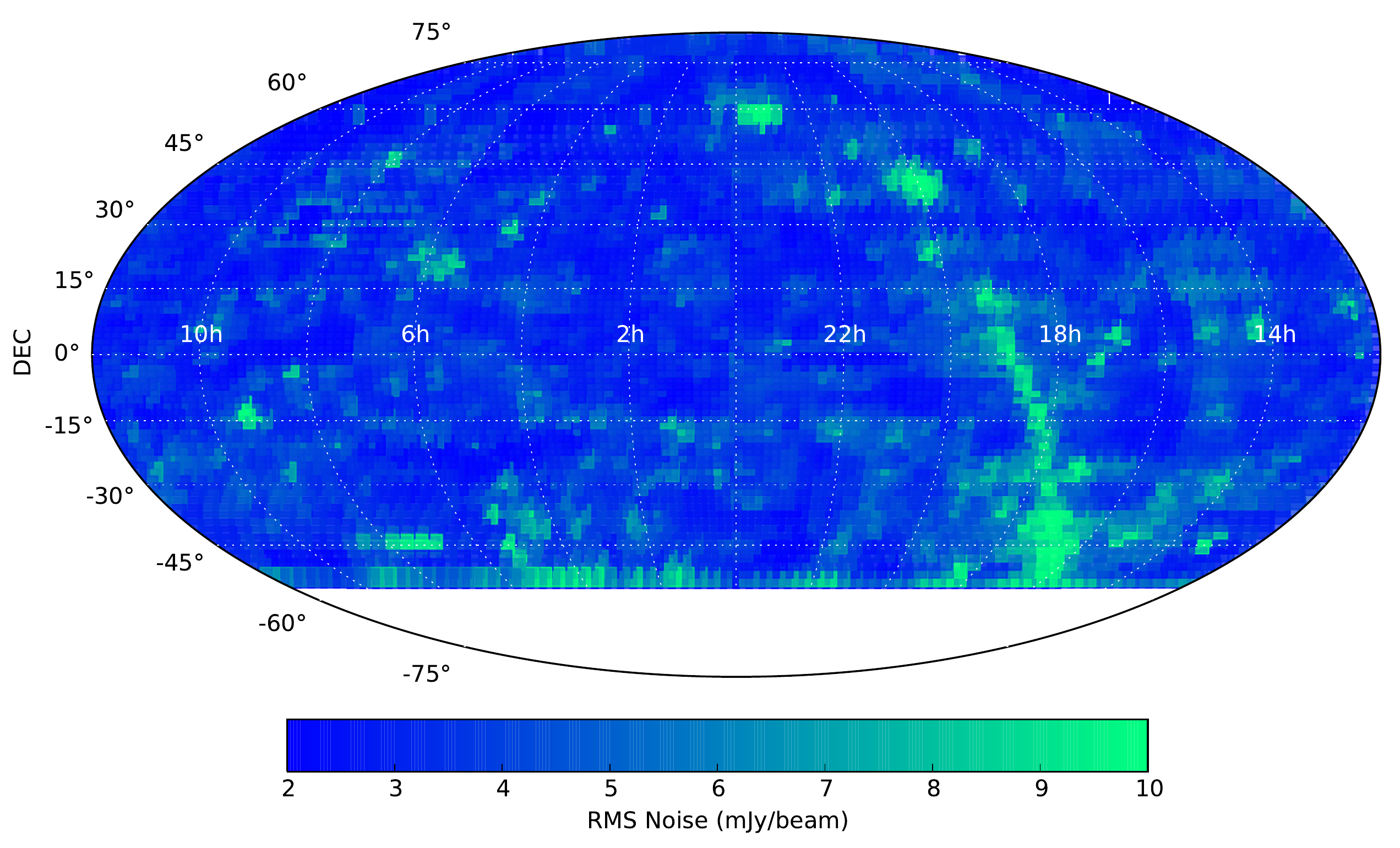}}
\caption{A Mollweide projection plot in equatorial coordinates showing the distribution of the mosaic background RMS noise as a function of the sky position for the TGSS ADR1. The RMS noise is mostly smooth with typical deviations between 2.5 and 5 mJy beam$^{-1}$, with a median of 3.5 mJy beam$^{-1}$. The noise is higher in the Galactic plane due to increased receiver brightness temperature from the diffuse synchrotron emission of the Galaxy. There is a increase in the RMS noise at low declinations since these observations were taken at lower elevation angles than the rest of the data.  Localized regions of higher RMS noise are due to known bright radio sources at this frequency such as Cas~A, Cen~A and Cyg~A.}
\label{fig:rms_map}
\end{center}
\end{figure*}

The first alternative data release (ADR1) of the TGSS consists of 5336~mosaicked images of \pbeam{5}{5} each, centered on the observing pointing grid, (close to) continuously covering the 150~MHz radio sky north of DEC \adeg{-53}. A small fraction (3.3~percent) of problematic pointings were left out from this data release. Figure~\ref{fig:rms_histo} shows the distribution of RMS background noise as measured \changetwo{over} these images. The median survey sensitivity is \mjybeam{3.5}, and over 80~percent of the sky area has a sensitivity below \mjybeam{5}. The sensitivity distribution across the sky is depicted in Figure~\ref{fig:rms_map}.

The RMS noise is likely dominated by thermal receiver noise and image noise due sparse UV coverage in most directions, except towards the galactic plane where the noise temperature is raised by the bright, diffuse synchrotron background of the Galaxy. Confusion noise from unresolved sources in the beam is unlikely to be significant in most directions. \changetwo{\citet{2015A&A...582A.123H} provide two different equations for estimating confusion noise based on VLA measurements. The first equation is based on a scaling of B configuration images at 74 MHz, while the second is based on scaling of deep C configuration images at 3 GHz. Based on these equations, we estimate the TGSS confusion noise at 150 MHz and with a \asec{25} beam range to be \mjybeam{0.44} and \mjybeam{2.5}, respectively.} The RMS noise distribution for most directions in Figure~\ref{fig:rms_map} lies above the largest of these estimates of confusion noise.

Snapshot observations, due to their sparse UV coverage, can result in unwelcome sidelobes upon Fourier transform of the UV data. Typically while imaging, the CLEAN algorithm inadvertently subtracts the dirty beam from these sidelobes just as it does from real sources. When the dirty beam is subtracted from sidelobes, flux is taken away also from the real sources in the image, and as a result, the flux densities of sources reported in the image plane are smaller than their actual values. This artificial decrease in the flux densities of real sources is known as CLEAN bias. For the NVSS and FIRST surveys carried out at the VLA, the bias reduced the flux density of sources by \mjybeam{0.3} and \mjybeam{0.25} mJy, respectively
\citep{1998AJ....115.1693C,2015ApJ...801...26H}.
We plan to carry out a detailed simulation in a future release in order to estimate the effect of this bias in the TGSS catalog. For the moment, We quote an estimate for the GMRT from 
\citet{2013MNRAS.435..650M}
who carried out a snapshot survey at over 90 deg$^2$ with 15-minute pointings, consisting of two 7.5 minute scans a different hour angles. They injected 500 false sources into their calibrated visibility data and re-imaged the data, estimating a reduction in the peak flux densities by \mjybeam{0.9}. While further testing is required, this suggests that any correction for the CLEAN bias for TGSS will be below the median RMS noise.


\subsection{Catalog Completeness and Reliability}
\label{sec:sp_cr}

The quality of a radio source catalog is typically characterized by its completeness and reliability as a function of flux density threshold. The completeness defines which fraction of available radio sources in our survey area have ended up in our catalog. The reliability defines which fraction of radio sources in our catalog are real. These measures are mainly determined by the radio image properties and the details of the source extraction.

The false detection fraction is complementary to the reliability, as it defines the fraction of sources in our catalog is that false at a given flux density. We estimated the false detection fraction by running the exact same source extraction and catalog merging as described in Section~\ref{sec:sp_se} on inverted versions of the images. Inverted means that the image pixel values are changed sign, so that all source emission (that is intrinsically positive) becomes negative, and will no longer be picked up by the source extraction. Assuming that the image background properties (noise and artifacts) are approximately symmetric around the background mean, the local detection thresholds are exactly the same as for the regular (non-inverted) source extraction, and this results in equal number of false detections in the inverted images as in the regular images. 

\begin{figure}
\begin{center}
\resizebox{\hsize}{!}{
\includegraphics[angle=0]{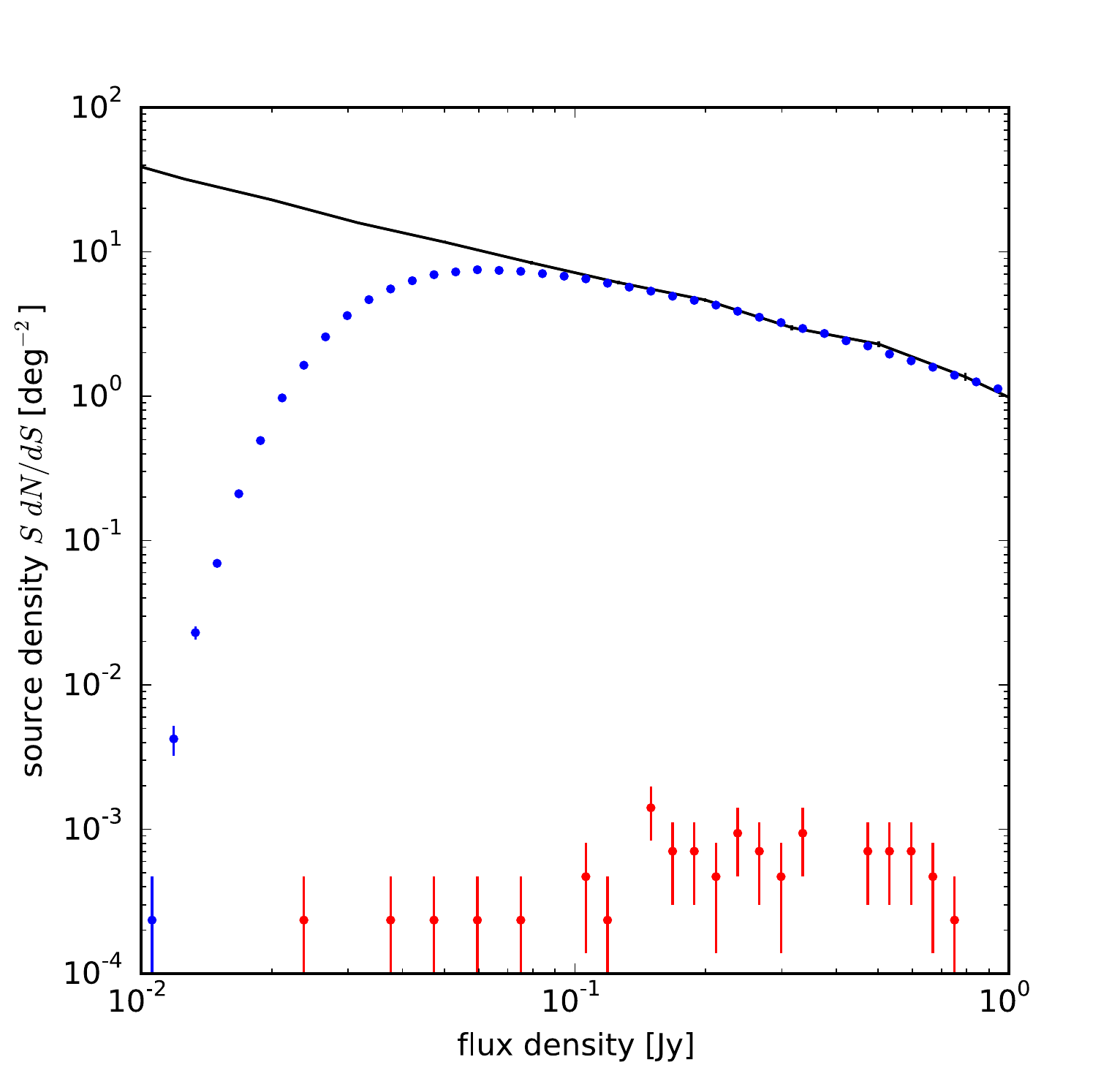}}
\caption{Average number of sources per square degree as a function of flux density, plotted over a relevant flux range. The y-axis is scaled with an extra flux density factor $S$ to have equal counts in logarithmic bins come out at equal height. Above 100~mJy, the source counts of TGSS (blue dots, with Poissonian error bars smaller than the dot size) closely follow the model 151~MHz source counts derived from the SKADS simulation by
\citet{2008MNRAS.388.1335W} 
(black line). Below 100~mJy, the TGSS number counts increasingly deviate from the SKADS number counts due to incompleteness of the TGSS catalog. The false counts (red dots) are several orders of magnitude smaller than the source counts.}
\label{fig:counts}
\end{center}
\end{figure}

\begin{figure}
\begin{center}
\resizebox{\hsize}{!}{
\includegraphics[angle=0]{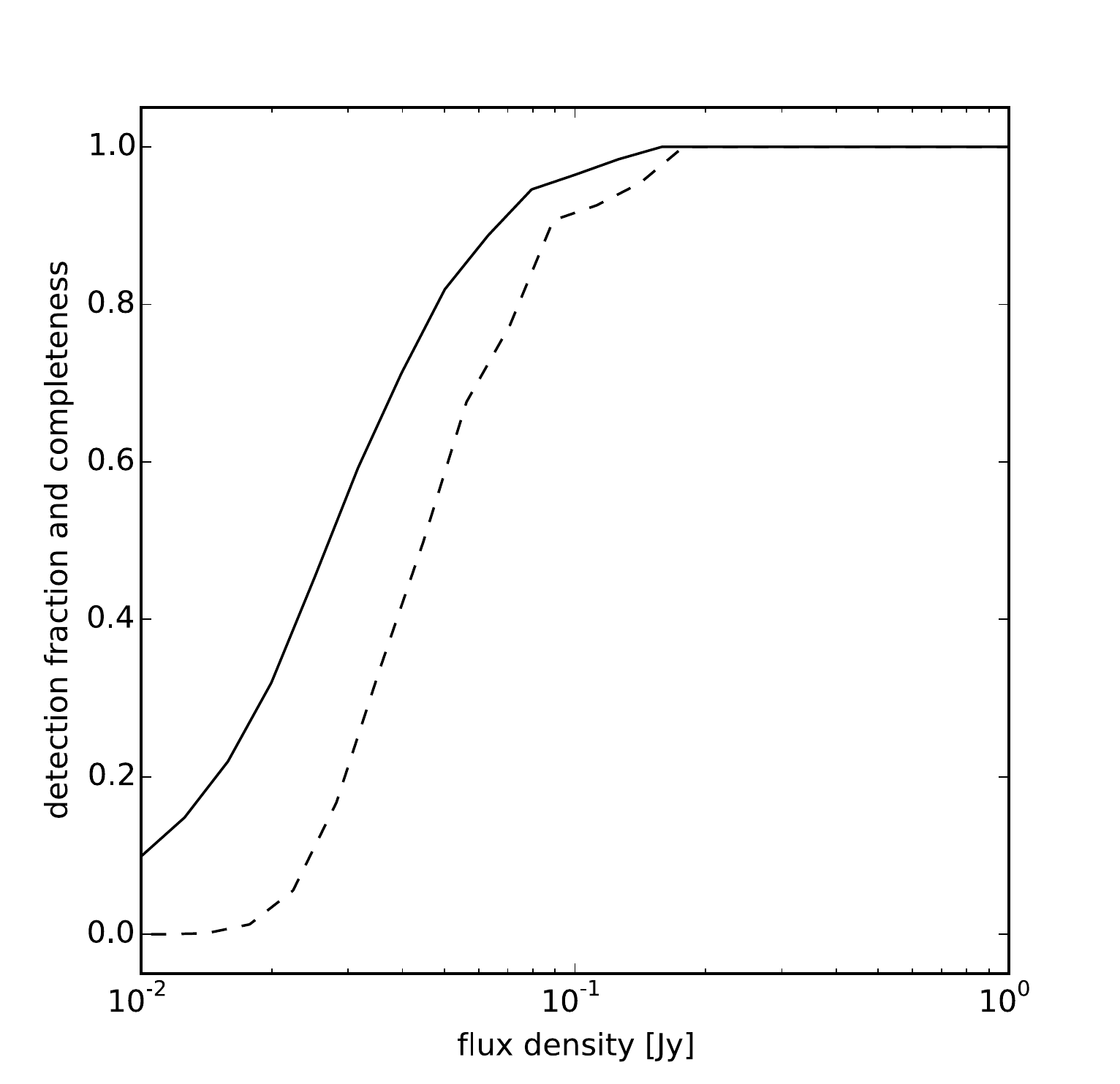}}
\caption{\changetwo{Estimated completeness (solid line) of the TGSS catalog as a function of flux density limit. The completeness is derived from the detection fraction (dashed line) and the SKADS model source counts.}}
\label{fig:completeness}
\end{center}
\end{figure}

The false detection density plotted in Figure~\ref{fig:counts} shows that there are at most a few counts in each flux bin. 
\changetwo{The false detection fraction is very low in all flux density bins, with typical values well below 0.1~percent. The reliability, defined as the fraction of true sources in our catalog above a certain flux density threshold, is derived by integration and normalisation of the product of one minus the false detection fraction and the TGSS detected source counts. Because of the very low false detection fractions, the reliability is very high over the full flux range, with typical values above 99.9~percent.}

Given a large enough area so that cosmic variance is negligible, the differential number of radio sources $dN(S)/dS$ per solid angle as a function of flux density $S$ is fixed. There is a rich literature on measured radio source counts at various frequencies
\citep[see][and references therein] {2012ApJ...758...23C}, 
including the 100--200~MHz frequency range
\citep[e.g.,][]{1983MNRAS.204..151L,1988MNRAS.234..919H,1990MNRAS.246..110M,2008MNRAS.390..741G,2010MNRAS.405..436I,2011A&A...535A..38I,2012MNRAS.426.3295G}. 
Recently, using the LOFAR telescope, source counts at 150~MHz have been extended for the first time down to 1--2 mJy, which is an order of magnitude deeper than previous results (Williams et al., submitted). The semi-empirical source counts derived from a simulated \pbeam{20}{20} area of radio sky at 151~MHz
\citep{2008MNRAS.388.1335W} 
have been shown to reproduce the observed 150~MHz counts to high accuracy over large part of the observed flux density range 
\citep[e.g.,][Williams et al., submitted]{2013A&A...549A..55W}. 
This simulation was performed within the scope of the SKA Design Study (SKADS), and includes radio sources with flux densities far below the detection threshold of current instruments, including GMRT.

We use the TGSS measured source counts and SKADS model source counts to assess the completeness of the TGSS ADR1 source catalog. Sources in both catalogs are counted using the same set of flux density bins. For the TGSS catalog, we do not perform any corrections for possible biases introduced by the TGSS image properties or source extraction. In Figure~\ref{fig:counts} we show the results for a relevant flux density range. It demonstrates an excellent agreement between SKADS and TGSS above 100~mJy, while below 100~mJy the TGSS source counts drop below the SKADS source counts due to incompleteness.

The detection fraction plotted in Figure~\ref{fig:completeness} is simply the ratio of the TGSS counts over the SKADS source counts, fixing the ratio to one above 200~mJy to overcome low number statistics in the SKADS simulation. 
The completeness, defined as the recovered fraction of the absolute number of sources above a certain flux density threshold, is derived by integration and normalisation of the product of the detection fraction (dashed line) and the SKADS model source counts, and is also plotted in Figure~\ref{fig:completeness}. Following the practice of previous surveys like VLSS, NVSS and FIRST, we define a TGSS point source survey threshold at 50~percent completeness which corresponds to approximately 25~mJy (or 7-sigma for point sources, with sigma being the median survey noise of \mjybeam{3.5}).


\subsection{Source Sizes and Smearing}
\label{sec:sp_sss}

In wide-field interferometric imaging, the finite visibility resolution in both frequency and time will lead to the well-known effects of bandwidth and time-averaging smearing, causing a radio source to appear larger in size with reduced peak flux while the total flux density is conserved
\citep[e.g., see][]{1999ASPC..180..371B}. 
For individual TGSS survey pointing images, this effect is zero at the phase center (which equals the pointing center) and grows linearly with distance from the phase center. The magnitude of the effects depend also on time and frequency resolution, and in a complex way on UV-coverage, frequency bandpass shape, and visibility \& image weights. The TGSS pointing images were typically created using visibilities with a time resolution of 16.1~seconds and a frequency channel width of 0.26~MHz. At the primary beam radius of \adeg{1.6}, using the formulae given in
\citet{1999ASPC..180..371B}, 
we expect that the peak flux of a \asec{25} point source to be reduced by 3--4~percent due to bandwidth smearing, and by 2--3~percent due to time-averaging smearing.

The mosaic images used to extract the TGSS source catalog are a noise-weighted sum of multiple pointing images. The strongest smearing occurs in the pointing overlap regions. Qualitatively, it is expected that the resulting smearing of the composite sources is no worse than what is calculated above. An additional smearing contribution may arise from small misalignments between overlapping pointings. From simple Monte-Carlo simulations based on the relative astrometric accuracy determined in Section~\ref{sec:dpp_mos_ac}, we expect up to 6~percent reduction in peak flux due to relative astrometric errors. Combined with bandwidth and time-averaging smearing, this may lead to a 5--13~percent drop in peak flux, and a broadening of the source size with 3--7~percent. Note that this result applies to sources in the overlap areas; in other areas the effect is expected to be much less.

As described in Section~\ref{sec:dpp_mp_ddc}, the TGSS processing pipeline corrects for time- and spatially variable ionospheric phase effects. One of the outputs of the ionospheric modeling is an estimate of the residual phase error per antenna. Typically, these residual phase errors were found to be in the range \adeg{10}--\adeg{20}. Following \citet{2004SPIE.5489..180C}, assuming a Gaussian distribution of phase errors, the resulting reduction in peak flux is estimated to be 3--6~percent, which corresponds to a source broadening of 2--3~percent. 

\begin{figure}
\begin{center}
\resizebox{\hsize}{!}{
\includegraphics[angle=0]{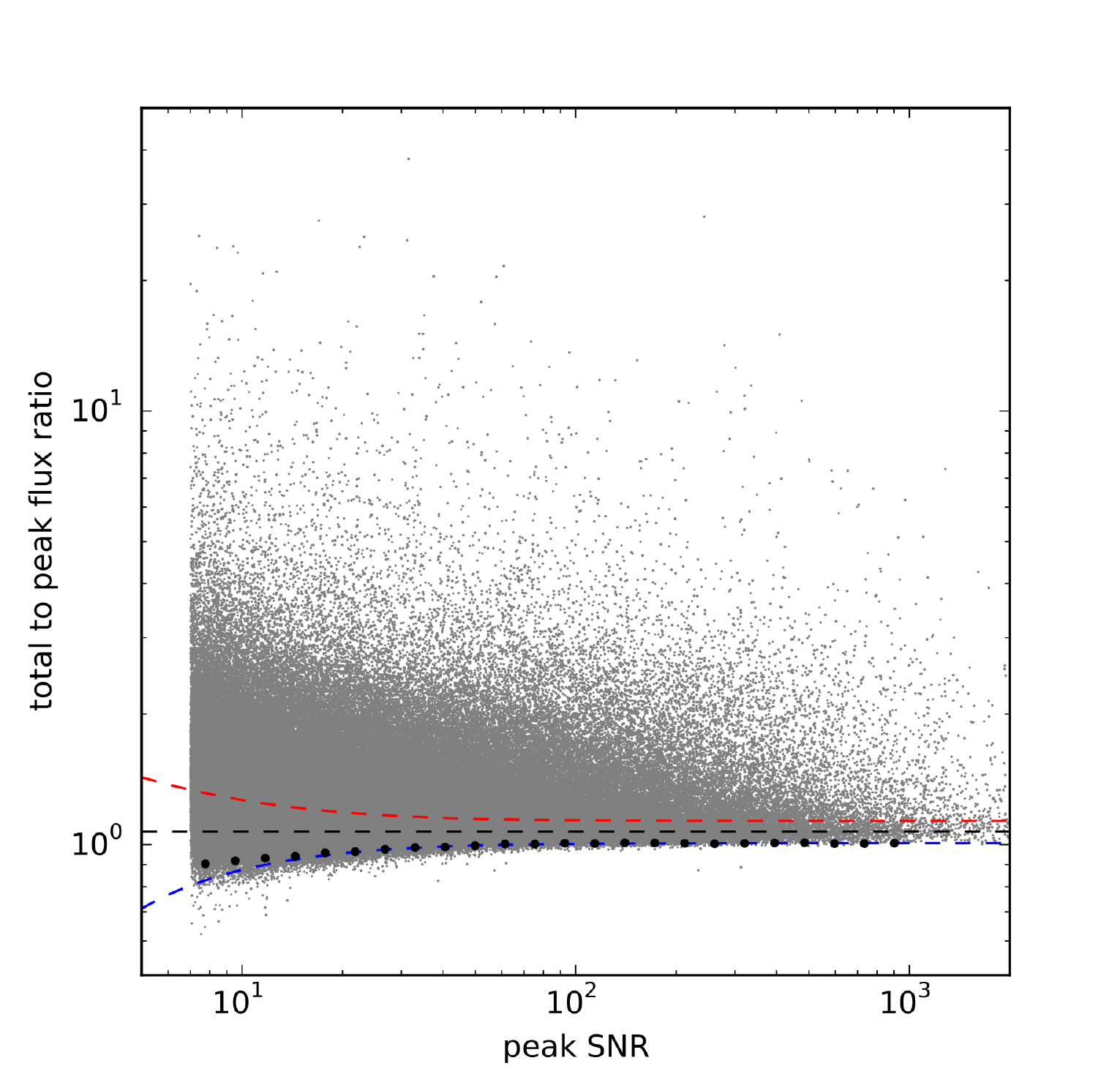}}
\caption{Ratio of the TGSS source (total) flux density over the peak flux a function of peak flux signal-to-noise (SNR). The unresolved source distribution is characterized by a median (dashed black line) slightly above unity and -2 sigma deviations (black dots, measured in logarithmic bins) that increase with lower peak SNR, which is compatible with the effects of smearing (see text). The large scatter in the upper part of the plot are resolved sources, that can be separated from the unresolved sources by the +2~sigma line (dashed red line) based on a model fit (dashed blue line) to the -2~sigma deviations.}
\label{fig:tgss_source_sizes}
\end{center}
\end{figure}

To properly quantify the true smearing of TGSS sources, we analyzed the source properties in the catalog above DEC \adeg{19} (where we used a circular \asec{25} restoring beam). Figure~\ref{fig:tgss_source_sizes} shows for all TGSS sources the ratio of the (total) flux density over the peak flux ($R_\mathrm{ip}$) as a function of peak flux signal-to-noise (SNR$_\mathrm{p}$). For unresolved sources without smearing, we expect per SNR$_\mathrm{p}$ bin a $R_\mathrm{ip}$ distribution centered on unity with a width $\sigma_\mathrm{ip}$ depending solely on SNR$_\mathrm{p}$. Smearing will cause the peak flux to drop while the flux density stays constant, therefore the (median) distribution center $\bar{R}_\mathrm{ip}$ shifts upwards from unity. A fraction of the sources will be resolved, which causes the distribution to have a long tail upwards from the distribution center.

Per logarithmic SNR$_\mathrm{p}$ bin, we estimated the center $\bar{R}_\mathrm{ip}$ and width of the distribution $\sigma_\mathrm{ip}$ using only the lower part of the distribution so not to be biased by resolved sources. We find the center and width by iteratively rejecting sources above $\bar{R}_\mathrm{ip} + 2\,\sigma_\mathrm{ip}$ until  the center and width values converge. At the high SNR$_\mathrm{p}$ end above $\sim 100$, the center and width of the distribution were found to be constant, namely $\bar{R}_\mathrm{ip} = 1.071$ and $\sigma_\mathrm{ip} = 0.027$, respectively. This indicates an average reduction in peak flux of $6.6 \pm 2.4$~percent for all sources above DEC \adeg{19}, which agrees well with the smearing effects described above. The average smearing for sources below DEC \adeg{19} is expected to be less, since the restoring beam size is larger (see Section~\ref{sec:dpp_mos_bm}). 

Towards lower SNR, relatively higher image noise causes a larger scatter in the flux measurements, which creates a widening of the distribution. To enable identification of truly resolved sources, we characterize the widening of the distribution with a power-law by fitting the measured values of $\sigma_\mathrm{ip}$ with the following empirical expression \citep[e.g.,][]{2013A&A...549A..55W}:
\begin{equation}
\sigma_\mathrm{ip}^2 = \left(0.027\right)^2 + \left(0.784 \, \mathrm{SNR}_\mathrm{p}^{-0.925}\right)^2
\label{eq:envelope}
\end{equation}
This expression can be used to assess whether sources are resolved. For instance, with $R_\mathrm{ip} > \bar{R}_\mathrm{ip} + 2\,\sigma_\mathrm{ip}$ there is a $\sim 98$~percent chance that a source is resolved, which is true for about 50~percent of all sources above DEC \adeg{19}. Below DEC \adeg{19} this is about 48~percent, but here we note that a proper assessment is complicated by the varying restoring beam size.


\subsection{Global Astrometric Accuracy: Comparison to RFC}
\label{sec:sp_gaa}

\begin{figure}
\begin{center}
\resizebox{\hsize}{!}{
\includegraphics[angle=0]{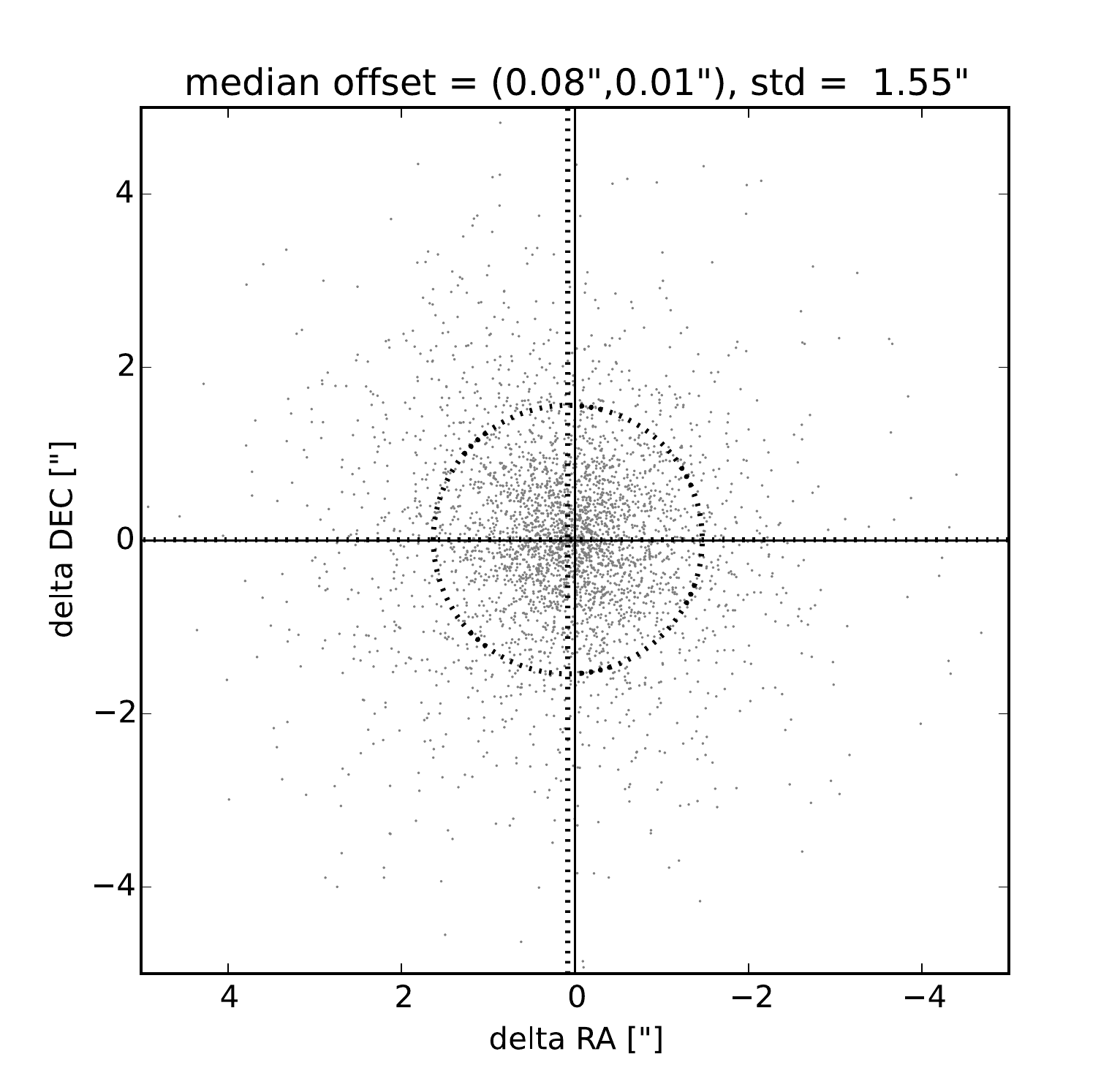}}
\caption{Offsets between VLBA calibrator positions and their TGSS counterparts. The dotted lines indicate the position of the median offset in RA and DEC, while the dotted circle indicates the extend of the standard deviation of the offset radii.}
\label{fig:absolute_astrometry}
\end{center}
\end{figure}

In Section~\ref{sec:dpp_mos_ac} we made an initial estimate of the astrometric accuracy of our catalog, using position differences of multiply observed sources detected in overlapping fields. This test shows that the positions of even faint TGSS sources are accurate to $\lesssim{2}^{\prime\prime}$ (see Figure~\ref{fig:relative_astrometry}). To better quantify any systematic deviations from our astrometric reference frame, we carried out a second test using the Radio Fundamental Catalog (RFC\footnote{\url{http://astrogeo.org/rfc/}}). The RFC is a collection of nearly 10,000 radio sources with positions  measured from numerous astrometric and geodetic campaigns accurate to several milliarcseconds 
\cite[e.g.,][]{2011AJ....142...89P}. 
The sky coverage is good; there are 8573 of 9120 RFC sources down to the declination limit of the TGSS. We cross-matched the RFC catalog for TGSS counterparts and determined the position offsets for each of the 3530~matches. The results are shown in Figure~\ref{fig:absolute_astrometry}. The average offset of the delta-RA and delta-DEC positions is small, with limits delta-RA$\simeq{0.08}^{\prime\prime}$ and delta-DEC$\simeq{0.01}^{\prime\prime}$, for any systematic offset of the survey from the radio reference frame. The 68 and 90~percent confidence error circles of the RA and DEC differences is \asec{1.55} and \asec{2.6}, respectively. 

This should be considered an upper limit on the errors. While the sources in the RFC catalog are compact at VLBI baselines and GHz frequencies, faint, extended steep-spectrum emission may dominate AGN morphology at 150~MHz. To counteract any centroid shifts from the core to the lobes, we used only bright TGSS sources ($>100$~mJy) without multiple components. Summarizing, the global astrometric accuracy of the TGSS is excellent with systematic errors of \asec{$\lesssim 0.1$} and random errors \asec{$\lesssim 2$}. We quadratically added a \asec{2} contribution to our catalog source position uncertainties in both RA and DEC to account for this global effect.



\subsection{Flux Density Accuracy: Comparison to 7C}
\label{sec:sp_7c}

Our flux density scale is tied to the low-frequency point source models of 
\citet[][Section~\ref{sec:dpp_pp}]{2012MNRAS.423L..30S}. 
Except for 3C\,286, we reproduce the flux density of the flux calibrators within a few percent. Given the various amplitude corrections introduced  in the data processing (e.g., system temperature, pointing errors, primary beam), it is important to cross check the consistency of our flux scale against a large group of sources over a large sky area.

\begin{figure}
\begin{center}
\resizebox{\hsize}{!}{
\includegraphics[angle=0]{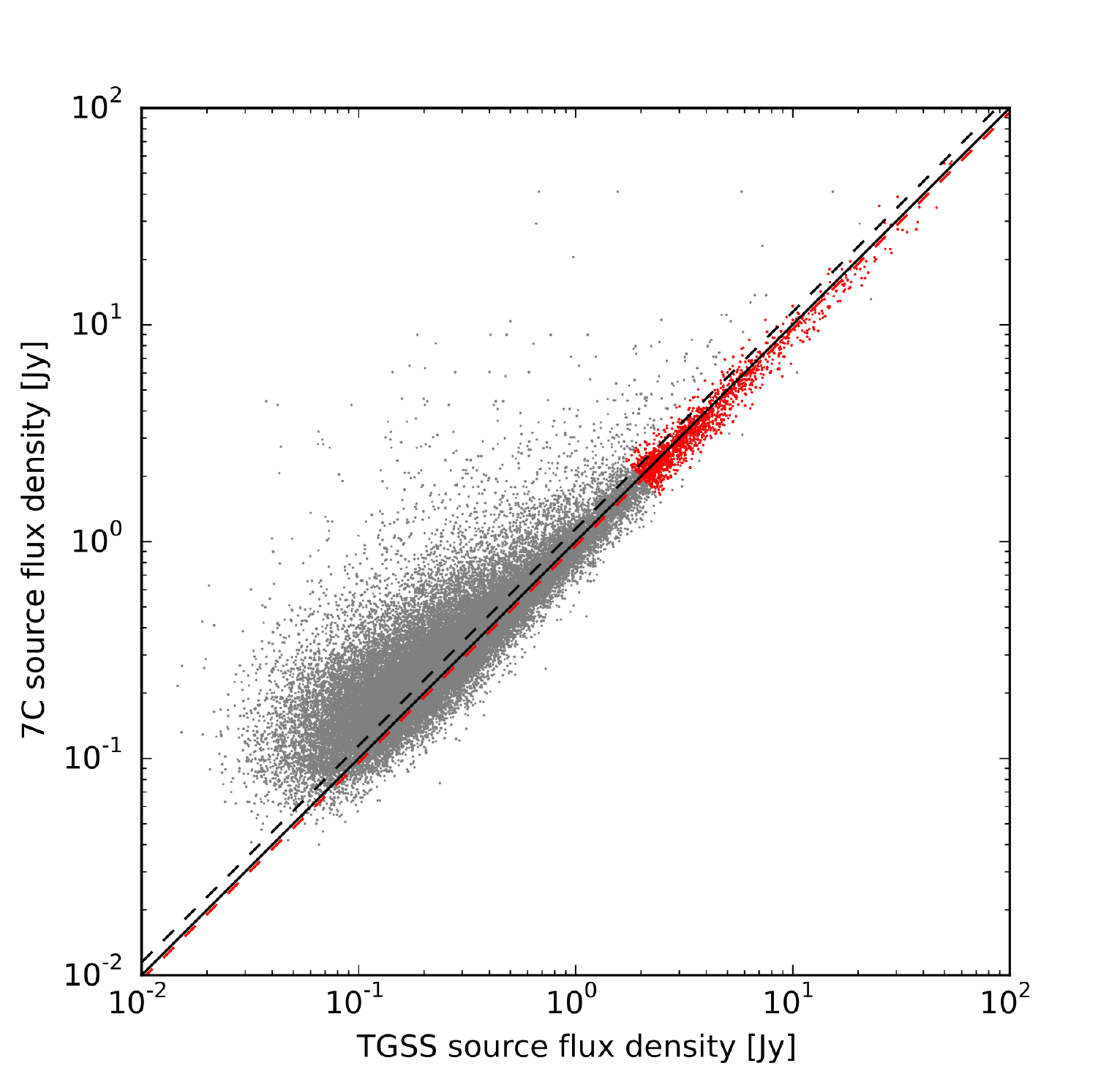}}
\caption{Flux density comparison for 41,605 matched sources between TGSS and 7C (grey dots). Without correcting for difference in resolution, the median 7C to TGSS flux ratio (black dashed line) is biased upwards. When selecting a higher flux sample and applying outlier rejection (red dots), the median flux ratio (red dashed line) reduces to close to unity.}
\label{fig:adr1vs7c_totalflux}
\end{center}
\end{figure}

The 7C survey was conducted with the Cambridge Low Frequency Synthesis Telescope at a frequency of 151~MHz with resolution \sbeam{70}{70}$/\sin({\mathrm{DEC}})$. Over a region of 1.7~sr north of \adeg{30} DEC, 43,683 radio sources were cataloged down to a typical threshold of 0.1~Jy. The 7C flux scale is \changetwo{indirectly} tied to
\citet{1973AJ.....78.1030R}, 
which is the same flux scale that
\citet{2012MNRAS.423L..30S} 
is based on. \changetwo{The accuracy of the flux density scale of 7C is reported to be $\sim 5$~percent relative to 6C
\citep[e.g.,][]{1995MNRAS.276..614L,1995A&AS..110..419V}, 
which itself is believed to have a $\sim 10$~percent absolute accuraty
\citep{1985MNRAS.217..717B}. 
The absolute accuracy of the 7C flux scale is therefore likely to be in the range of 10--20~percent.}
Following 
\citet{2015A&A...582A.123H} 
we compared the TGSS against the combined 7C catalog from
\citet{2007MNRAS.382.1639H}. 

While 7C is not an absolute reference with respect to flux density accuracy, it is one of very few large-area surveys available at virtually the same frequency. We cross-matched the TGSS and 7C catalogs over the common area using a matching radius of \asec{70}, yielding 41,605~\changetwo{unique} matches (95~percent of the 7C sources). The results of this source match are shown in Figure \ref{fig:adr1vs7c_totalflux}. 

The effect of the resolution bias of the larger 7C beam is clearly visible especially at lower flux densities. When including all matches, the median 7C flux density is 14~percent higher than TGSS, but this result is strongly affected by resolution bias. When using a subset of brighter sources (above 2~Jy) and outlier rejection, we find that the median 7C to TGSS flux density ratio is 0.95, which is close to the perfect case of unity. The magnitude of the deviation from unity ($\sim 5$~percent) is similar to the variation in recovered flux densities of the
\citet{2012MNRAS.423L..30S} 
calibrators. To account for these, we quadratically added a 10~percent flux scale uncertainty to the flux uncertainties (peak and total) in our catalog.

\changetwo{In Section~\ref{sec:dpp_mos_fdc} we discussed the system temperature corrections and how they introduce flux scale uncertainties depending on the magnitude of the correction. For the GMRT at 150~MHz, the sky temperature is the major contributer to the system temperature. It varies significantly across the sky and therefore so does the flux scale correction and its uncertainty. For TGSS ADR1, this uncertainty is absorbed into the global 10~percent flux scale uncertainty. We plan to improve on this and include sky-dependent source flux density uncertainties in subsequent data releases.}


\subsection{Comparison to \changetwo{MSSS HBA}}
\label{sec:sp_msss}

\begin{figure*}[!ht]
\begin{center}
\resizebox{\hsize}{!}{
\includegraphics[angle=0]{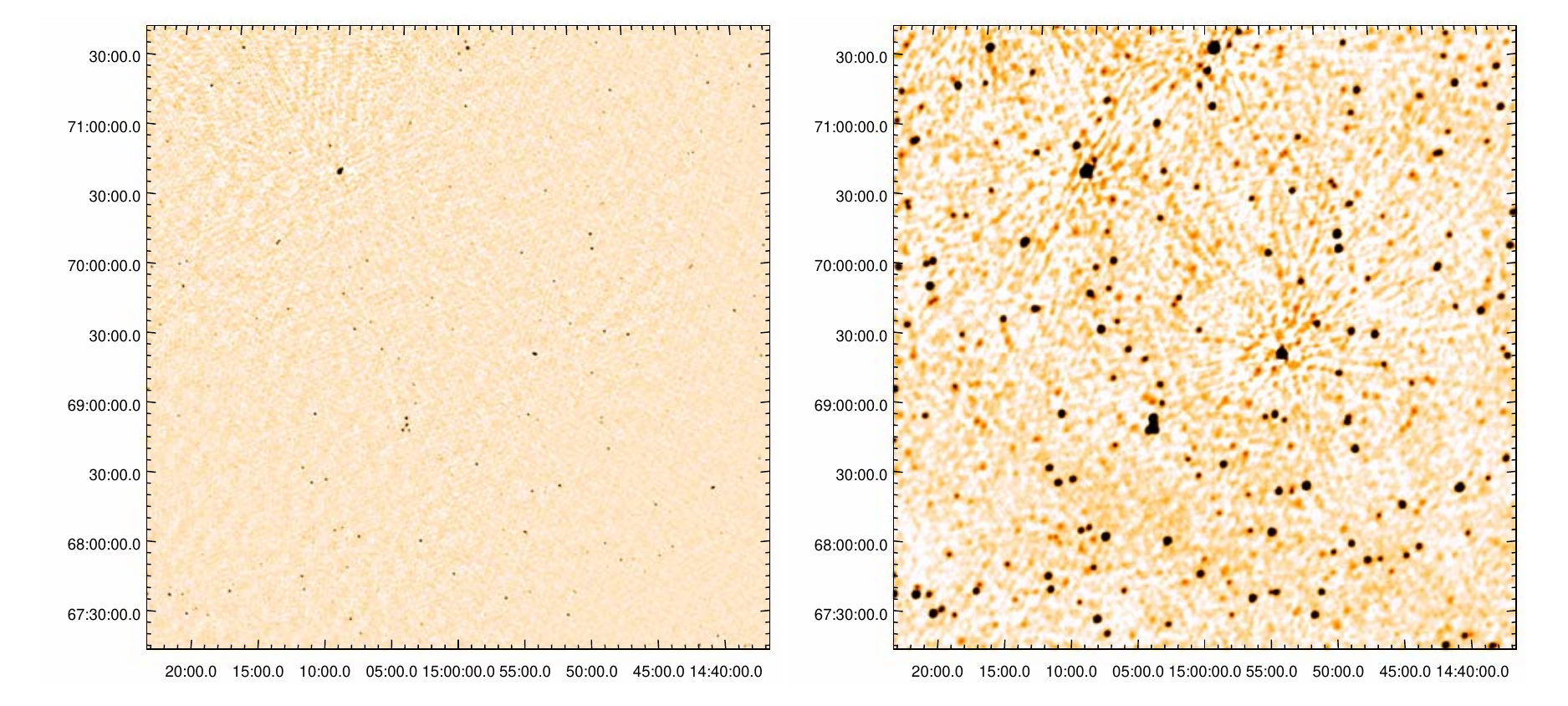}}
\resizebox{\hsize}{!}{
\includegraphics[angle=0]{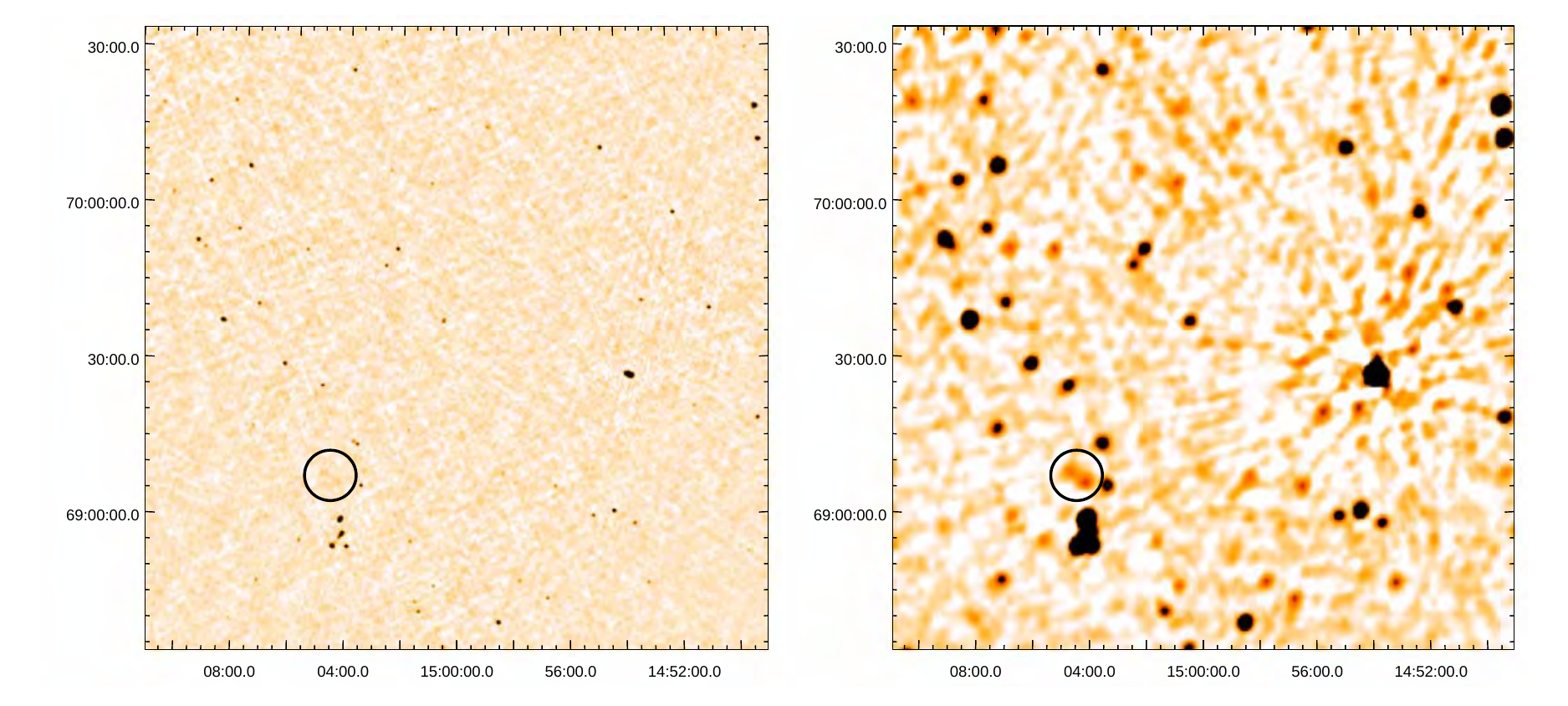}}
\caption{\emph{Top row:} Example \pbeam{4.5}{4.5} mosaic image centered on the TGSS pointing R23D74 as found in ADR1 (\emph{left}; \mjybeam{3.6} noise for a \sbeam{25}{25} beam), and the corresponding area in the frequency-averaged LOFAR HBA MSSS verification field (MVF \emph{right}; \mjybeam{6.4} noise for a \sbeam{108}{108} beam). Both images are displayed using the same color scale. \emph{Bottom row:} A \pbeam{2}{2} zoomed-in version of the top row images, showing more detail. The ADR1 and MVF images are quite complementary, with the MVF image demonstrating good low surface brightness sensitivity, while the TGSS image demonstrates its resolving power and lack of sidelobe structure near the brightest sources. \changetwo{The black circle marks the location of a double source with low surface brightness, illustrating the difference in detectability of such sources in the two surveys.}}  
\label{fig:adr1vsmsss_images}
\end{center}
\end{figure*}

To get a sense of the relative performance of two new low-frequency surveys, we compare our TGSS results against the small HBA part of the LOFAR multi-snapshot sky survey (MSSS) that is publicly available. The MSSS verification field (MVF) image includes data from $8 \times 2$~MHz subbands spaced out over 119--158~MHz (filling factor of 0.41). Although planned for later releases, the current MSSS data processing pipeline does not yet include direction-dependent calibration. For this and other reasons the MSSS images are generated from a subset of shorter baseline data, resulting for the HBA part in a resolution of \asec{108}
\citep{2015A&A...582A.123H}. 
The frequency-averaged, most sensitive MVF image available is a \pbeam{10}{10} image centered on \thour{15}~\adeg{+69}, and has an effective frequency of $\sim 138.4$~MHz. The small difference in frequency between TGSS and MSSS may cause up to a few percent difference in flux density. We created a matching TGSS mosaic at \sbeam{25}{25} resolution, of which parts are shown in Figure~\ref{fig:adr1vsmsss_images} together with the MVF image. Additionally, we created a low-resolution mosaic by convolving and regridding the TGSS mosaic to match the MSSS resolution and pixel grid (`TGSSc' from here on).

\ctable[botcap,center,star,
caption = {Properties of the source extraction based comparison between TGSS and \changetwo{MSSS HBA}.},
label = tab:tgss_vs_msss
]{l c c c}{
}{
\FL Property & TGSS & TGSSc & MSSS
\ML Average RMS noise & \mjybeam{5.3} & \mjybeam{28} & \mjybeam{7.3}
\NN Total source flux & 580~Jy & 523~Jy & 622~Jy
\NN Source detections & 2198 & 724 & 1591
\NN False detections & 49 & 2 & 1
\NN NVSS unique matches & 1988 & 719 & 1561
\NN NVSS no matches & 123 & 5 & 30
\LL}

The average image noise properties given in Table~\ref{tab:tgss_vs_msss} \changetwo{(as measured over the full MVF area)} show that TGSS has a point source sensitivity that is $\sim 30$~percent better than MSSS, while the surface brightness sensitivity is $\sim 4$ times worse. The latter is illustrated in Figure~\ref{fig:adr1vsmsss_images}, where faint, low surface brightness features in TGSS are \changetwo{more} clearly visible in MSSS. 

\changetwo{The PyBDSM source extraction was run on the three images using the standard settings, which includes multi-Gaussian source extraction down to the threshold of 5 times the local peak-to-noise ratio. We intentionally use a lower threshold here (and also in Section~\ref{sec:sp_dr5}) than for the source catalog generation (see Section~\ref{sec:sp_se}) to enable a deeper quality comparison between the surveys in terms of of reliability and completeness.} The results in Table~\ref{tab:tgss_vs_msss} are hardly surprising: most sources (or source components) are detected in the TGSS image because of its higher resolution, while MSSS retrieves most flux density because of its better low surface brightness sensitivity. It is interesting to notice that, despite the significantly worse sensitivity, TGSSc still retrieves 80--90~percent of the total flux density as compared to TGSS and MSSS.

\begin{figure}
\begin{center}
\resizebox{\hsize}{!}{
\includegraphics[angle=0]{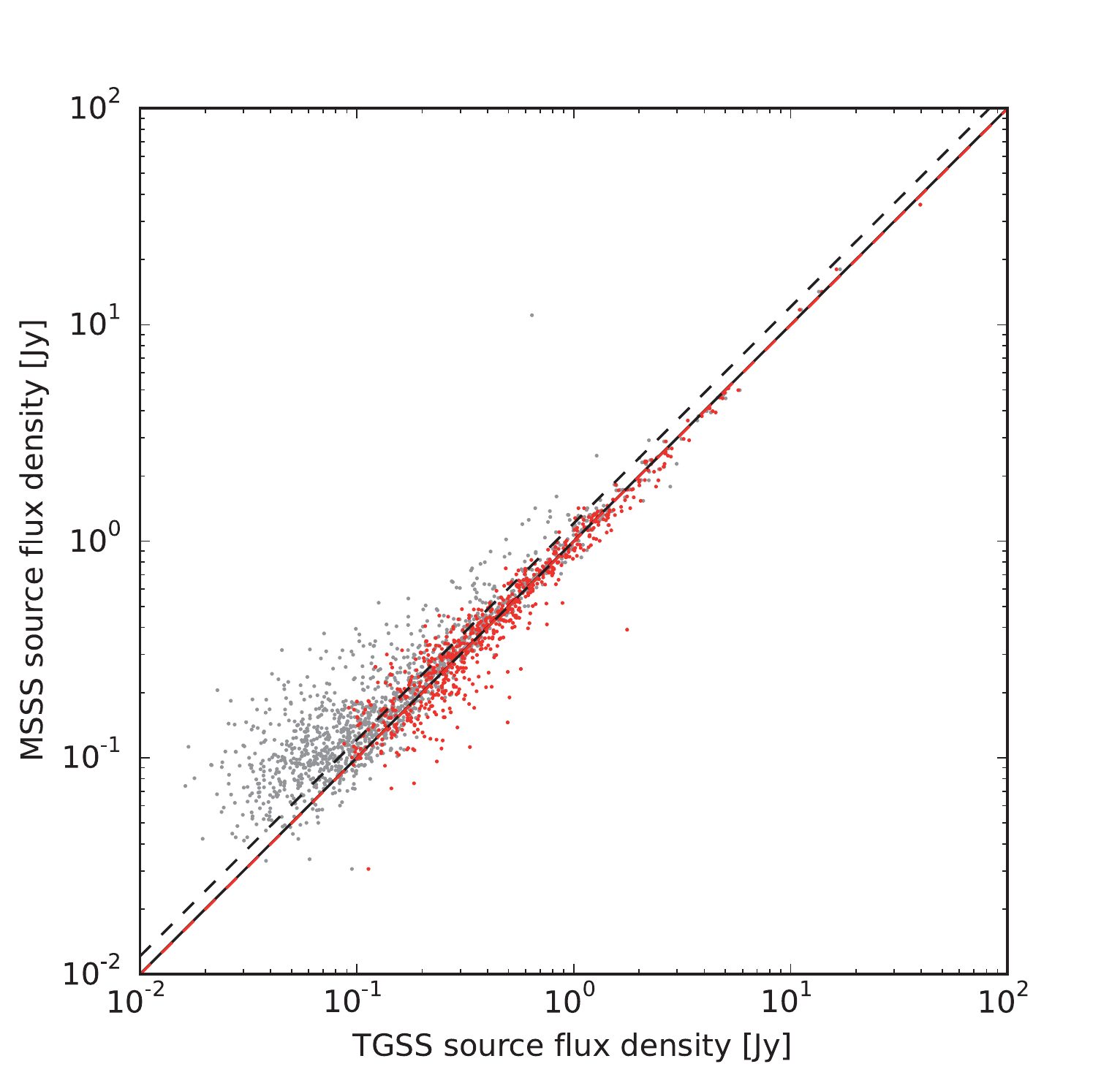}}
\caption{Flux density comparison for 1,437 matched sources between TGSS and LOFAR MSSS HBA (grey dots). Due to the difference in resolution, the median MSSS to TGSS flux ratio (black dashed line) is biased upwards. When convolving the TGSS images to the MSSS resolution and repeating the matching (712~matches; red dots), the median MSSS to TGSSc flux ratio (red dashed line) reduces to close to unity. Flux densities for matched sources between TGSS and LOFAR HBA MSSS. }
\label{fig:adr1vsmsss_totalflux}
\end{center}
\end{figure}

We uniquely matched the TGSS source catalog to the MSSS catalog using a \asec{108} search radius, yielding 1437~matches (90~percent of the MSSS sources). The measured flux densities as plotted in Figure~\ref{fig:adr1vsmsss_totalflux} show an overall good match for many sources, but are biased towards higher MSSS flux density because of resolution differences. The median flux density ratio of 1.21 in favor of MSSS reduces to near-unity (0.99) when comparing flux densities of matched TGSSc and MSSS sources (712 matches, or 98~percent of the TGSSc sources). 

As in Section~\ref{sec:sp_dr5}, we ran the source extraction process on the inverted images to test the reliability. 

The results in Table~\ref{tab:tgss_vs_msss} show a remarkably low number of false detections in the MSSS and TGSSc images ($\ll 1$~percent). The fraction of false detections in the TGSS image is 2.2~percent, somewhat higher than what we found in Section~\ref{sec:sp_dr5}, but still a minor contribution to the total number of sources detected down to the detection threshold of 5 times the local noise. We also matched the catalogs against NVSS, using a \asec{45} search radius for the TGSS catalog and \asec{108} for the TGSSc and MSSS catalogs. The results in Table~\ref{tab:tgss_vs_msss} show that 5.6~percent of TGSS sources have no NVSS counterpart, while for TGSSc and MSSS it is 0.7 and 1.9~percent, respectively. 

Summarizing, we find a very good match in flux density between the TGSS and LOFAR MSSS HBA. TGSS has better resolution and point source sensitivity, while MSSS has better surface brightness sensitivity. Both surveys are complementary in resolution, yet well matched in sensitivity. A similar comparison between TGSS images and MWA images is underway (Tingay et al., in prep.).


\subsection{Comparison to TGSS DR5}
\label{sec:sp_dr5}

\begin{figure*}[!ht]
\begin{center}
\resizebox{\hsize}{!}{
\includegraphics[angle=0]{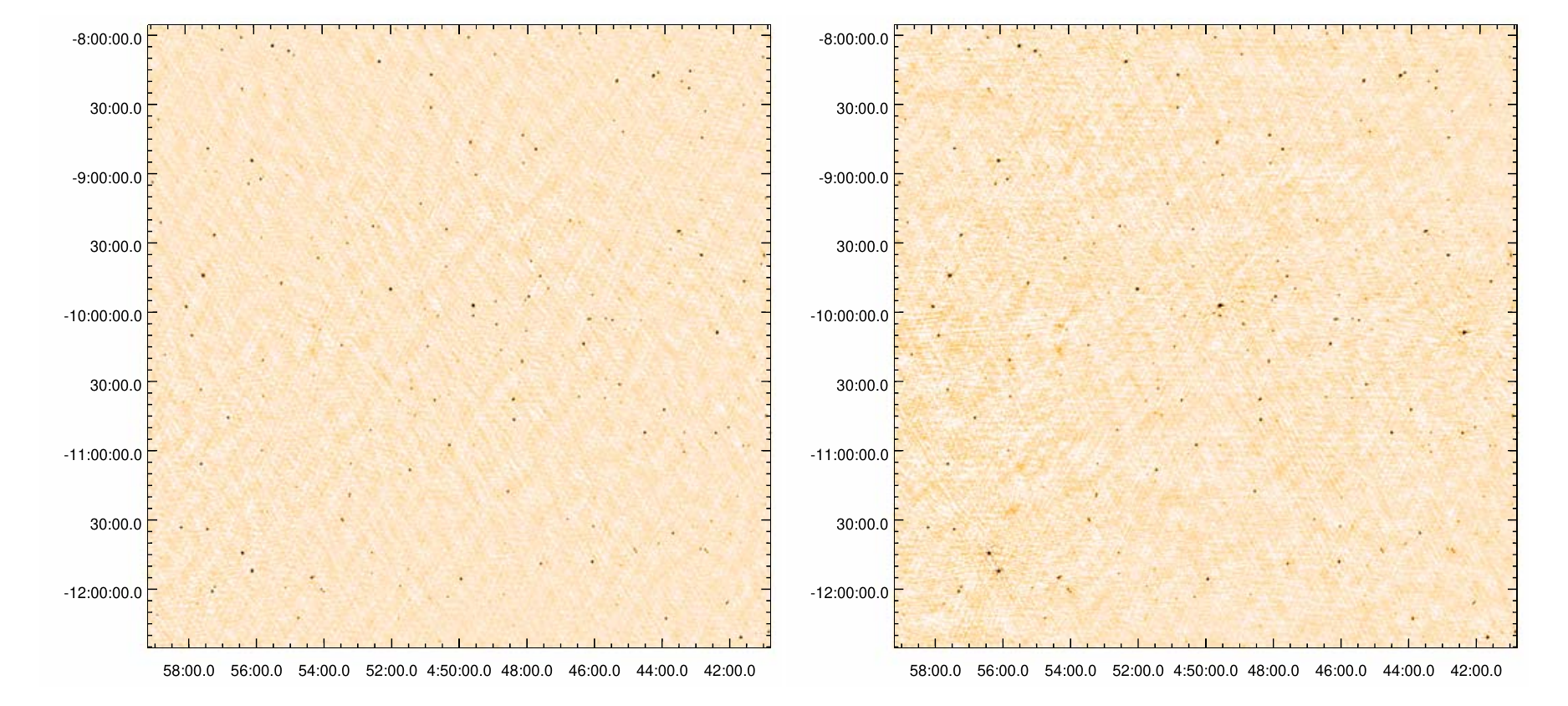}}
\resizebox{\hsize}{!}{
\includegraphics[angle=0]{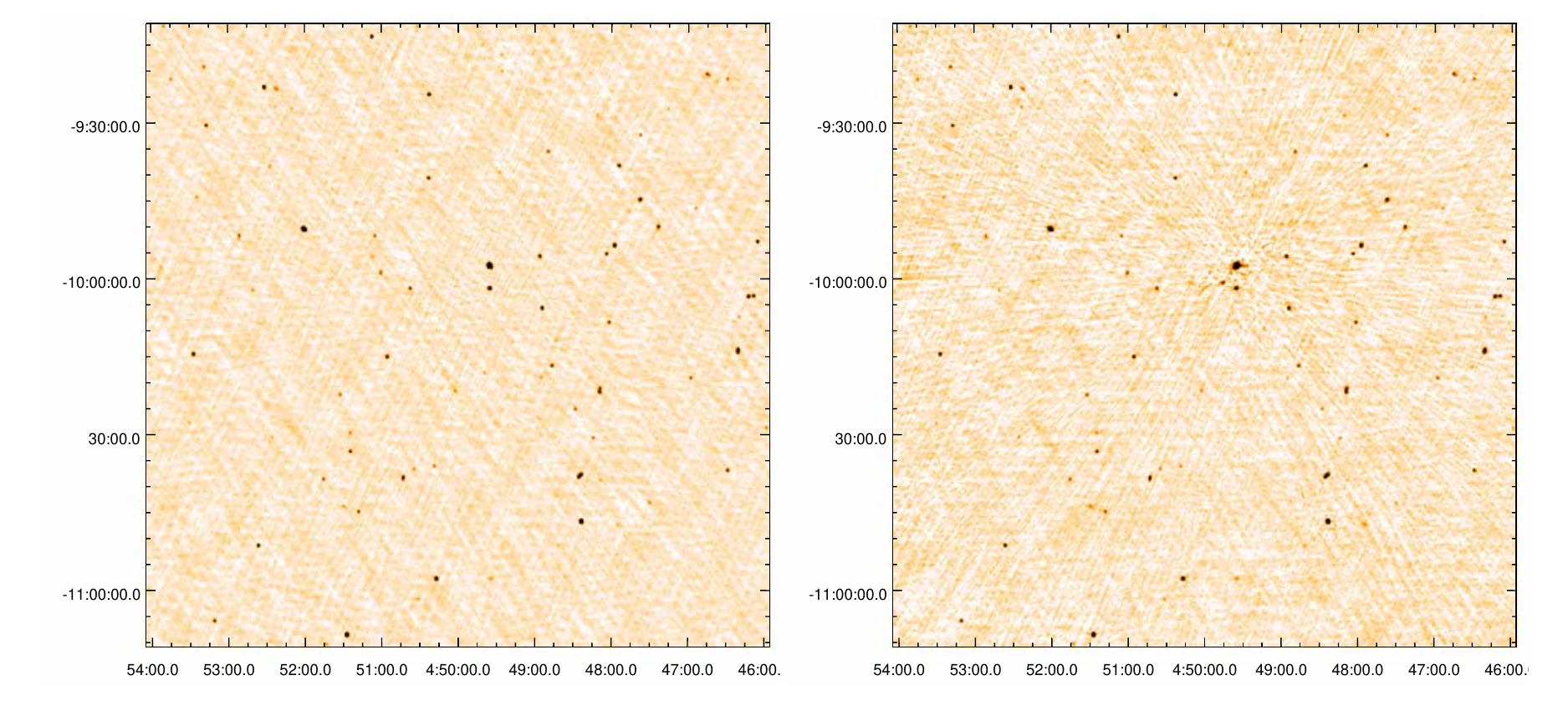}}
\caption{\emph{Top row:} Example \pbeam{4.5}{4.5} \changetwo{mosaics} centered on the TGSS pointing R15D26 as found in ADR1 (\emph{left}; \mjybeam{3.5} noise) and in DR5 (\emph{right}; \mjybeam{6.7} noise). Both \changetwo{mosaics} are displayed using the same color scale. Besides a higher overall noise level, the DR5 \changetwo{mosaic} also shows residual sidelobe structure near brighter sources ($\gtrsim 1$~Jy) as compared to the ADR1 \changetwo{mosaic}. \emph{Bottom row:} A \pbeam{2}{2} zoomed-in version of the top row \changetwo{mosaics}, showing more detail.}  
\label{fig:adr1vsdr5_images}
\end{center}
\end{figure*}

Our re-processing effort of the archival TGSS data was motivated by having available a robust and fast pipeline that includes direction-dependent ionospheric calibration. Especially the latter is essential for processing radio interferometry data at frequencies below a few hundred MHz in order to obtain good image fidelity.
\changetwo{The pipeline used for producing the original TGSS} data releases 1 through 5 does not include direction-dependent calibration. In this section we investigate the relative performance of the TGSS DR5 pipeline versus our (ADR1) pipeline by inspecting the resulting image quality.

To facilitate a detailed, objective comparison between ADR1 and DR5, we selected a connected region of 32~pointings with the following properties: (i) \changetwo{the DR5 mosaics centered on the pointing positions} are all publicly available\footnote{\url{http://tgss.ncra.tifr.res.in/150MHz/radec.html}}, (ii) \changetwo{the DR5 mosaics are created from a complete set of overlapping pointing images, ensuring maximum sensitivity}, (iii) \changetwo{the region contains} no excessively bright sources, and (iv) all the pointings were observed only once therefore the images result from the exact same visibility data. The \changetwo{set of mosaics covers} an approximate rectangular $\sim 300$~deg$^2$ area centered on \thour{4}\tmin{50}~\adeg{-6} with approximate dimensions of \pbeam{15}{20}. We convolved and regridded the selected DR5 \changetwo{mosaics} to match the ADR1 resolution and pixel grid, and blanked the outer edge of the ADR1 \changetwo{mosaics} to match the DR5 mosaic size (\pbeam{4.5}{4.5}). At this stage the pointing images are as equal as possible. 

We started the comparison by measuring the \changetwo{central background RMS noise in all mosaics}. The average RMS values listed in Table~\ref{tab:adr1_vs_dr5} match the general noise properties of ADR1 and DR5 as advertized, and can therefore be considered a representative sample. For this subset, the ADR1 noise levels are 10--50~percent lower than the DR5 noise levels.

\changetwo{Similar to Section~\ref{sec:sp_msss} we ran the PyBDSM source extraction in default mode on all prepared ADR1 and DR5 mosaics}. The resulting catalogs per \changetwo{mosaic} were merged into single, unambiguous catalogs for ADR1 and DR5, respectively. Table~\ref{tab:adr1_vs_dr5} shows that the number of extracted sources is virtually equal between ADR1 and DR5, although the total source flux in DR5 is about 9~percent higher than ADR1. 
\changetwo{Merely based on noise levels, we would expect for the ADR1 images to yield more extracted sources, which is not what we observe. This can only be caused by flux scale differences between DR5 and ADR1, or by false source detections in DR5 (or a combination of both).}

\ctable[botcap,center,
caption = {Properties of the source extraction based comparison between ADR1 and DR5.},
label = tab:adr1_vs_dr5
]{l c c}{
}{
\FL Property & ADR1 & DR5
\ML Average RMS noise & \mjybeam{3.6} & \mjybeam{6.0}
\NN Total source flux & 1484~Jy & 1615~Jy
\NN Source detections & 6573 & 6443
\NN False detections & 17 & 566
\NN NVSS unique matches & 6138 & 5384
\NN NVSS no matches & 204 & 786
\LL}

\begin{figure}
\begin{center}
\resizebox{\hsize}{!}{
\includegraphics[angle=0]{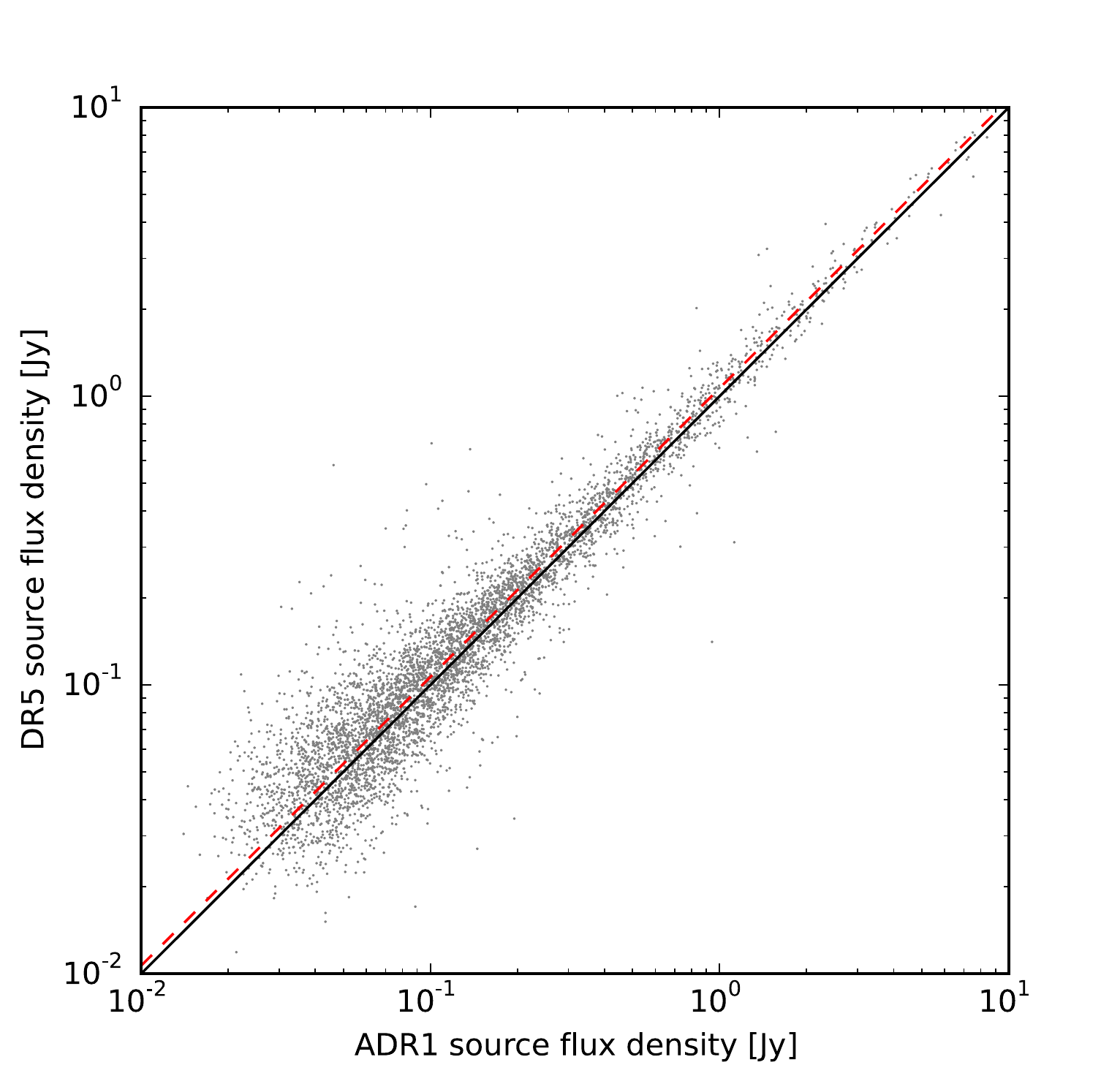}}
\caption{Flux density comparison for 5,288 matched sources between TGSS ADR1 and DR5 (grey dots). The median DR5 to ADR1 flux ratio (red dashed line) is very close to unity.}
\label{fig:adr1vsdr5_totalflux}
\end{center}
\end{figure}

We check for potential flux scale differences by uniquely matching the ADR1 and DR5 catalogs with a \asec{25} search radius, yielding 5288~matches. Figure~\ref{fig:adr1vsdr5_totalflux} shows that there is an overall good agreement between the source flux densities. We find a median DR5 to ADR1 flux density ratio of 1.07. Differences between ADR1 and DR5 for the choice of flux scale and \temp{sys}{} corrections can easily give rise to a systematic difference of this (small) magnitude
\citep[e.g.,][]{2009MNRAS.398..853S,2012MNRAS.423L..30S}. 
\changetwo{Although little information available in literature on the DR5 processing details, they most likely used the now deprecated Perley--Taylor 1999 flux scale as was done in 
\citet{2009MNRAS.392.1403S}, 
which itself is based on the
\citet{1977A&A....61...99B} 
flux scale. The observed flux scale difference} is too small to explain the apparent discrepancy between noise levels and extracted sources. And although it has a very minor impact on the rest of this comparison, we scale the ADR1 fluxes by 1.07 to better match the DR5 fluxes.

There is a relatively large number of unmatched sources ($\sim 1200$) between ADR1 and DR5, about 18~percent of the total source count in both ADR1 and DR5. This does not change when using a matching radius twice as large, indicating these detections are uncorrelated. To investigate the reliability of the source detections we performed two tests. In the first test we estimated the false detection rate per image by following the same procedure as described in Section~\ref{sec:sp_cr}, and combined the results into single catalogs for ADR1 and DR5.The results given in Table~\ref{tab:adr1_vs_dr5} illustrate a rather striking difference: for ADR1 the estimated false detection fraction is 0.3~percent, while for DR5 it is 8.8~percent. 

\begin{figure}
\begin{center}
\resizebox{\hsize}{!}{
\includegraphics[angle=0]{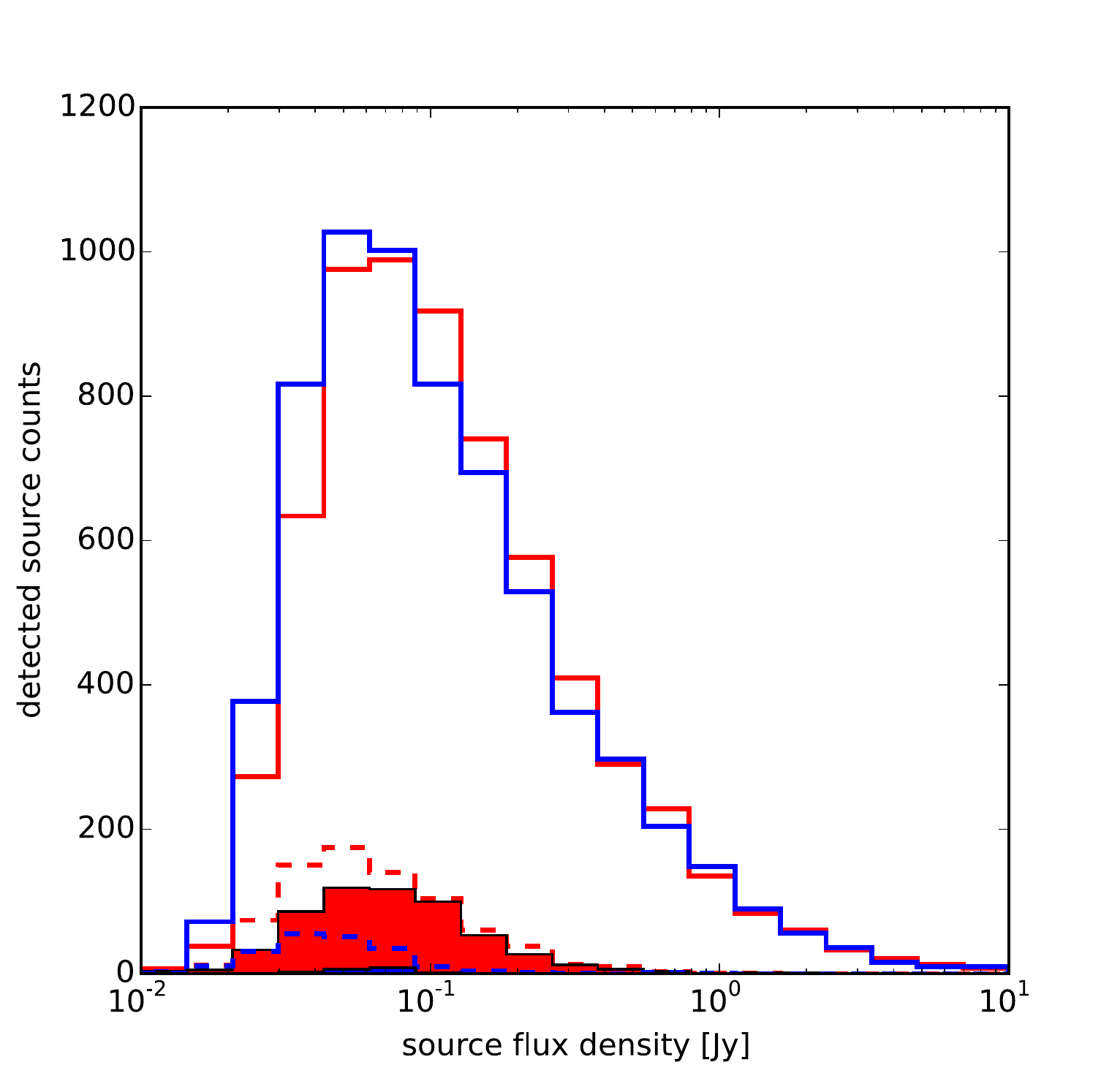}}
\caption{The flux density distribution of (solid blue) ADR1 and DR5 (solid red) TGSS sources using \changetwo{identical mosaics} and source extraction method. The dashed lines show the flux density distribution of the same sources with {\it no} NVSS counterpart, and the solid red and blue histograms show the results of source-finding on the inverted images (i.e. negative).}
\label{fig:adr1vsdr5_counts}
\end{center}
\end{figure}

Figure~\ref{fig:adr1vsdr5_counts} shows the distribution of detected sources as a function of source flux density. For both ADR1 and DR5 the number of detected sources increases towards lower flux density due to the nature of radio source counts, where DR5 counts appear to increase more rapidly than ADR1 count. This may be explained by the similarly increasing number of false detections in DR5. Below 100~mJy, the source counts of both ADR1 and DR5 decrease due to incompleteness (see also Section~\ref{sec:sp_cr}). Here, DR5 counts decrease more rapidly than ADR1 counts, despite the continuing presence of false detections in this flux density range. This shows that ADR1 detects more faint sources than DR5, as is expected given the lower average background noise.

In the second test we separately matched the ADR1 and DR5 catalogs to the NVSS catalog using a \asec{45} search radius (the NVSS resolution), allowing for multiple matches of TGSS sources to NVSS sources to overcome resolution differences. For unresolved sources with spectral index $\alpha \gtrsim -0.9$ (with $S_\nu \propto \nu^\alpha$) the NVSS is more sensitive than TGSS. For resolved sources, the better surface brightness sensitivity of NVSS makes it more sensitive than TGSS for emission with spectral index $\alpha \gtrsim -1.4$. Therefore we expect the majority of TGSS sources to have an NVSS counterpart. From the results in Table~\ref{tab:adr1_vs_dr5} we see that 12~percent the DR5 sources have no NVSS counterparts versus 3.1~percent for the ADR1 sources. With a roughly equal number of detected sources in ADR1 and DR5, this strongly suggests that the higher number of NVSS non-matches in DR5 relative to ADR1 is due to false source detections in DR5. Note that the excess of NVSS non-detections (582) is roughly similar to the excess in false detections (549), which could be the same population. Figure~\ref{fig:adr1vsdr5_counts} shows the distribution of TGSS sources without an NVSS counterpart as a function of source flux density.

The most likely cause of false detections is the presence of significant residual sidelobe structure around sources due to calibration errors, which is common in wide-field low-frequency radio interferometry images (e.g., see Figure~\ref{fig:adr1vsdr5_images}). Visual inspection of the \changetwo{mosaics} and the detected source locations show that indeed many DR5-specific source detections occur in the near-vicinity of bright sources. These may be suppressed by fine-tuning the source extraction process, but will always negatively impact the completeness of the source catalog. Summarizing, we find strong indications that the fidelity of DR5 images is negatively affected by the absence of direction-dependent calibration in the processing pipeline.


\subsection{Normalized Source Counts and Spectral Indices}
\label{sec:sp_scsi}

\begin{figure}
\begin{center}
\resizebox{\hsize}{!}{
\includegraphics[angle=0]{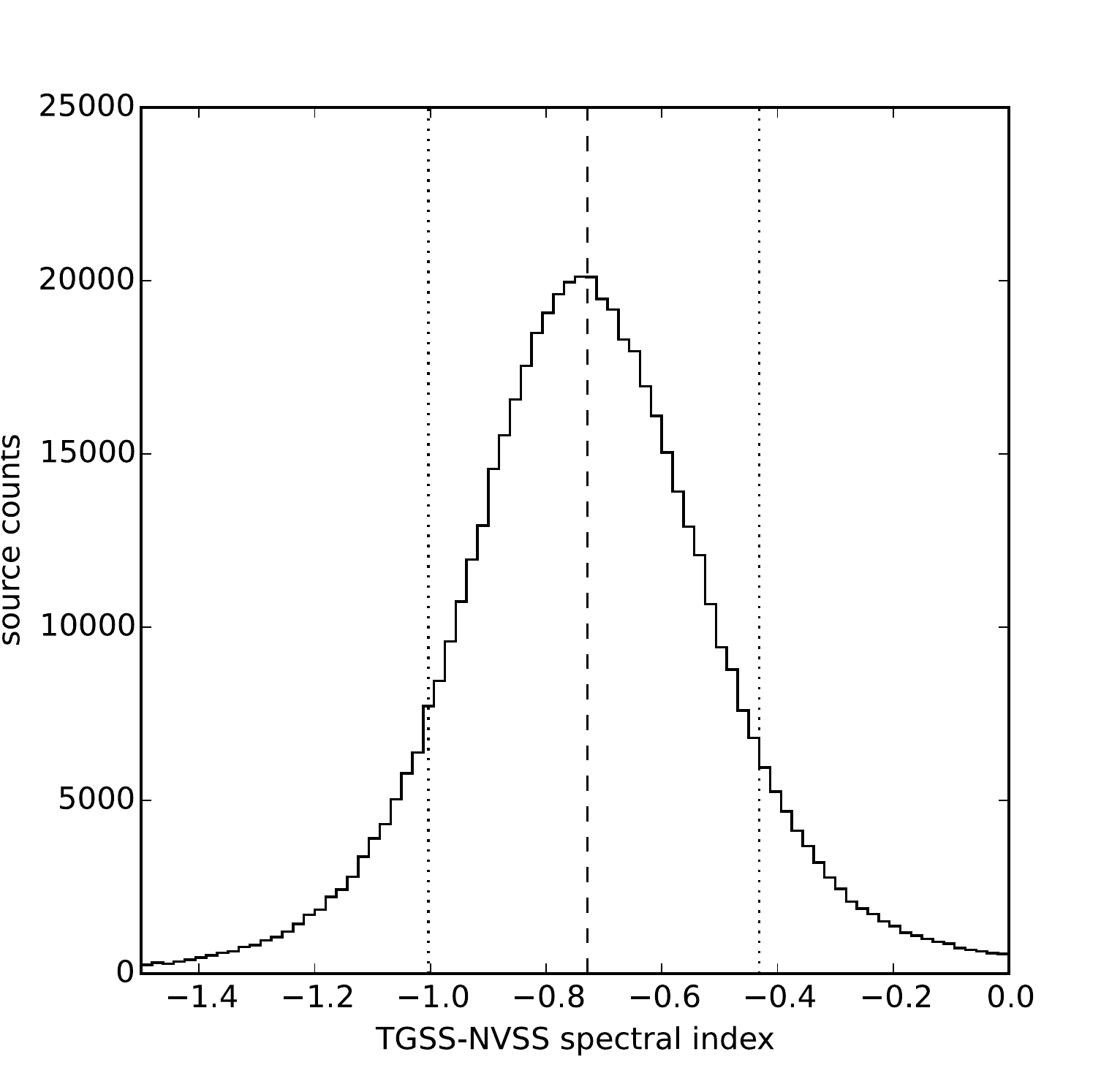}}
\caption{Spectral index histogram for over 550,000~matched sources between NSSS and TGSS using a \asec{22.5} search radius (the NVSS beam radius). The median spectral index is \changeone{-0.73} (dashed line). The top and bottom 10~percent of the sources lie above \changeone{-0.43} and below \changeone{-1.00}, respectively (dotted lines). Note that this source match is not corrected for any selection effects.}
\label{fig:tgss_nvss_spix}
\end{center}
\end{figure}

\begin{figure*}[!ht]
\begin{center}
\resizebox{\hsize}{!}{
\includegraphics[angle=0]{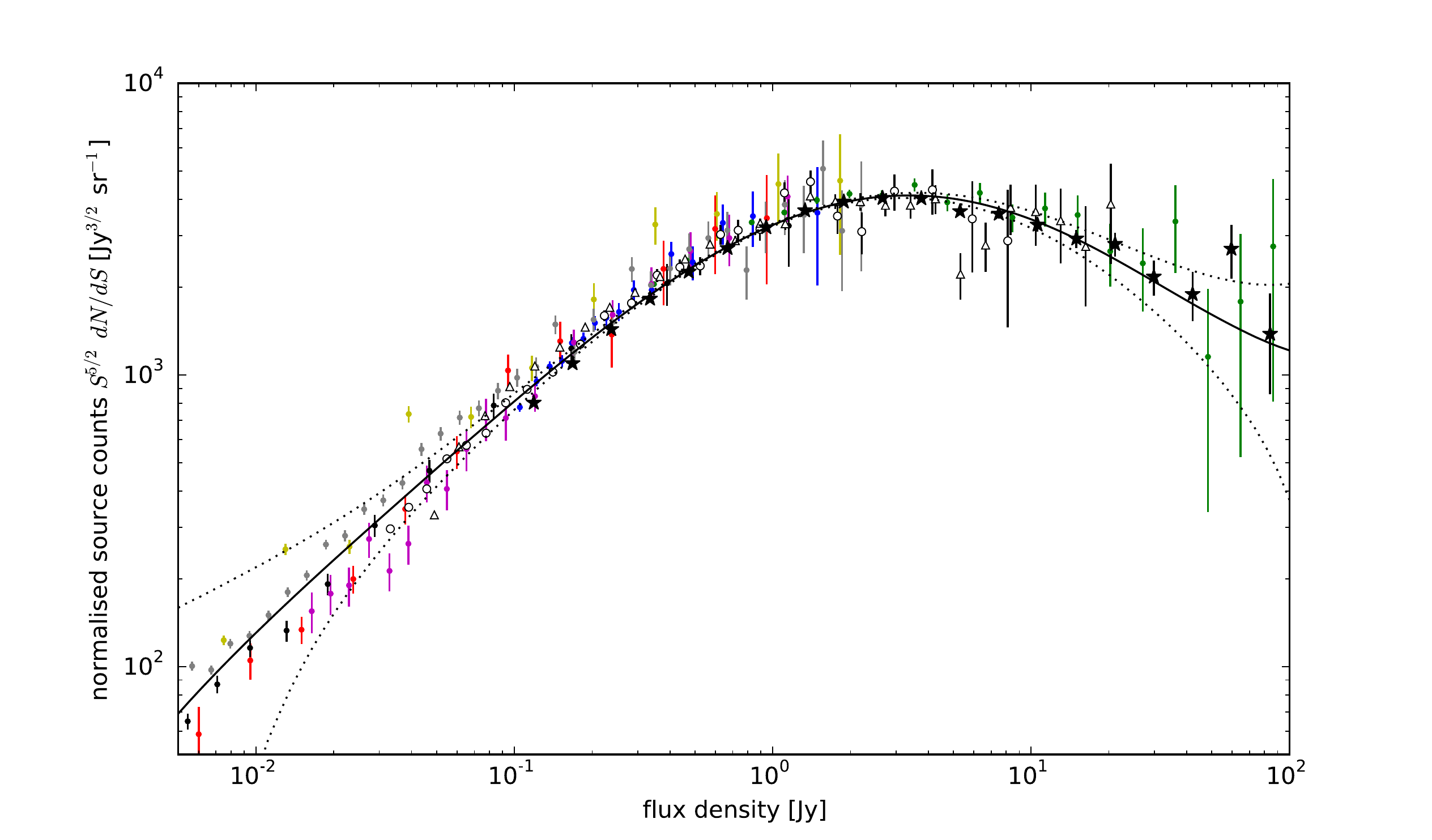}}
\caption{
150~MHz Euclidean normalized differential source counts as derived from the TGSS ADR1 source catalog (black \changetwo{stars} with 1-sigma poissonian error bars), covering the highly complete flux range from 100~mJy to 100~Jy with 20~logarithmic flux bins. Overplotted are various other source counts for this frequency from literature, namely source counts from both a single deep GMRT integration and a larger-area GMRT survey centered on the Bo{\"o}tes field
\citep[][red and magenta dots, respectively]{2011A&A...535A..38I,2013A&A...549A..55W}, 
source counts from the 7C survey
\citep[][blue and green dots, respectively]{1990MNRAS.246..110M,2007MNRAS.382.1639H}, 
\changetwo{recent source counts from deep, small-area surveys with the LOFAR HBA system
\citep[][black, yellow, and gray dots, respectively]{2016MNRAS.460.2385W,2016MNRAS.tmp.1337M,2016MNRAS.462.1910H}, 
and source counts from the MWA GLEAM survey as well as deep, single-pointing MWA survey
\citep[][open triangles and dots, respectively]{2016MNRAS.tmp.1444H,2016MNRAS.459.3314F}. 
Some literature points with very large uncertainties have been omitted. Over the flux density range of the plot, the observed source counts can be described by an empirical model as given in Equation~\ref{eq:counts_model} (black solid line, with 1-sigma deviations plotted as dotted lines).}
}
\label{fig:source_counts}
\end{center}
\end{figure*}

\changetwo{The large TGSS ADR1 source catalog lends itself well for statistical studies, and allows for the selection of statistically significant subsets of sources with interesting properties. In this section we provide some (preliminary) examples of ways in which this data can be used for scientific explorations.}

\changetwo{A simple yet powerful way to characterize radio source populations is by means of radio source counts as a function of flux density. These source counts are for instance used in estimates of foreground contamination for EoR experiments
\citep[e.g.,][]{2016PASA...33...19T}. 
It also provides a fairly objective way to compare results from different surveys. E.g., in Section~\ref{sec:sp_cr} we compared TGSS ADR1 radio source counts to an empirical model from literature to test completeness. Euclidean normalized differential source counts at $\sim 150$~MHz from various surveys are available, although still somewhat sparse. Figure~\ref{fig:source_counts} shows the ADR1 differential source counts above 100~mJy, for which the catalog is complete. In general, we find a good match between our counts and other observational results from literature. Because of the large sky coverage (and therefore large number statistics), the TGSS ADR1 provides the most strict constraints on the shape of the source count distribution over the relevant flux density range.}

\changetwo{The combined measurements provide the best direct constraints on the 150~MHz source counts to date. Similar to \citet[][]{1991PhDT.......241W}, 
we fit the source counts with a polynomial in log-log space over the flux density range of 5~mJy to 100~Jy:
\begin{equation}
\log_{10}{\left( S^{5/2} d\mathrm{N}/d\mathrm{S} \right)} = C_0 + \sum_{i=1}^{5} C_i \times \left[ \log_{10}{(S)} \right]^{i}
\label{eq:counts_model}
\end{equation}
where $S$ is the flux density in Jy, and $C_i$ the model coefficients. The resulting coefficient values and their uncertainties are given in Table~\ref{tab:counts_model}, while the model is overplotted in Figure~\ref{fig:source_counts}. The model is best constrained over the flux density range 100~mJy to 10~Jy. Above 10~Jy the model uncertainty increases due to low number counts. Below 100~mJy the uncertainty increases due to larger differences between the measured source counts.}

\ctable[botcap,center,
caption = {\changetwo{Model coefficients for the radio source counts at 150~MHz, as defined in Equation~\ref{eq:counts_model}.}},
label = tab:counts_model
]{c c}{
}{
\FL Coefficient & Value
\ML $C_0$ &  $3.5142 \pm 0.0016$
\NN $C_1$ &  $0.3738 \pm 0.0039$
\NN $C_2$ & $-0.3138 \pm 0.0075$
\NN $C_3$ & $-0.0717 \pm 0.0050$
\NN $C_4$ &  $0.0213 \pm 0.0044$
\NN $C_5$ &  $0.0097 \pm 0.0018$
\LL}

\changetwo{A popular way of characterizing and selecting radio sources is by means of their spectral index. For example, high-redshift radio galaxies
\citep[$z \gtrsim 2$; e.g.,][]{2008A&ARv..15...67M} 
are efficiently selected by identifying compact sources with steep radio spectra. Surveys like TGSS are relatively sensitive to steep-spectrum radio sources. For example, based on point source sensitivity, TGSS ADR1 is more sensitive than the NVSS survey at 1.4~GHz for compact sources with a spectral index $\alpha \lesssim -0.87$. The spectral index distribution in Figure~\ref{fig:tgss_nvss_spix} is derived from a global source match between TGSS ADR1 and NVSS, finding NVSS counterparts for 96~percent of the ADR1 sources above DEC \adeg{-40} (the NVSS cutoff DEC). The median spectral index of -0.73 is comparable to other studies using the same frequency range
\citep[e.g.,][]{2011A&A...535A..38I,2013A&A...549A..55W,2016MNRAS.460.2385W} 
1.3~percent of the spectral indices are lower than -1.3, and 0.3~percent are below -1.5. Although small percentages, this still corresponds to statistically significant numbers of sources (about 7000 and 1900, respectively). Further classification of these sources requires matching to other wavelength data, which is beyond the scope of this article. A more elaborate analysis of the spectral index distribution between TGSS ADR1 and NVSS is given in \citet{2016arXiv160901308T}. 
}


\section{Public Survey Data Products}
\label{sec:dp}

We release this TGSS ADR1 as a service to the astronomical community in a format that should make it straightforward for non-specialists to use these data in their research. The public data comes in two main forms: images and source catalogs. All-sky representations of the images and source distribution are given in Figures~\ref{fig:tgss_map} and \ref{fig:source_map}


\subsection{Images}
\label{sec:dp_im}

The TGSS ADR1 images are being released as a series $5^\circ\times{5}^\circ$ mosaics with a pixel size of \asec{6.2}. The construction of the mosaic images is described in more detail in \S\ref{sec:dpp_mos_bm}. These mosaic images are available to be directly downloaded as FITS files. The size of full dataset of 5336 mosaic images is 180~Gbytes. The TGSS images and source catalog are available through our project website\footnote{\url{http://tgssadr.strw.leidenuniv.nl/}}. We also provide a cutout server for those researchers wanting to extract a small region around some sky position (up to \pbeam{1}{1} guaranteed). We have no immediate plans to release the individual pointing images or the visibility data.


\subsection{Source Catalog}
\label{sec:dp_sc}

The TGSS ADR1 source catalog contains J2000 positions, Stokes~I flux densities (peak and total) and angular sizes, along with error estimates for \changeone{0.62}~Million sources brighter than 7~sigma. In addition, for each source we give the name of the mosaic image that the parameters were extracted from, as well as a Gaussian source structure flag.

Table~\ref{tab:catalog} is a small sample of the full table. Complete details of how the catalog was generated are found in \S\ref{sec:sp_se}. The complete FITS format table is available on the same server as the images (see above).  Upon publication of this paper we will release a searchable catalog to services such as VizieR and HEASARC.


\section{Conclusions and Future Work}
\label{sec:concl}

We provide the first full public release of the 150~MHz continuum survey (TGSS) from the Giant Metrewave Radio Telescope. This was possible because we developed powerful, automatic tools to deal with widespread radio interference and challenges to calibration, including direction-dependent ionospheric effects. The resulting images cover the full sky north of DEC~\adeg{-53} (or 36,900 deg$^2$) with a median RMS background noise of \mjybeam{3.5} and a typical resolution of \asec{25}. We achieved an improvement over earlier data releases in terms of noise properties (lower background RMS) and image fidelity (fewer residual sidelobes), yielding more robust detections at low flux densities during the source extraction process. Our final catalog contains \changeone{0.62}~Million sources with flux densities ranging from 11.1~mJy to 9.22~kJy.

\begin{figure}
\begin{center}
\resizebox{\hsize}{!}{
\includegraphics[angle=0]{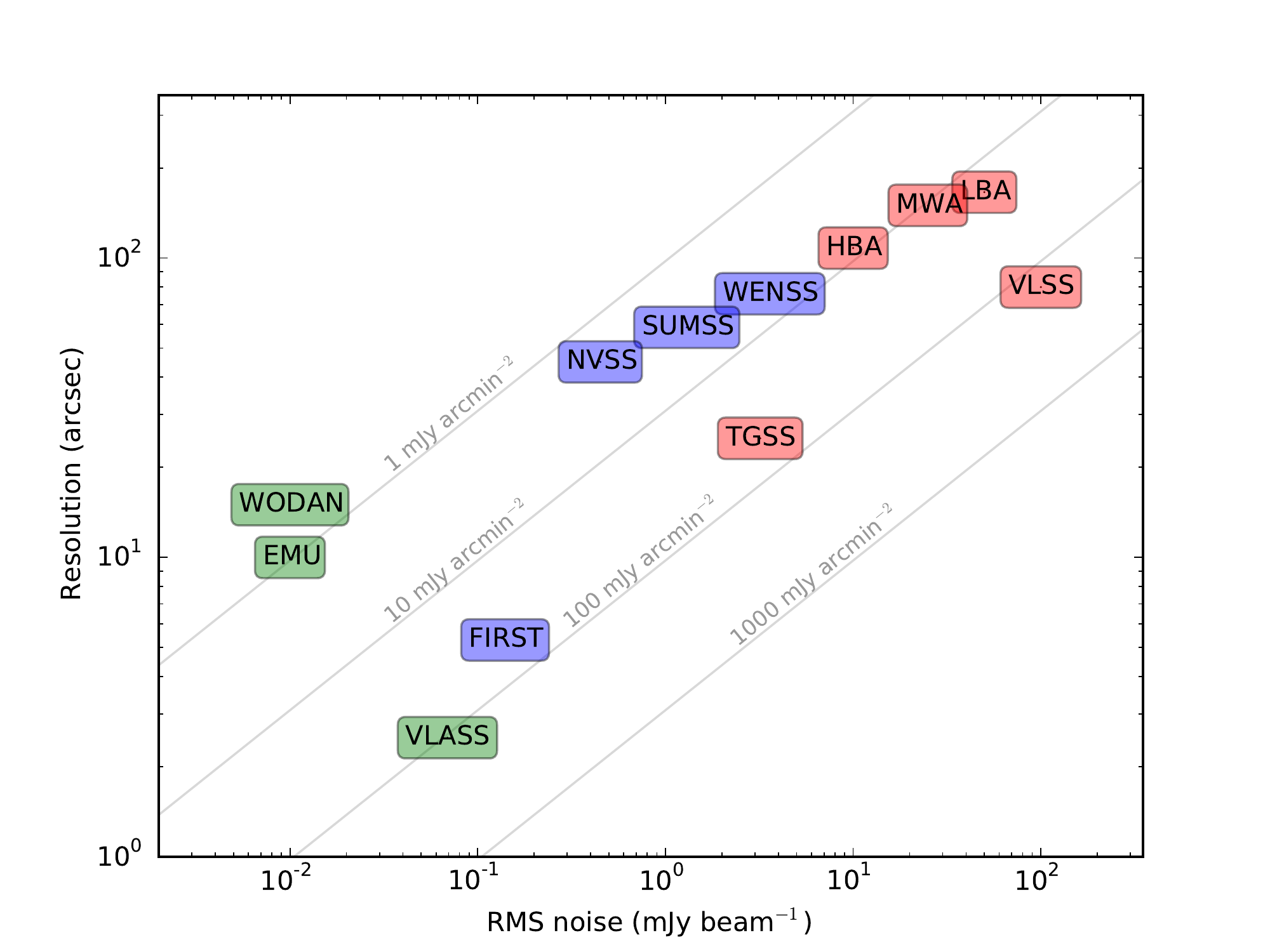}}
\caption{Comparison between the angular resolution and RMS noise (sensitivity) of the TGSS and wide-field meter and centimeter wavelength surveys. We define a wide-field survey as having observed more than 10,000~deg$^2$ or more of the sky. Although the survey area of SUMSS is only 8,100~deg$^2$, it is included here since it provided much needed southern hemisphere coverage. Red symbols indicate meter wavelength surveys ($\nu<300$~MHz), while blue and green symbols are existing and planned centimeter surveys ($\nu>$300~MHz), respectively. Lines of constant surface brightness are shown in gray for a surface brightness sensitivity calculated as a function of resolution for 3$\,\times\,$RMS noise.}
\label{fig:surveys}
\end{center}
\end{figure}

In Figure~\ref{fig:surveys} we show a comparison of the performance between the TGSS with other existing or planned wide-field radio surveys at meter and centimeter wavelengths. At meter wavelengths we include the low (LBA) and high (HBA) frequency versions of the LOFAR MSSS
\citep{2015A&A...582A.123H}, 
the revised VLSS 
\citep{2014MNRAS.440..327L}, 
and the MWA GLEAM.
\citep{2015PASA...32...25W}. 
For centimeter wavelength surveys we show the WENSS
\citep{1997A&AS..124..259R}, 
SUMSS
\citep{2003MNRAS.342.1117M}, 
NVSS
\citep{1998AJ....115.1693C}, 
and FIRST 
\citep{1995ApJ...450..559B}. 
We also show several centimeter surveys that will be undertaken before the end of this decade including the VLA Sky Survey (VLASS)\footnote{\url{https://science.nrao.edu/science/surveys/vlass}},
the Apertif WODAN, and ASKAP EMU surveys
\citep{2011PASA...28..215N,2013PASA...30...20N}. 

TGSS stands out among the meter-wavelength surveys ($\nu < 300$~MHz) as having the lowest noise and highest resolution. In these respects, and in its large number of cataloged sources, TGSS is a meter-wavelength equivalent to the centimeter NVSS. The good sensitivity at meter-wavelengths will be useful since the TGSS provides a large lever arm ($10\,\times$) when compared to existing centimeter surveys like NVSS. More than 95~percent of TGSS sources within the NVSS survey area have NVSS counterparts. Steep-spectrum sources that are faint or undetectable in centimeter surveys will stand out in the TGSS. Pulsars, high-redshift radio galaxies, fossil radio galaxies, and giant radio halos and relics in galaxy clusters are all sources that are found at the steep-spectrum end of the spectral index distribution.

The large area, low noise and good angular resolution makes TGSS a useful dataset in the calibration of instrumentation and propagation effects for other low-frequency telescopes. We also anticipate that a TGSS sky model could be used for foreground subtraction in the search for the Epoch of Reionization. TGSS does not have the same sensitivity to extended emission (Section~\ref{sec:dpp_mp_dic}) as LOFAR and MWA with their larger beam size, and more centrally concentrated array configurations (Table \ref{tab:tgss_vs_msss}). In this sense TGSS could be seen as complimentary to the MSSS and GLEAM surveys, but we note that in the second data release of MSSS there is a plan to use the longer LOFAR baselines to achieve resolutions of \asec{60} and \asec{30} in MSSS-LBA and MSSS-HBA, respectively
\citep{2015A&A...582A.123H}. 

A small synthesized beam is particularly important for identifications at other wavelengths. The TGSS synthesized beam is too large for reliable matches of large-scale optical and NIR catalogs
\citep{2015ApJ...801...26H}, 
as the false positive rates likely swamp identification at the faint magnitudes of such surveys. However, TGSS will likely do well in with identifying rare objects, particularly those that are bright at meter wavelengths. The matching of high energy sources at X-ray and gamma-rays will likely be a productive activity, as might searches for coherent radio emission from nearby stars.

The alternative data release of TGSS is an ongoing project, in which we want to continuously improve the quality of the survey products. News and updates will be presented through the project webpage. Maximizing the sky coverage has top priority. Next to revisiting the reduction of problematic pointings, we \changetwo{proposed} for new GMRT observations of the pointings that persistently failed. 
\changetwo{Furthermore, we will address the issues that causes a small fraction of the observations to have larger-than-average systematic flux density offsets, such as simultaneous, persistent phase delay jumps on multiple GMRT antennas.}
We also plan to obtain and include better quality images of several of the very brightest yet (mostly) morphologically complex sources, like Cas~A, Cen~A, Cyg~A, Her~A, Hya~A, Per~B, Pic~A, Sgr~A, Tau~A, and Vir~A. Several of these have existing data in the GMRT archive, and for some others we have obtained targeted observations. While TGSS ADR1 is a compact source survey by design, we will continue to experiment with data reduction strategies to improve the representation of large-scale emission. 



\begin{acknowledgements}
\changetwo{We thank the referee for providing many useful comments.}. We are grateful for the discussions, feedback and advice we received from our colleagues, who include \changetwo{in no particular order} Bill Cotton, Nirupam Roy, Niruj Mohan, Heinz Andernach, Sanjay Bhatnagar, Wendy Williams, Tim Shimwell, George Heald, Steven Tingay, Rachel Webster, \changeone{Rick Perley,} \changetwo{Francesco de Gasperin, Tim Shimwell,} and many others.
We would like to thank the TIFR GMRT Sky Survey team for planning and carry out the original observations, and we offer a special thanks to the GMRT staff for their on-going support. GMRT is run by the National Centre for Radio Astrophysics of the Tata Institute of Fundamental Research.
For processing the TGSS data we used the NRAO NMPOST compute cluster, and appreciate the support we received for this effort from James Robnett, Erik Bryer, K. Scott Rowe, and Andrew Murtland.
We are very grateful to ASTRON, and in particular Mike Sipior, for hosting our survey products on the ASTRON VO system.
This research has made use of the NASA/IPAC Extragalactic Database (NED) which is operated by the Jet Propulsion Laboratory, California Institute of Technology, under contract with the National Aeronautics and Space Administration.
This research has made use of SAOImage DS9, developed by Smithsonian Astrophysical Observatory.
This research has made use of \emph{Aladin Sky Atlas} developed at CDS, Strasbourg Observatory, France.
HTI and PJ, acknowledge financial support by the National Radio Astronomy Observatory, a facility of the National Science Foundation operated under cooperative agreement by Associated Universities, Inc. HTI acknowledges financial support through the NL-SKA roadmap project funded by the NWO.
KPM acknowledges support from the Hintze Foundation.
\end{acknowledgements}


\bibliographystyle{aa}
\bibliography{tgss}

\begin{thebibliography}{97}
\expandafter\ifx\csname natexlab\endcsname\relax\def\natexlab#1{#1}\fi

\bibitem[{{Arora} {et~al.}(2015){Arora}, {Morgan}, {Ord}, {Tingay},
  {Hurley-Walker}, {Bell}, {Bernardi}, {Bhat}, {Briggs}, {Callingham},
  {Deshpande}, {Dwarakanath}, {Ewall-Wice}, {Feng}, {For}, {Hancock},
  {Hazelton}, {Hindson}, {Jacobs}, {Johnston-Hollitt}, {Kapi{\'n}ska},
  {Kudryavtseva}, {Lenc}, {McKinley}, {Mitchell}, {Oberoi}, {Offringa},
  {Pindor}, {Procopio}, {Riding}, {Staveley-Smith}, {Wayth}, {Wu}, {Zheng},
  {Bowman}, {Cappallo}, {Corey}, {Emrich}, {Goeke}, {Greenhill}, {Kaplan},
  {Kasper}, {Kratzenberg}, {Lonsdale}, {Lynch}, {McWhirter}, {Morales},
  {Morgan}, {Prabu}, {Rogers}, {Roshi}, {Shankar}, {Srivani}, {Subrahmanyan},
  {Waterson}, {Webster}, {Whitney}, {Williams}, \&
  {Williams}}]{2015PASA...32...29A}
{Arora}, B.~S., {Morgan}, J., {Ord}, S.~M., {et~al.} 2015, \pasa, 32, e029

\bibitem[{{Baars} {et~al.}(1977){Baars}, {Genzel}, {Pauliny-Toth}, \&
  {Witzel}}]{1977A&A....61...99B}
{Baars}, J.~W.~M., {Genzel}, R., {Pauliny-Toth}, I.~I.~K., \& {Witzel}, A.
  1977, \aap, 61, 99

\bibitem[{{Bagchi} {et~al.}(2011){Bagchi}, {Sirothia}, {Werner}, {Pandge},
  {Kantharia}, {Ishwara-Chandra}, {Gopal-Krishna}, {Paul}, \&
  {Joshi}}]{2011ApJ...736L...8B}
{Bagchi}, J., {Sirothia}, S.~K., {Werner}, N., {et~al.} 2011, \apjl, 736, L8

\bibitem[{{Baldwin} {et~al.}(1985){Baldwin}, {Boysen}, {Hales}, {Jennings},
  {Waggett}, {Warner}, \& {Wilson}}]{1985MNRAS.217..717B}
{Baldwin}, J.~E., {Boysen}, R.~C., {Hales}, S.~E.~G., {et~al.} 1985, \mnras,
  217, 717

\bibitem[{{Becker} {et~al.}(1995){Becker}, {White}, \&
  {Helfand}}]{1995ApJ...450..559B}
{Becker}, R.~H., {White}, R.~L., \& {Helfand}, D.~J. 1995, \apj, 450, 559

\bibitem[{{Bilous} {et~al.}(2015){Bilous}, {Kondratiev}, {Kramer}, {Keane},
  {Hessels}, {Stappers}, {Malofeev}, {Sobey}, {Breton}, {Cooper}, {Falcke},
  {Karastergiou}, {Michilli}, {Os{\l}owski}, {Sanidas}, {ter Veen}, {van
  Leeuwen}, {Verbiest}, {Weltevrede}, {Zarka}, {Grie{\ss}meier}, {Serylak},
  {Bell}, {Broderick}, {Eisl{\"o}ffel}, {Markoff}, \&
  {Rowlinson}}]{2015arXiv151101767B}
{Bilous}, A., {Kondratiev}, V., {Kramer}, M., {et~al.} 2015, ArXiv e-prints
  [\eprint[arXiv]{1511.01767}]

\bibitem[{{Bonafede} {et~al.}(2015){Bonafede}, {Intema}, {Br{\"u}ggen},
  {Vazza}, {Basu}, {Sommer}, {Ebeling}, {de Gasperin}, {R{\"o}ttgering}, {van
  Weeren}, \& {Cassano}}]{2015MNRAS.454.3391B}
{Bonafede}, A., {Intema}, H., {Br{\"u}ggen}, M., {et~al.} 2015, \mnras, 454,
  3391

\bibitem[{{Bonafede} {et~al.}(2014){Bonafede}, {Intema}, {Br{\"u}ggen},
  {Russell}, {Ogrean}, {Basu}, {Sommer}, {van Weeren}, {Cassano}, {Fabian}, \&
  {R{\"o}ttgering}}]{2014MNRAS.444L..44B}
{Bonafede}, A., {Intema}, H.~T., {Br{\"u}ggen}, M., {et~al.} 2014, \mnras, 444,
  L44

\bibitem[{{Bonnarel} {et~al.}(2000){Bonnarel}, {Fernique}, {Bienaym{\'e}},
  {Egret}, {Genova}, {Louys}, {Ochsenbein}, {Wenger}, \&
  {Bartlett}}]{2000A&AS..143...33B}
{Bonnarel}, F., {Fernique}, P., {Bienaym{\'e}}, O., {et~al.} 2000, \aaps, 143,
  33

\bibitem[{{Bridle} \& {Schwab}(1999)}]{1999ASPC..180..371B}
{Bridle}, A.~H. \& {Schwab}, F.~R. 1999, in Astronomical Society of the Pacific
  Conference Series, Vol. 180, Synthesis Imaging in Radio Astronomy II, ed.
  G.~B. {Taylor}, C.~L. {Carilli}, \& R.~A. {Perley}, 371

\bibitem[{{Brienza} {et~al.}(2016){Brienza}, {Godfrey}, {Morganti}, {Vilchez},
  {Maddox}, {Murgia}, {Orru}, {Shulevski}, {Best}, {Br{\"u}ggen}, {Harwood},
  {Jamrozy}, {Jarvis}, {Mahony}, {McKean}, \&
  {R{\"o}ttgering}}]{2016A&A...585A..29B}
{Brienza}, M., {Godfrey}, L., {Morganti}, R., {et~al.} 2016, \aap, 585, A29

\bibitem[{{Cohen} {et~al.}(2007){Cohen}, {Lane}, {Cotton}, {Kassim}, {Lazio},
  {Perley}, {Condon}, \& {Erickson}}]{2007AJ....134.1245C}
{Cohen}, A.~S., {Lane}, W.~M., {Cotton}, W.~D., {et~al.} 2007, \aj, 134, 1245

\bibitem[{{Condon}(1997)}]{1997PASP..109..166C}
{Condon}, J.~J. 1997, \pasp, 109, 166

\bibitem[{{Condon}(1999)}]{1999PNAS...96.4756C}
{Condon}, J.~J. 1999, Proceedings of the National Academy of Science, 96, 4756

\bibitem[{{Condon} {et~al.}(2012){Condon}, {Cotton}, {Fomalont}, {Kellermann},
  {Miller}, {Perley}, {Scott}, {Vernstrom}, \& {Wall}}]{2012ApJ...758...23C}
{Condon}, J.~J., {Cotton}, W.~D., {Fomalont}, E.~B., {et~al.} 2012, \apj, 758,
  23

\bibitem[{{Condon} {et~al.}(1998){Condon}, {Cotton}, {Greisen}, {Yin},
  {Perley}, {Taylor}, \& {Broderick}}]{1998AJ....115.1693C}
{Condon}, J.~J., {Cotton}, W.~D., {Greisen}, E.~W., {et~al.} 1998, \aj, 115,
  1693

\bibitem[{{Cotton}(1999)}]{1999ASPC..180..357C}
{Cotton}, W.~D. 1999, in ASPC Series, Vol. 180, Synthesis Imaging in Radio
  Astronomy II, ed. G.~B. {Taylor}, C.~L. {Carilli}, \& R.~A. {Perley}, 357

\bibitem[{{Cotton}(2005)}]{2005ASPC..345..337C}
{Cotton}, W.~D. 2005, in Astronomical Society of the Pacific Conference Series,
  Vol. 345, From Clark Lake to the Long Wavelength Array: Bill Erickson's Radio
  Science, ed. N.~{Kassim}, M.~{Perez}, W.~{Junor}, \& P.~{Henning}, 337

\bibitem[{{Cotton}(2008)}]{2008PASP..120..439C}
{Cotton}, W.~D. 2008, \pasp, 120, 439

\bibitem[{{Cotton} {et~al.}(2004){Cotton}, {Condon}, {Perley}, {Kassim},
  {Lazio}, {Cohen}, {Lane}, \& {Erickson}}]{2004SPIE.5489..180C}
{Cotton}, W.~D., {Condon}, J.~J., {Perley}, R.~A., {et~al.} 2004, in \procspie,
  Vol. 5489, Ground-based Telescopes, ed. J.~M. {Oschmann}, Jr., 180--189

\bibitem[{{De Breuck} {et~al.}(2002){De Breuck}, {Tang}, {de Bruyn},
  {R{\"o}ttgering}, \& {van Breugel}}]{2002A&A...394...59D}
{De Breuck}, C., {Tang}, Y., {de Bruyn}, A.~G., {R{\"o}ttgering}, H., \& {van
  Breugel}, W. 2002, \aap, 394, 59

\bibitem[{{de Gasperin} {et~al.}(2015){de Gasperin}, {Intema}, {van Weeren},
  {Dawson}, {Golovich}, {Wittman}, {Bonafede}, \&
  {Br{\"u}ggen}}]{2015MNRAS.453.3483D}
{de Gasperin}, F., {Intema}, H.~T., {van Weeren}, R.~J., {et~al.} 2015, \mnras,
  453, 3483

\bibitem[{{de Gasperin} {et~al.}(2014){de Gasperin}, {Intema}, {Williams},
  {Br{\"u}ggen}, {Murgia}, {Beck}, \& {Bonafede}}]{2014MNRAS.440.1542D}
{de Gasperin}, F., {Intema}, H.~T., {Williams}, W., {et~al.} 2014, \mnras, 440,
  1542

\bibitem[{{Feretti} {et~al.}(2012){Feretti}, {Giovannini}, {Govoni}, \&
  {Murgia}}]{2012A&ARv..20...54F}
{Feretti}, L., {Giovannini}, G., {Govoni}, F., \& {Murgia}, M. 2012, \aapr, 20,
  54

\bibitem[{{Franzen} {et~al.}(2016){Franzen}, {Jackson}, {Offringa}, {Ekers},
  {Wayth}, {Bernardi}, {Bowman}, {Briggs}, {Cappallo}, {Deshpande}, {Gaensler},
  {Greenhill}, {Hazelton}, {Johnston-Hollitt}, {Kaplan}, {Lonsdale},
  {McWhirter}, {Mitchell}, {Morales}, {Morgan}, {Morgan}, {Oberoi}, {Ord},
  {Prabu}, {Seymour}, {Shankar}, {Srivani}, {Subrahmanyan}, {Tingay}, {Trott},
  {Webster}, {Williams}, \& {Williams}}]{2016MNRAS.459.3314F}
{Franzen}, T.~M.~O., {Jackson}, C.~A., {Offringa}, A.~R., {et~al.} 2016,
  \mnras, 459, 3314

\bibitem[{{Garn} {et~al.}(2007){Garn}, {Green}, {Hales}, {Riley}, \&
  {Alexander}}]{2007MNRAS.376.1251G}
{Garn}, T., {Green}, D.~A., {Hales}, S.~E.~G., {Riley}, J.~M., \& {Alexander},
  P. 2007, \mnras, 376, 1251

\bibitem[{{Garrett}(2013)}]{2013arXiv1307.0386G}
{Garrett}, M.~A. 2013, ArXiv e-prints [\eprint[arXiv]{1307.0386}]

\bibitem[{{George} \& {Stevens}(2008)}]{2008MNRAS.390..741G}
{George}, S.~J. \& {Stevens}, I.~R. 2008, \mnras, 390, 741

\bibitem[{{Ghosh} {et~al.}(2012){Ghosh}, {Prasad}, {Bharadwaj}, {Ali}, \&
  {Chengalur}}]{2012MNRAS.426.3295G}
{Ghosh}, A., {Prasad}, J., {Bharadwaj}, S., {Ali}, S.~S., \& {Chengalur}, J.~N.
  2012, \mnras, 426, 3295

\bibitem[{{Gopal-Krishna} {et~al.}(2012){Gopal-Krishna}, {Mhaskey}, {Wiita},
  {Sirothia}, {Kantharia}, \& {Ishwara-Chandra}}]{2012MNRAS.423.1053G}
{Gopal-Krishna}, {Mhaskey}, M., {Wiita}, P.~J., {et~al.} 2012, \mnras, 423,
  1053

\bibitem[{{Greisen}(2003)}]{2003ASSL..285..109G}
{Greisen}, E.~W. 2003, Information Handling in Astronomy - Historical Vistas,
  285, 109

\bibitem[{{Hales} {et~al.}(1988){Hales}, {Baldwin}, \&
  {Warner}}]{1988MNRAS.234..919H}
{Hales}, S.~E.~G., {Baldwin}, J.~E., \& {Warner}, P.~J. 1988, \mnras, 234, 919

\bibitem[{{Hales} {et~al.}(2007){Hales}, {Riley}, {Waldram}, {Warner}, \&
  {Baldwin}}]{2007MNRAS.382.1639H}
{Hales}, S.~E.~G., {Riley}, J.~M., {Waldram}, E.~M., {Warner}, P.~J., \&
  {Baldwin}, J.~E. 2007, \mnras, 382, 1639

\bibitem[{{Hales} {et~al.}(1995){Hales}, {Waldram}, {Rees}, \&
  {Warner}}]{1995MNRAS.274..447H}
{Hales}, S.~E.~G., {Waldram}, E.~M., {Rees}, N., \& {Warner}, P.~J. 1995,
  \mnras, 274, 447

\bibitem[{{Hancock} {et~al.}(2012){Hancock}, {Murphy}, {Gaensler}, {Hopkins},
  \& {Curran}}]{2012MNRAS.422.1812H}
{Hancock}, P.~J., {Murphy}, T., {Gaensler}, B.~M., {Hopkins}, A., \& {Curran},
  J.~R. 2012, \mnras, 422, 1812

\bibitem[{{Hardcastle} {et~al.}(2016){Hardcastle}, {G{\"u}rkan}, {van Weeren},
  {Williams}, {Best}, {de Gasperin}, {Rafferty}, {Read}, {Sabater}, {Shimwell},
  {Smith}, {Tasse}, {Bourne}, {Brienza}, {Br{\"u}ggen}, {Brunetti},
  {Chy{\.z}y}, {Conway}, {Dunne}, {Eales}, {Maddox}, {Jarvis}, {Mahony},
  {Morganti}, {Prandoni}, {R{\"o}ttgering}, {Valiante}, \&
  {White}}]{2016MNRAS.462.1910H}
{Hardcastle}, M.~J., {G{\"u}rkan}, G., {van Weeren}, R.~J., {et~al.} 2016,
  \mnras, 462, 1910

\bibitem[{{Haslam} {et~al.}(1982){Haslam}, {Salter}, {Stoffel}, \&
  {Wilson}}]{1982A&AS...47....1H}
{Haslam}, C.~G.~T., {Salter}, C.~J., {Stoffel}, H., \& {Wilson}, W.~E. 1982,
  \aaps, 47, 1

\bibitem[{{Heald} {et~al.}(2015){Heald}, {Pizzo}, {Orr{\'u}}, {Breton},
  {Carbone}, {Ferrari}, {Hardcastle}, {Jurusik}, {Macario}, {Mulcahy},
  {Rafferty}, {Asgekar}, {Brentjens}, {Fallows}, {Frieswijk}, {Toribio},
  {Adebahr}, {Arts}, {Bell}, {Bonafede}, {Bray}, {Broderick}, {Cantwell},
  {Carroll}, {Cendes}, {Clarke}, {Croston}, {Daiboo}, {de Gasperin}, {Gregson},
  {Harwood}, {Hassall}, {Heesen}, {Horneffer}, {van der Horst}, {Iacobelli},
  {Jeli{\'c}}, {Jones}, {Kant}, {Kokotanekov}, {Martin}, {McKean}, {Morabito},
  {Nikiel-Wroczy{\'n}ski}, {Offringa}, {Pandey}, {Pandey-Pommier}, {Pietka},
  {Pratley}, {Riseley}, {Rowlinson}, {Sabater}, {Scaife}, {Scheers},
  {Sendlinger}, {Shulevski}, {Sipior}, {Sobey}, {Stewart}, {Stroe}, {Swinbank},
  {Tasse}, {Tr{\"u}stedt}, {Varenius}, {van Velzen}, {Vilchez}, {van Weeren},
  {Wijnholds}, {Williams}, {de Bruyn}, {Nijboer}, {Wise}, {Alexov}, {Anderson},
  {Avruch}, {Beck}, {Bell}, {van Bemmel}, {Bentum}, {Bernardi}, {Best},
  {Breitling}, {Brouw}, {Br{\"u}ggen}, {Butcher}, {Ciardi}, {Conway}, {de
  Geus}, {de Jong}, {de Vos}, {Deller}, {Dettmar}, {Duscha}, {Eisl{\"o}ffel},
  {Engels}, {Falcke}, {Fender}, {Garrett}, {Grie{\ss}meier}, {Gunst},
  {Hamaker}, {Hessels}, {Hoeft}, {H{\"o}randel}, {Holties}, {Intema},
  {Jackson}, {J{\"u}tte}, {Karastergiou}, {Klijn}, {Kondratiev}, {Koopmans},
  {Kuniyoshi}, {Kuper}, {Law}, {van Leeuwen}, {Loose}, {Maat}, {Markoff},
  {McFadden}, {McKay-Bukowski}, {Mevius}, {Miller-Jones}, {Morganti}, {Munk},
  {Nelles}, {Noordam}, {Norden}, {Paas}, {Polatidis}, {Reich}, {Renting},
  {R{\"o}ttgering}, {Schoenmakers}, {Schwarz}, {Sluman}, {Smirnov}, {Stappers},
  {Steinmetz}, {Tagger}, {Tang}, {ter Veen}, {Thoudam}, {Vermeulen}, {Vocks},
  {Vogt}, {Wijers}, {Wucknitz}, {Yatawatta}, \& {Zarka}}]{2015A&A...582A.123H}
{Heald}, G.~H., {Pizzo}, R.~F., {Orr{\'u}}, E., {et~al.} 2015, \aap, 582, A123

\bibitem[{{Helfand} {et~al.}(2015){Helfand}, {White}, \&
  {Becker}}]{2015ApJ...801...26H}
{Helfand}, D.~J., {White}, R.~L., \& {Becker}, R.~H. 2015, \apj, 801, 26

\bibitem[{{Hopkins} {et~al.}(2015){Hopkins}, {Whiting}, {Seymour}, {Chow},
  {Norris}, {Bonavera}, {Breton}, {Carbone}, {Ferrari}, {Franzen}, {Garsden},
  {Gonz{\'a}lez-Nuevo}, {Hales}, {Hancock}, {Heald}, {Herranz}, {Huynh},
  {Jurek}, {L{\'o}pez-Caniego}, {Massardi}, {Mohan}, {Molinari}, {Orr{\`u}},
  {Paladino}, {Pestalozzi}, {Pizzo}, {Rafferty}, {R{\"o}ttgering}, {Rudnick},
  {Schisano}, {Shulevski}, {Swinbank}, {Taylor}, \& {van der
  Horst}}]{2015PASA...32...37H}
{Hopkins}, A.~M., {Whiting}, M.~T., {Seymour}, N., {et~al.} 2015, \pasa, 32,
  e037

\bibitem[{{Hurley-Walker} {et~al.}(2016){Hurley-Walker}, {Callingham},
  {Hancock}, {Franzen}, {Hindson}, {Kapi{\'n}ska}, {Morgan}, {Offringa},
  {Wayth}, {Wu}, {Zheng}, {Murphy}, {Bell}, {Dwarakanath}, {For}, {Gaensler},
  {Johnston-Hollitt}, {Lenc}, {Procopio}, {Staveley-Smith}, {Ekers}, {Bowman},
  {Briggs}, {Cappallo}, {Deshpande}, {Greenhill}, {Hazelton}, {Kaplan},
  {Lonsdale}, {McWhirter}, {Mitchell}, {Morales}, {Morgan}, {Oberoi}, {Ord},
  {Prabu}, {Shankar}, {Srivani}, {Subrahmanyan}, {Tingay}, {Webster},
  {Williams}, \& {Williams}}]{2016MNRAS.tmp.1444H}
{Hurley-Walker}, N., {Callingham}, J.~R., {Hancock}, P.~J., {et~al.} 2016,
  \mnras

\bibitem[{{Hurley-Walker} {et~al.}(2015){Hurley-Walker}, {Johnston-Hollitt},
  {Ekers}, {Hunstead}, {Sadler}, {Hindson}, {Hancock}, {Bernardi}, {Bowman},
  {Briggs}, {Cappallo}, {Corey}, {Deshpande}, {Emrich}, {Gaensler}, {Goeke},
  {Greenhill}, {Hazelton}, {Hewitt}, {Kaplan}, {Kasper}, {Kratzenberg},
  {Lonsdale}, {Lynch}, {Mitchell}, {McWhirter}, {Morales}, {Morgan}, {Oberoi},
  {Offringa}, {Ord}, {Prabu}, {Rogers}, {Roshi}, {Shankar}, {Srivani},
  {Subrahmanyan}, {Tingay}, {Waterson}, {Wayth}, {Webster}, {Whitney},
  {Williams}, \& {Williams}}]{2015MNRAS.447.2468H}
{Hurley-Walker}, N., {Johnston-Hollitt}, M., {Ekers}, R., {et~al.} 2015,
  \mnras, 447, 2468

\bibitem[{{Intema}(2009)}]{2009PhDT........24I}
{Intema}, H.~T. 2009, PhD thesis, PhD Thesis, Leiden University, 2009

\bibitem[{{Intema}(2014{\natexlab{a}})}]{2014arXiv1402.4889I}
{Intema}, H.~T. 2014{\natexlab{a}}, ArXiv e-prints [\eprint[arXiv]{1402.4889}]

\bibitem[{{Intema}(2014{\natexlab{b}})}]{2014ascl.soft08006I}
{Intema}, H.~T. 2014{\natexlab{b}}, {SPAM: Source Peeling and Atmospheric
  Modeling}, Astrophysics Source Code Library

\bibitem[{{Intema} {et~al.}(2009){Intema}, {van der Tol}, {Cotton}, {Cohen},
  {van Bemmel}, \& {R{\"o}ttgering}}]{2009A&A...501.1185I}
{Intema}, H.~T., {van der Tol}, S., {Cotton}, W.~D., {et~al.} 2009, \aap, 501,
  1185

\bibitem[{{Intema} {et~al.}(2011){Intema}, {van Weeren}, {R{\"o}ttgering}, \&
  {Lal}}]{2011A&A...535A..38I}
{Intema}, H.~T., {van Weeren}, R.~J., {R{\"o}ttgering}, H.~J.~A., \& {Lal},
  D.~V. 2011, \aap, 535, A38

\bibitem[{{Ishwara-Chandra} {et~al.}(2010){Ishwara-Chandra}, {Sirothia},
  {Wadadekar}, {Pal}, \& {Windhorst}}]{2010MNRAS.405..436I}
{Ishwara-Chandra}, C.~H., {Sirothia}, S.~K., {Wadadekar}, Y., {Pal}, S., \&
  {Windhorst}, R. 2010, \mnras, 405, 436

\bibitem[{{Jauncey}(1977)}]{1977IAUS...74.....J}
{Jauncey}, D.~L., ed. 1977, IAU Symposium, Vol.~74, {Radio astronomy and
  cosmology; Proceedings of the Symposium, Cambridge University, Cambridge,
  England, August 16-20, 1976}

\bibitem[{{Kettenis} {et~al.}(2006){Kettenis}, {van Langevelde}, {Reynolds}, \&
  {Cotton}}]{2006ASPC..351..497K}
{Kettenis}, M., {van Langevelde}, H.~J., {Reynolds}, C., \& {Cotton}, B. 2006,
  in Astronomical Society of the Pacific Conference Series, Vol. 351,
  Astronomical Data Analysis Software and Systems XV, ed. C.~{Gabriel},
  C.~{Arviset}, D.~{Ponz}, \& S.~{Enrique}, 497

\bibitem[{{Krishna} {et~al.}(2014){Krishna}, {Sirothia}, {Mhaskey}, {Ranadive},
  {Wiita}, {Goyal}, {Kantharia}, \& {Ishwara-Chandra}}]{2014MNRAS.443.2824K}
{Krishna}, G., {Sirothia}, S.~K., {Mhaskey}, M., {et~al.} 2014, \mnras, 443,
  2824

\bibitem[{{Lacy} {et~al.}(1995){Lacy}, {Riley}, {Waldram}, {McMahon}, \&
  {Warner}}]{1995MNRAS.276..614L}
{Lacy}, M., {Riley}, J.~M., {Waldram}, E.~M., {McMahon}, R.~G., \& {Warner},
  P.~J. 1995, \mnras, 276, 614

\bibitem[{{Laing} {et~al.}(1983){Laing}, {Riley}, \&
  {Longair}}]{1983MNRAS.204..151L}
{Laing}, R.~A., {Riley}, J.~M., \& {Longair}, M.~S. 1983, \mnras, 204, 151

\bibitem[{{Lane} {et~al.}(2012){Lane}, {Cotton}, {Helmboldt}, \&
  {Kassim}}]{2012RaSc...47.0K04L}
{Lane}, W.~M., {Cotton}, W.~D., {Helmboldt}, J.~F., \& {Kassim}, N.~E. 2012,
  Radio Science, 47, 0

\bibitem[{{Lane} {et~al.}(2014){Lane}, {Cotton}, {van Velzen}, {Clarke},
  {Kassim}, {Helmboldt}, {Lazio}, \& {Cohen}}]{2014MNRAS.440..327L}
{Lane}, W.~M., {Cotton}, W.~D., {van Velzen}, S., {et~al.} 2014, \mnras, 440,
  327

\bibitem[{{Mahony} {et~al.}(2016){Mahony}, {Morganti}, {Prandoni}, {van
  Bemmel}, {Shimwell}, {Brienza}, {Best}, {Br{\"u}ggen}, {Rivera}, {de
  Gasperin}, {Hardcastle}, {Harwood}, {Heald}, {Jarvis}, {Mandal}, {Miley},
  {Retana-Montenegro}, {R{\"o}ttgering}, {Sabater}, {Tasse}, {van Velzen}, {van
  Weeren}, {Williams}, \& {White}}]{2016MNRAS.tmp.1337M}
{Mahony}, E.~K., {Morganti}, R., {Prandoni}, I., {et~al.} 2016, \mnras
  [\eprint[arXiv]{1609.00537}]

\bibitem[{{Marcote} {et~al.}(2015){Marcote}, {Rib{\'o}}, {Paredes}, \&
  {Ishwara-Chandra}}]{2015MNRAS.451...59M}
{Marcote}, B., {Rib{\'o}}, M., {Paredes}, J.~M., \& {Ishwara-Chandra}, C.~H.
  2015, \mnras, 451, 59

\bibitem[{{Mauch} {et~al.}(2013){Mauch}, {Kl{\"o}ckner}, {Rawlings}, {Jarvis},
  {Hardcastle}, {Obreschkow}, {Saikia}, \& {Thompson}}]{2013MNRAS.435..650M}
{Mauch}, T., {Kl{\"o}ckner}, H.-R., {Rawlings}, S., {et~al.} 2013, \mnras, 435,
  650

\bibitem[{{Mauch} {et~al.}(2003){Mauch}, {Murphy}, {Buttery}, {Curran},
  {Hunstead}, {Piestrzynski}, {Robertson}, \& {Sadler}}]{2003MNRAS.342.1117M}
{Mauch}, T., {Murphy}, T., {Buttery}, H.~J., {et~al.} 2003, \mnras, 342, 1117

\bibitem[{{McGilchrist} {et~al.}(1990){McGilchrist}, {Baldwin}, {Riley},
  {Titterington}, {Waldram}, \& {Warner}}]{1990MNRAS.246..110M}
{McGilchrist}, M.~M., {Baldwin}, J.~E., {Riley}, J.~M., {et~al.} 1990, \mnras,
  246, 110

\bibitem[{{Miley} \& {De Breuck}(2008)}]{2008A&ARv..15...67M}
{Miley}, G. \& {De Breuck}, C. 2008, \aapr, 15, 67

\bibitem[{{Mitchell} {et~al.}(2008){Mitchell}, {Greenhill}, {Wayth}, {Sault},
  {Lonsdale}, {Cappallo}, {Morales}, \& {Ord}}]{2008ISTSP...2..707M}
{Mitchell}, D.~A., {Greenhill}, L.~J., {Wayth}, R.~B., {et~al.} 2008, IEEE
  Journal of Selected Topics in Signal Processing, 2, 707

\bibitem[{{Mohan} \& {Rafferty}(2015)}]{2015ascl.soft02007M}
{Mohan}, N. \& {Rafferty}, D. 2015, {PyBDSM: Python Blob Detection and Source
  Measurement}, Astrophysics Source Code Library

\bibitem[{{Mooley} {et~al.}(2013){Mooley}, {Frail}, {Ofek}, {Miller},
  {Kulkarni}, \& {Horesh}}]{2013ApJ...768..165M}
{Mooley}, K.~P., {Frail}, D.~A., {Ofek}, E.~O., {et~al.} 2013, \apj, 768, 165

\bibitem[{{Murphy} {et~al.}(2007){Murphy}, {Mauch}, {Green}, {Hunstead},
  {Piestrzynska}, {Kels}, \& {Sztajer}}]{2007MNRAS.382..382M}
{Murphy}, T., {Mauch}, T., {Green}, A., {et~al.} 2007, \mnras, 382, 382

\bibitem[{{Noordam}(2004)}]{2004SPIE.5489..817N}
{Noordam}, J.~E. 2004, in \procspie, Vol. 5489, Ground-based Telescopes, ed.
  J.~M. {Oschmann}, Jr., 817--825

\bibitem[{{Norris} {et~al.}(2013){Norris}, {Afonso}, {Bacon}, {Beck}, {Bell},
  {Beswick}, {Best}, {Bhatnagar}, {Bonafede}, {Brunetti}, {Budav{\'a}ri},
  {Cassano}, {Condon}, {Cress}, {Dabbech}, {Feain}, {Fender}, {Ferrari},
  {Gaensler}, {Giovannini}, {Haverkorn}, {Heald}, {Van der Heyden}, {Hopkins},
  {Jarvis}, {Johnston-Hollitt}, {Kothes}, {Van Langevelde}, {Lazio}, {Mao},
  {Mart{\'{\i}}nez-Sansigre}, {Mary}, {Mcalpine}, {Middelberg}, {Murphy},
  {Padovani}, {Paragi}, {Prandoni}, {Raccanelli}, {Rigby}, {Roseboom},
  {R{\"o}ttgering}, {Sabater}, {Salvato}, {Scaife}, {Schilizzi}, {Seymour},
  {Smith}, {Umana}, {Zhao}, \& {Zinn}}]{2013PASA...30...20N}
{Norris}, R.~P., {Afonso}, J., {Bacon}, D., {et~al.} 2013, \pasa, 30, e020

\bibitem[{{Norris} {et~al.}(2011){Norris}, {Hopkins}, {Afonso}, {Brown},
  {Condon}, {Dunne}, {Feain}, {Hollow}, {Jarvis}, {Johnston-Hollitt}, {Lenc},
  {Middelberg}, {Padovani}, {Prandoni}, {Rudnick}, {Seymour}, {Umana},
  {Andernach}, {Alexander}, {Appleton}, {Bacon}, {Banfield}, {Becker}, {Brown},
  {Ciliegi}, {Jackson}, {Eales}, {Edge}, {Gaensler}, {Giovannini}, {Hales},
  {Hancock}, {Huynh}, {Ibar}, {Ivison}, {Kennicutt}, {Kimball}, {Koekemoer},
  {Koribalski}, {L{\'o}pez-S{\'a}nchez}, {Mao}, {Murphy}, {Messias},
  {Pimbblet}, {Raccanelli}, {Randall}, {Reiprich}, {Roseboom},
  {R{\"o}ttgering}, {Saikia}, {Sharp}, {Slee}, {Smail}, {Thompson}, {Urquhart},
  {Wall}, \& {Zhao}}]{2011PASA...28..215N}
{Norris}, R.~P., {Hopkins}, A.~M., {Afonso}, J., {et~al.} 2011, \pasa, 28, 215

\bibitem[{{Obenberger} {et~al.}(2014){Obenberger}, {Taylor}, {Hartman},
  {Dowell}, {Ellingson}, {Helmboldt}, {Henning}, {Kavic}, {Schinzel},
  {Simonetti}, {Stovall}, \& {Wilson}}]{2014ApJ...788L..26O}
{Obenberger}, K.~S., {Taylor}, G.~B., {Hartman}, J.~M., {et~al.} 2014, \apjl,
  788, L26

\bibitem[{{Petrov} \& {Taylor}(2011)}]{2011AJ....142...89P}
{Petrov}, L. \& {Taylor}, G.~B. 2011, \aj, 142, 89

\bibitem[{{Rees}(1990)}]{1990MNRAS.244..233R}
{Rees}, N. 1990, \mnras, 244, 233

\bibitem[{{Rengelink} {et~al.}(1997){Rengelink}, {Tang}, {de Bruyn}, {Miley},
  {Bremer}, {Roettgering}, \& {Bremer}}]{1997A&AS..124..259R}
{Rengelink}, R.~B., {Tang}, Y., {de Bruyn}, A.~G., {et~al.} 1997, \aaps, 124,
  259

\bibitem[{{Roger} {et~al.}(1973){Roger}, {Costain}, \&
  {Bridle}}]{1973AJ.....78.1030R}
{Roger}, R.~S., {Costain}, C.~H., \& {Bridle}, A.~H. 1973, \aj, 78, 1030

\bibitem[{{Roger} {et~al.}(1999){Roger}, {Costain}, {Landecker}, \&
  {Swerdlyk}}]{1999A&AS..137....7R}
{Roger}, R.~S., {Costain}, C.~H., {Landecker}, T.~L., \& {Swerdlyk}, C.~M.
  1999, \aaps, 137, 7

\bibitem[{{Scaife} \& {Heald}(2012)}]{2012MNRAS.423L..30S}
{Scaife}, A.~M.~M. \& {Heald}, G.~H. 2012, \mnras, 423, L30

\bibitem[{{Schwab}(1984)}]{1984AJ.....89.1076S}
{Schwab}, F.~R. 1984, 89, 1076

\bibitem[{{Sirothia}(2009)}]{2009MNRAS.398..853S}
{Sirothia}, S.~K. 2009, \mnras, 398, 853

\bibitem[{{Sirothia} {et~al.}(2014){Sirothia}, {Lecavelier des Etangs},
  {Gopal-Krishna}, {Kantharia}, \& {Ishwar-Chandra}}]{2014A&A...562A.108S}
{Sirothia}, S.~K., {Lecavelier des Etangs}, A., {Gopal-Krishna}, {Kantharia},
  N.~G., \& {Ishwar-Chandra}, C.~H. 2014, \aap, 562, A108

\bibitem[{{Sirothia} {et~al.}(2009){Sirothia}, {Saikia}, {Ishwara-Chandra}, \&
  {Kantharia}}]{2009MNRAS.392.1403S}
{Sirothia}, S.~K., {Saikia}, D.~J., {Ishwara-Chandra}, C.~H., \& {Kantharia},
  N.~G. 2009, \mnras, 392, 1403

\bibitem[{{Smirnov}(2011)}]{2011A&A...527A.107S}
{Smirnov}, O.~M. 2011, \aap, 527, A107

\bibitem[{{Stovall} {et~al.}(2015){Stovall}, {Ray}, {Blythe}, {Dowell},
  {Eftekhari}, {Garcia}, {Lazio}, {McCrackan}, {Schinzel}, \&
  {Taylor}}]{2015ApJ...808..156S}
{Stovall}, K., {Ray}, P.~S., {Blythe}, J., {et~al.} 2015, \apj, 808, 156

\bibitem[{{Swarup}(1991)}]{1991ASPC...19..376S}
{Swarup}, G. 1991, in Astronomical Society of the Pacific Conference Series,
  Vol.~19, IAU Colloq. 131: Radio Interferometry. Theory, Techniques, and
  Applications, ed. T.~J. {Cornwell} \& R.~A. {Perley}, 376--380

\bibitem[{{Tingay} {et~al.}(2013){Tingay}, {Goeke}, {Bowman}, {Emrich}, {Ord},
  {Mitchell}, {Morales}, {Booler}, {Crosse}, {Wayth}, {Lonsdale}, {Tremblay},
  {Pallot}, {Colegate}, {Wicenec}, {Kudryavtseva}, {Arcus}, {Barnes},
  {Bernardi}, {Briggs}, {Burns}, {Bunton}, {Cappallo}, {Corey}, {Deshpande},
  {Desouza}, {Gaensler}, {Greenhill}, {Hall}, {Hazelton}, {Herne}, {Hewitt},
  {Johnston-Hollitt}, {Kaplan}, {Kasper}, {Kincaid}, {Koenig}, {Kratzenberg},
  {Lynch}, {Mckinley}, {Mcwhirter}, {Morgan}, {Oberoi}, {Pathikulangara},
  {Prabu}, {Remillard}, {Rogers}, {Roshi}, {Salah}, {Sault}, {Udaya-Shankar},
  {Schlagenhaufer}, {Srivani}, {Stevens}, {Subrahmanyan}, {Waterson},
  {Webster}, {Whitney}, {Williams}, {Williams}, \&
  {Wyithe}}]{2013PASA...30....7T}
{Tingay}, S.~J., {Goeke}, R., {Bowman}, J.~D., {et~al.} 2013, \pasa, 30, 7

\bibitem[{{Tiwari} \& {Nusser}(2016)}]{2016arXiv160901308T}
{Tiwari}, P. \& {Nusser}, A. 2016, ArXiv e-prints [\eprint[arXiv]{1609.01308}]

\bibitem[{{Trott} \& {Wayth}(2016)}]{2016PASA...33...19T}
{Trott}, C.~M. \& {Wayth}, R.~B. 2016, \pasa, 33, e019

\bibitem[{{van Haarlem} {et~al.}(2013){van Haarlem}, {Wise}, {Gunst}, {Heald},
  {McKean}, {Hessels}, {de Bruyn}, {Nijboer}, {Swinbank}, {Fallows},
  {Brentjens}, {Nelles}, {Beck}, {Falcke}, {Fender}, {H{\"o}randel},
  {Koopmans}, {Mann}, {Miley}, {R{\"o}ttgering}, {Stappers}, {Wijers},
  {Zaroubi}, {van den Akker}, {Alexov}, {Anderson}, {Anderson}, {van Ardenne},
  {Arts}, {Asgekar}, {Avruch}, {Batejat}, {B{\"a}hren}, {Bell}, {Bell}, {van
  Bemmel}, {Bennema}, {Bentum}, {Bernardi}, {Best}, {B{\^i}rzan}, {Bonafede},
  {Boonstra}, {Braun}, {Bregman}, {Breitling}, {van de Brink}, {Broderick},
  {Broekema}, {Brouw}, {Br{\"u}ggen}, {Butcher}, {van Cappellen}, {Ciardi},
  {Coenen}, {Conway}, {Coolen}, {Corstanje}, {Damstra}, {Davies}, {Deller},
  {Dettmar}, {van Diepen}, {Dijkstra}, {Donker}, {Doorduin}, {Dromer}, {Drost},
  {van Duin}, {Eisl{\"o}ffel}, {van Enst}, {Ferrari}, {Frieswijk}, {Gankema},
  {Garrett}, {de Gasperin}, {Gerbers}, {de Geus}, {Grie{\ss}meier}, {Grit},
  {Gruppen}, {Hamaker}, {Hassall}, {Hoeft}, {Holties}, {Horneffer}, {van der
  Horst}, {van Houwelingen}, {Huijgen}, {Iacobelli}, {Intema}, {Jackson},
  {Jelic}, {de Jong}, {Juette}, {Kant}, {Karastergiou}, {Koers}, {Kollen},
  {Kondratiev}, {Kooistra}, {Koopman}, {Koster}, {Kuniyoshi}, {Kramer},
  {Kuper}, {Lambropoulos}, {Law}, {van Leeuwen}, {Lemaitre}, {Loose}, {Maat},
  {Macario}, {Markoff}, {Masters}, {McFadden}, {McKay-Bukowski}, {Meijering},
  {Meulman}, {Mevius}, {Middelberg}, {Millenaar}, {Miller-Jones}, {Mohan},
  {Mol}, {Morawietz}, {Morganti}, {Mulcahy}, {Mulder}, {Munk}, {Nieuwenhuis},
  {van Nieuwpoort}, {Noordam}, {Norden}, {Noutsos}, {Offringa}, {Olofsson},
  {Omar}, {Orr{\'u}}, {Overeem}, {Paas}, {Pandey-Pommier}, {Pandey}, {Pizzo},
  {Polatidis}, {Rafferty}, {Rawlings}, {Reich}, {de Reijer}, {Reitsma},
  {Renting}, {Riemers}, {Rol}, {Romein}, {Roosjen}, {Ruiter}, {Scaife}, {van
  der Schaaf}, {Scheers}, {Schellart}, {Schoenmakers}, {Schoonderbeek},
  {Serylak}, {Shulevski}, {Sluman}, {Smirnov}, {Sobey}, {Spreeuw}, {Steinmetz},
  {Sterks}, {Stiepel}, {Stuurwold}, {Tagger}, {Tang}, {Tasse}, {Thomas},
  {Thoudam}, {Toribio}, {van der Tol}, {Usov}, {van Veelen}, {van der Veen},
  {ter Veen}, {Verbiest}, {Vermeulen}, {Vermaas}, {Vocks}, {Vogt}, {de Vos},
  {van der Wal}, {van Weeren}, {Weggemans}, {Weltevrede}, {White}, {Wijnholds},
  {Wilhelmsson}, {Wucknitz}, {Yatawatta}, {Zarka}, {Zensus}, \& {van
  Zwieten}}]{2013A&A...556A...2V}
{van Haarlem}, M.~P., {Wise}, M.~W., {Gunst}, A.~W., {et~al.} 2013, \aap, 556,
  A2

\bibitem[{{van Weeren} {et~al.}(2014{\natexlab{a}}){van Weeren}, {Intema},
  {Lal}, {Andrade-Santos}, {Br{\"u}ggen}, {de Gasperin}, {Forman}, {Hoeft},
  {Jones}, {Nuza}, {R{\"o}ttgering}, \& {Stroe}}]{2014ApJ...786L..17V}
{van Weeren}, R.~J., {Intema}, H.~T., {Lal}, D.~V., {et~al.}
  2014{\natexlab{a}}, \apjl, 786, L17

\bibitem[{{van Weeren} {et~al.}(2014{\natexlab{b}}){van Weeren}, {Intema},
  {Lal}, {Bonafede}, {Jones}, {Forman}, {R{\"o}ttgering}, {Br{\"u}ggen},
  {Stroe}, {Hoeft}, {Nuza}, \& {de Gasperin}}]{2014ApJ...781L..32V}
{van Weeren}, R.~J., {Intema}, H.~T., {Lal}, D.~V., {et~al.}
  2014{\natexlab{b}}, \apjl, 781, L32

\bibitem[{{van Weeren} {et~al.}(2016){van Weeren}, {Williams}, {Hardcastle},
  {Shimwell}, {Rafferty}, {Sabater}, {Heald}, {Sridhar}, {Dijkema}, {Brunetti},
  {Br{\"u}ggen}, {Andrade-Santos}, {Ogrean}, {R{\"o}ttgering}, {Dawson},
  {Forman}, {de Gasperin}, {Jones}, {Miley}, {Rudnick}, {Sarazin}, {Bonafede},
  {Best}, {B{\^i}rzan}, {Cassano}, {Chy{\.z}y}, {Croston}, {Ensslin},
  {Ferrari}, {Hoeft}, {Horellou}, {Jarvis}, {Kraft}, {Mevius}, {Intema},
  {Murray}, {Orr{\'u}}, {Pizzo}, {Simionescu}, {Stroe}, {van der Tol}, \&
  {White}}]{2016arXiv160105422V}
{van Weeren}, R.~J., {Williams}, W.~L., {Hardcastle}, M.~J., {et~al.} 2016,
  ArXiv e-prints [\eprint[arXiv]{1601.05422}]

\bibitem[{{Visser} {et~al.}(1995){Visser}, {Riley}, {Roettgering}, \&
  {Waldram}}]{1995A&AS..110..419V}
{Visser}, A.~E., {Riley}, J.~M., {Roettgering}, H.~J.~A., \& {Waldram}, E.~M.
  1995, \aaps, 110, 419

\bibitem[{{Wayth} {et~al.}(2015){Wayth}, {Lenc}, {Bell}, {Callingham},
  {Dwarakanath}, {Franzen}, {For}, {Gaensler}, {Hancock}, {Hindson},
  {Hurley-Walker}, {Jackson}, {Johnston-Hollitt}, {Kapi{\'n}ska}, {McKinley},
  {Morgan}, {Offringa}, {Procopio}, {Staveley-Smith}, {Wu}, {Zheng}, {Trott},
  {Bernardi}, {Bowman}, {Briggs}, {Cappallo}, {Corey}, {Deshpande}, {Emrich},
  {Goeke}, {Greenhill}, {Hazelton}, {Kaplan}, {Kasper}, {Kratzenberg},
  {Lonsdale}, {Lynch}, {McWhirter}, {Mitchell}, {Morales}, {Morgan}, {Oberoi},
  {Ord}, {Prabu}, {Rogers}, {Roshi}, {Shankar}, {Srivani}, {Subrahmanyan},
  {Tingay}, {Waterson}, {Webster}, {Whitney}, {Williams}, \&
  {Williams}}]{2015PASA...32...25W}
{Wayth}, R.~B., {Lenc}, E., {Bell}, M.~E., {et~al.} 2015, \pasa, 32, e025

\bibitem[{{Wieringa}(1991)}]{1991PhDT.......241W}
{Wieringa}, M.~H. 1991, PhD thesis, , Rijksuniversiteit Leiden, (1991)

\bibitem[{{Wijnholds} {et~al.}(2010){Wijnholds}, {van der Tol}, {Nijboer}, \&
  {van der Veen}}]{2010ISPM...27...30W}
{Wijnholds}, S., {van der Tol}, S., {Nijboer}, R., \& {van der Veen}, A.-J.
  2010, IEEE Signal Processing Magazine, 27, 30

\bibitem[{{Williams} {et~al.}(2013){Williams}, {Intema}, \&
  {R{\"o}ttgering}}]{2013A&A...549A..55W}
{Williams}, W.~L., {Intema}, H.~T., \& {R{\"o}ttgering}, H.~J.~A. 2013, \aap,
  549, A55

\bibitem[{{Williams} {et~al.}(2016){Williams}, {van Weeren}, {R{\"o}ttgering},
  {Best}, {Dijkema}, {de Gasperin}, {Hardcastle}, {Heald}, {Prandoni},
  {Sabater}, {Shimwell}, {Tasse}, {van Bemmel}, {Br{\"u}ggen}, {Brunetti},
  {Conway}, {En{\ss}lin}, {Engels}, {Falcke}, {Ferrari}, {Haverkorn},
  {Jackson}, {Jarvis}, {Kapi{\'n}ska}, {Mahony}, {Miley}, {Morabito},
  {Morganti}, {Orr{\'u}}, {Retana-Montenegro}, {Sridhar}, {Toribio}, {White},
  {Wise}, \& {Zwart}}]{2016MNRAS.460.2385W}
{Williams}, W.~L., {van Weeren}, R.~J., {R{\"o}ttgering}, H.~J.~A., {et~al.}
  2016, \mnras, 460, 2385

\bibitem[{{Wilman} {et~al.}(2008){Wilman}, {Miller}, {Jarvis}, {Mauch},
  {Levrier}, {Abdalla}, {Rawlings}, {Kl{\"o}ckner}, {Obreschkow}, {Olteanu}, \&
  {Young}}]{2008MNRAS.388.1335W}
{Wilman}, R.~J., {Miller}, L., {Jarvis}, M.~J., {et~al.} 2008, \mnras, 388,
  1335

\bibitem[{{Wykes} {et~al.}(2014){Wykes}, {Intema}, {Hardcastle}, {Achterberg},
  {Jones}, {Jerjen}, {Orr{\'u}}, {Lazarian}, {Shimwell}, {Wise}, \&
  {Kronberg}}]{2014MNRAS.442.2867W}
{Wykes}, S., {Intema}, H.~T., {Hardcastle}, M.~J., {et~al.} 2014, \mnras, 442,
  2867

\end{thebibliography}


\clearpage

\begin{appendix}

\section{SPAM Pipeline}
\label{sec:spam_pipeline}

\changetwo{In this Appendix we provide implementation and computational details of the SPAM pipeline. This may be relevant for those readers who may wish to use it as a blue-print for the development of calibration strategies for new and future low-frequency telescopes
\citep[as was done for LOFAR facet calibration;][]{2016arXiv160105422V}. 
}

\changetwo{The pipeline is the result of many years of development and testing, and demonstrates that a high degree of automation is possible even for processing high-resolution, high-sensitivity wide-field radio interferometry data at sub-GHz frequencies. SPAM is a Python module that builds on ParselTongue
\citep{2006ASPC..351..497K}, 
a Python interface to the Astronomical Image Processing System
\citep[AIPS;][]{2003ASSL..285..109G}. 
ParselTongue itself makes use of parts of Obit
\citep{2008PASP..120..439C}, 
a stand-alone data reduction package compatible with AIPS data formats. The SPAM software (including the pipeline) can be obtained through \citet{2014ascl.soft08006I}. The SPAM pipeline has been successfully applied in many scientific investigations that make use of GMRT observations in any of the observing bands below 1~GHz,
\citep[150, 235, 325, and 610~MHz; e.g.,][]{2014ApJ...786L..17V,2014ApJ...781L..32V,2014MNRAS.442.2867W,2014MNRAS.444L..44B,2015MNRAS.454.3391B,2014MNRAS.440.1542D,2015MNRAS.453.3483D}. 
}

\begin{figure*}[!ht]
\begin{center}
\resizebox{\hsize}{!}{
\includegraphics[angle=0]{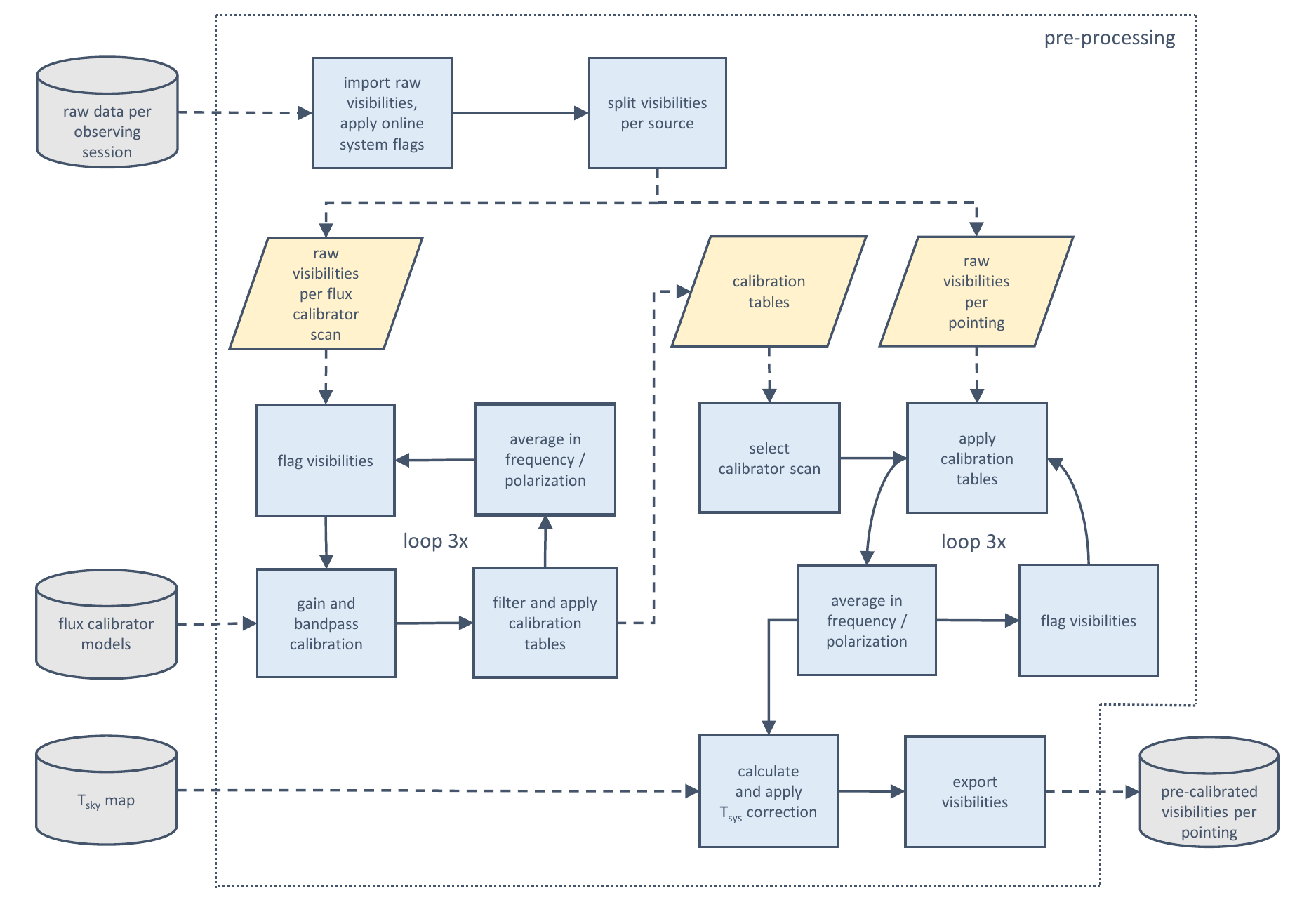}}
\caption{Flow diagram of the \emph{pre-processing} part of the SPAM pipeline, as defined in \changetwo{Section~\ref{sec:dpp_pp}. Gray cylinders depict permanent (disk-based) data storage, with dashed arrows marking the data flow to and from the data storage. Blue rectangles represent functions that operate on data. Yellow trapezoids represent temporary (ram-based) data storage, and solid arrows represent the data flow between functions.}}
\label{fig:pipeline_preprocessing}
\end{center}
\end{figure*}

\begin{figure*}[!ht]
\begin{center}
\resizebox{\hsize}{!}{
\includegraphics[angle=0]{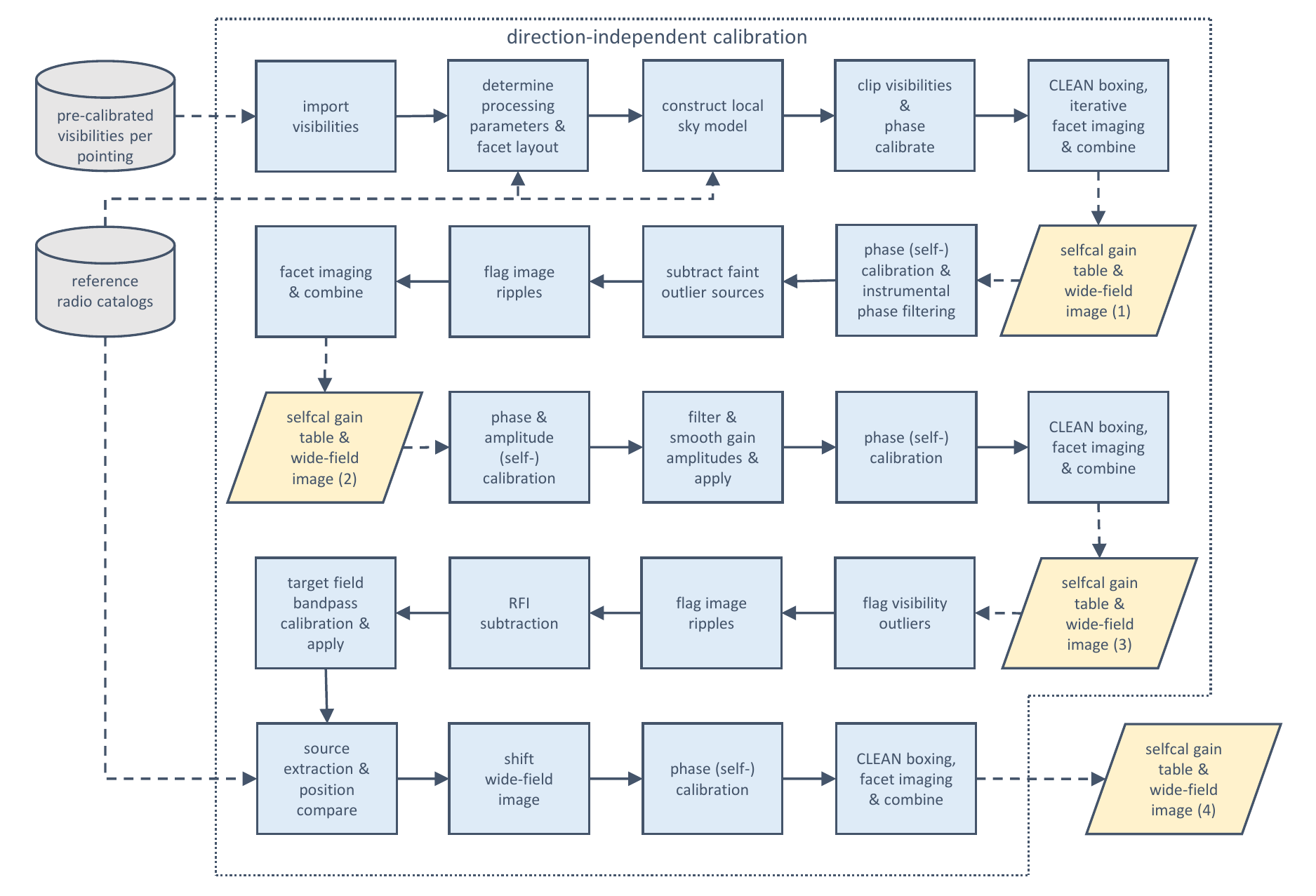}}
\caption{Flow diagram of the \emph{direction-independent calibration} part of the main pipeline, as defined in \changetwo{Section~\ref{sec:dpp_mp_dic}}. Symbols are defined in the caption of Figure~\ref{fig:pipeline_preprocessing}.}
\label{fig:pipeline_main_dic}
\end{center}
\end{figure*}

\begin{figure*}[!ht]
\begin{center}
\resizebox{\hsize}{!}{
\includegraphics[angle=0]{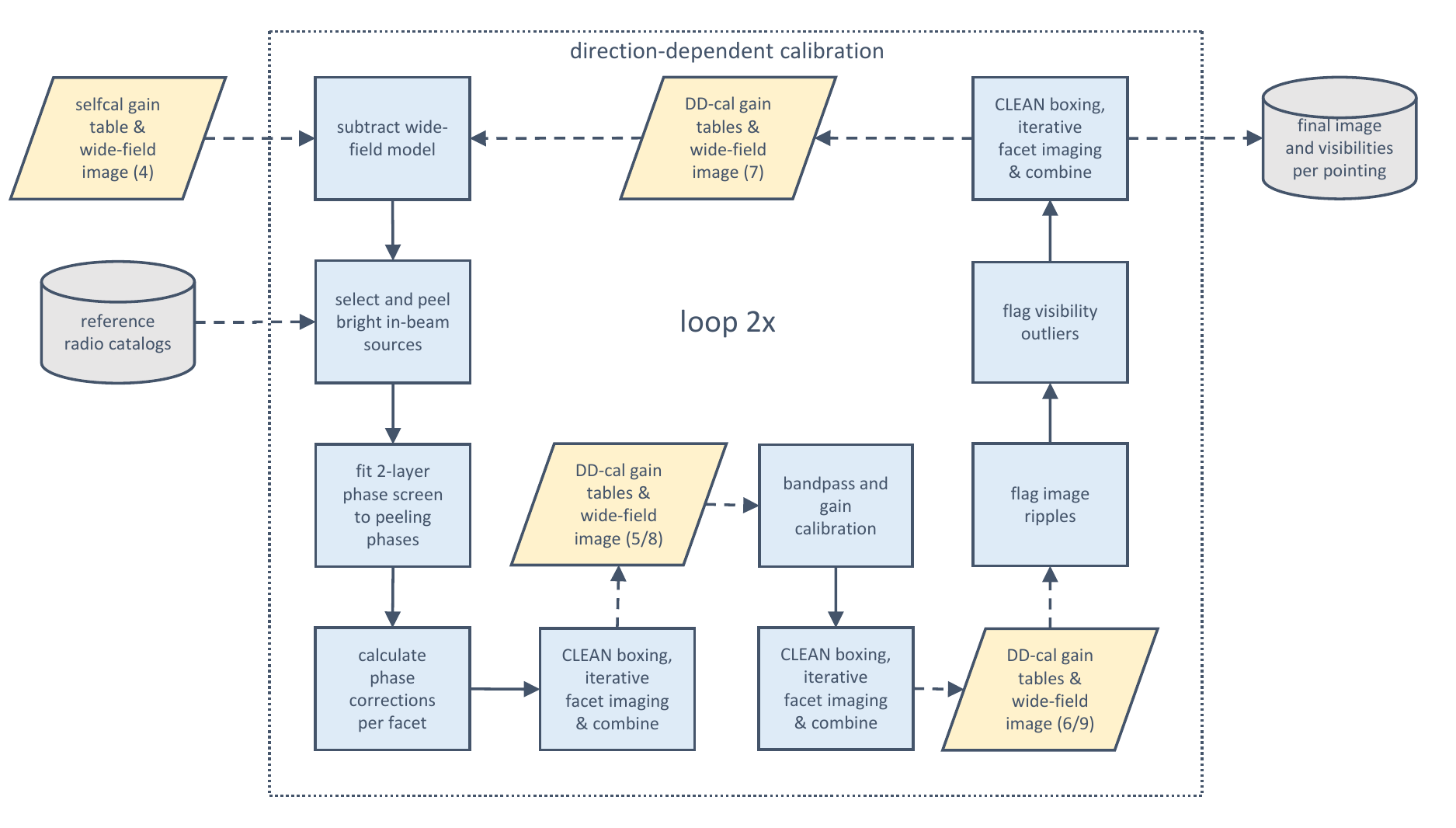}}
\caption{Flow diagram of the \emph{direction-dependent calibration} part of the main pipeline, as defined in \changetwo{Section~\ref{sec:dpp_mp_ddc}}. Symbols are defined in the caption of Figure~\ref{fig:pipeline_preprocessing}.}
\label{fig:pipeline_main_ddc}
\end{center}
\end{figure*}

\changetwo{
As noted in Section~\ref{sec:dpp}, the pipeline consists of two parts: a \emph{pre-processing} part (\S\ref{sec:dpp_pp}), and a \emph{main pipeline} part (\S\ref{sec:dpp_mp}), which itself is separable into direction-independent calibration (\S\ref{sec:dpp_mp_dic}) and direction-dependent calibration (\S\ref{sec:dpp_mp_ddc}). Figures \ref{fig:pipeline_preprocessing}, \ref{fig:pipeline_main_dic} and \ref{fig:pipeline_main_ddc} show flow diagrams for each of these stages. Both pipeline parts run as independent, single thread processes on multi-node, multi-core compute clusters, allowing for significant parallel processing of many observations and pointings at the same time. Making use of the NMPOST compute cluster\footnote{\url{https://info.nrao.edu/computing/guide/astronomySupport}} at the National Radio Astronomy Observatory (NRAO), bulk processing of all TGSS survey data was completed within a month. This involved converting the raw data files of 202~observe sessions into 5821~pre-calibrated visibility data sets for 5336~unique pointings, and converting these into calibrated and deconvolved radio images. About 95~percent of the data passed through the pipeline without manual interaction.}

\changetwo{
The flow diagram of the pre-processing step is depicted in Figure~\ref{fig:pipeline_preprocessing}. Per observing session, the TGSS data consists of one or multiple LTA files containing the visibility data, complemented with one or more FLAGS files containing telescope system health information. All LTA and FLAGS files were stored in a single directory on the shared LUSTRE filesystem, visible to all NMPOST cluster nodes. This filesystem is mostly optimized for large data reads and writes, but because of the high latency it is not well suited to do many small data reads and writes. The pre-processing step involves accessing a relatively small number of files, mostly large data reads and writes, and therefore was done directly on the LUSTRE filesystem.}

\changetwo{
A simple scheduler running on the cluster head took care of spawning pre-processing jobs to cluster nodes, monitoring the availability of unprocessed raw data and usage of assigned compute resources. To prevent exhaustion of the available cluster CPU power, node memory and node bandwidth, up to 8~pre-processing jobs were allowed to run per cluster node, and up to 16~pre-processing jobs in total. The average processing time per pre-processing job was about 30~minutes, therefore all pre-processing was effectively completed within a few hours.}

\changetwo{
Each spawned job processed all LTA and FLAGS files belonging to a single observing in the following manner. LTA files were individually converted to UVFITS format using the \texttt{listscan} and \texttt{gvfits} tools supplied by the NCRA\footnote{\url{http://www.ncra.tifr.res.in/ncra/gmrt/gmrt-users}}, dropping the cross-polarization visibilities (RL and LR) and the phase calibrator scans as they are not used in the processing. UVFITS files were loaded into AIPS, concatenated, and then split per source, yielding individual files for all observed primary calibrators and survey pointings}.

\changetwo{The pre-processed visibilities per pointing were exported to the LUSTRE filesystem in UVFITS format to be used by the main pipeline. Similar to the pre-processing as described in Section~\ref{sec:dpp_pp}, all UVFITS files of the pre-calibrated pointings (which incorporates some redundancy in the observations) were stored in a single directory on the LUSTRE filesystem, visible to all NMPOST cluster nodes. A simple scheduler running on the cluster head took care of spawning pipeline jobs to cluster nodes, monitoring the availability of unprocessed raw data and usage of assigned compute resources. Up to 12~pipeline jobs were allowed to run per cluster node, and up to 60~pipeline jobs in total. The average processing time per main pipeline job was about 2--3~hours, therefore all main pipeline processing was effectively completed within two weeks.}

\changetwo{The main pipeline step involves accessing a relatively large number of files, mostly small data reads and writes, and is therefore not well matched to the LUSTRE filesystem (see Section~\ref{sec:dpp_pp}). By optimizing the required dynamic disk space per job to be $\lesssim 2$~GB, the pipeline used the local RAM disk as its work area, only requiring access to the LUSTRE filesystem at the start and end of the pipeline job for copying data products. Simple tests showed that the RAM disk outperformed all other storage systems available by far, including local solid state drives, when running multiple pipeline jobs in parallel.} 

\changetwo{Each main pipeline job starts by creating a unique working directory on the cluster node's RAM disk. The UVFITS-format visibility data of the pointing is copied from the LUSTRE filesystem to this directory, after which the pipeline is executed. A log file is kept as part of the pipeline output. In case of a successful pipeline run, the log file together with the resulting (final) images and calibrated and flagged visibility data are copied back to the LUSTRE filesystem. In case of a failed pipeline run, only the log file is copied. Finally, the working directory and its contents are removed from the RAM disk.}

\begin{figure*}[!ht]
\begin{center}
\resizebox{\hsize}{!}{
\includegraphics[angle=0]{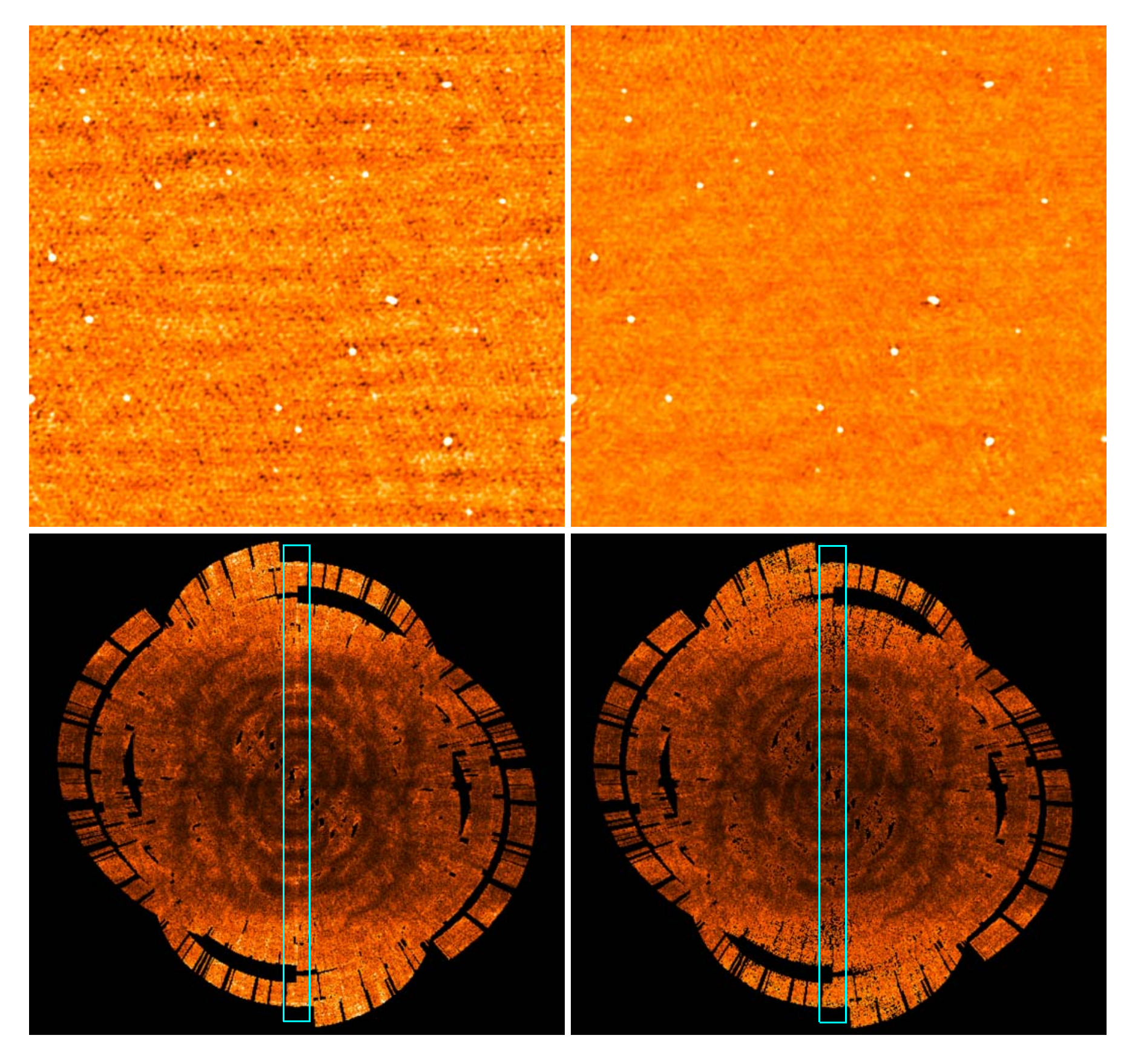}}
\caption{Example of the suppression of image background ripples. \emph{Top left:} Small part of a GMRT 150~MHz image based on an 8~hour observation, with a low-magnitude but clearly visible ripple in the background. \emph{Bottom left:} Rescaled Fourier transform of the image background, which reproduces the imprinted UV-coverage and reveals several bright spots as well as a bright vertical line (due to residual RFI; see inside cyan rectangle). \emph{Bottom right:} The same Fourier transform, but with the bright pixels filtered out. \emph{Top right:} Same GMRT 150~MHz image as before, but with all visibilities flagged that correspond to the bright, filtered pixels in the Fourier transform.}
\label{fig:ripple_killer}
\end{center}
\end{figure*}


\clearpage
\onecolumn
\section{Additional Images}
\label{sec:app_images}

\begin{figure*}[!h]
\begin{center}
\resizebox{\hsize}{!}{
\includegraphics[angle=0]{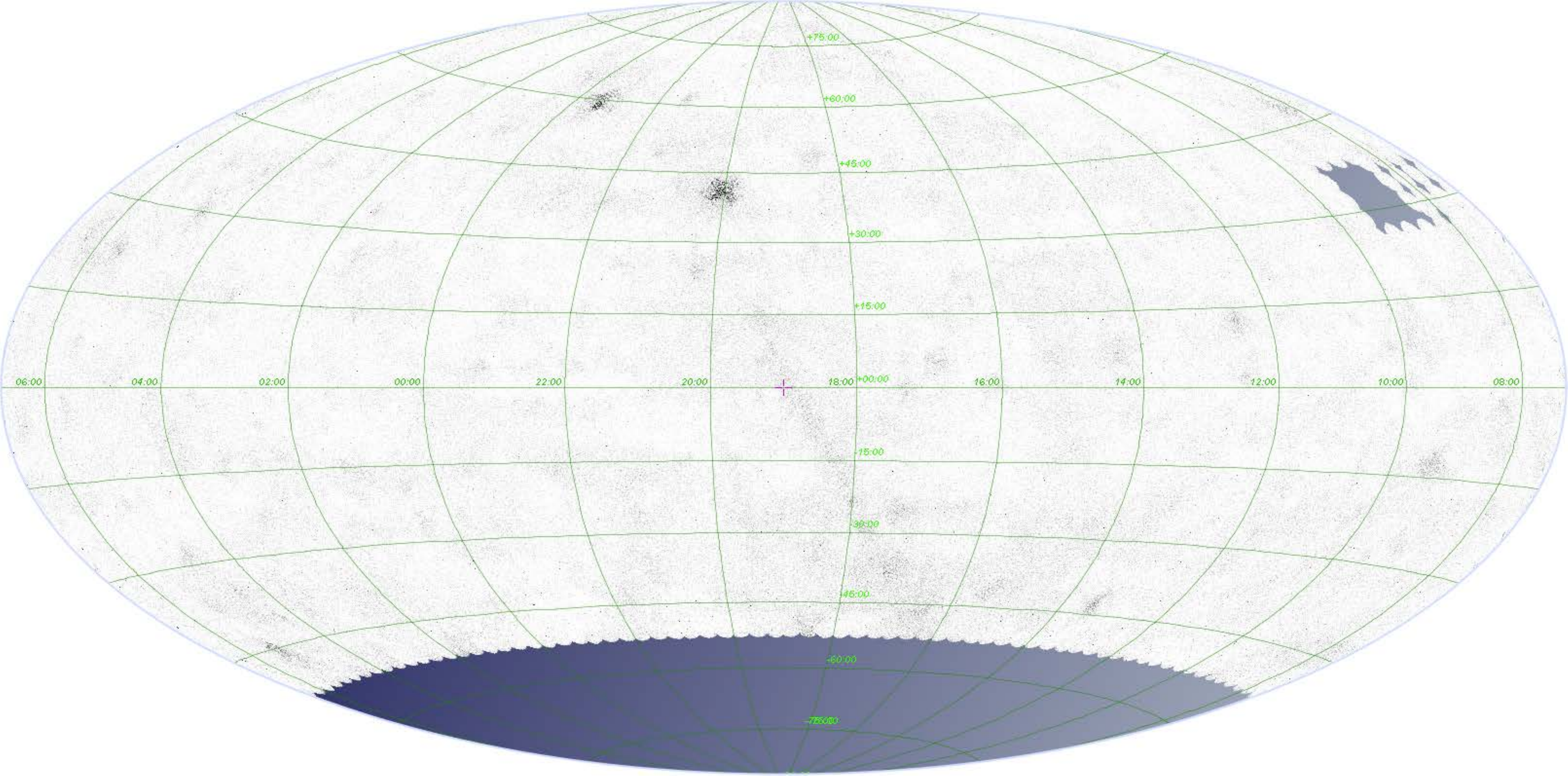}}
\caption{All-sky representation (grayscale indicates flux density) of the TGSS ADR survey in aitoff projection, marking the sky area included in this data release. The missing coverage towards the south celestial pole is due to the declination limit of GMRT, as well as the two lowest TGSS declination rows which were very difficult to process. The missing pointings of a single observing session under very difficult ionospheric conditions are visible towards the western edge. Also noticable are the regions near extremely bright sources such as Cas~A, Cen~A and Cyg~A, where imaging artifacts dominate the image background. This representation was created using \texttt{Aladin}
\citep{2000A&AS..143...33B}. 
}
\label{fig:tgss_map}
\end{center}
\end{figure*}

\begin{figure*}[!h]
\begin{center}
\resizebox{\hsize}{!}{
\includegraphics[angle=0]{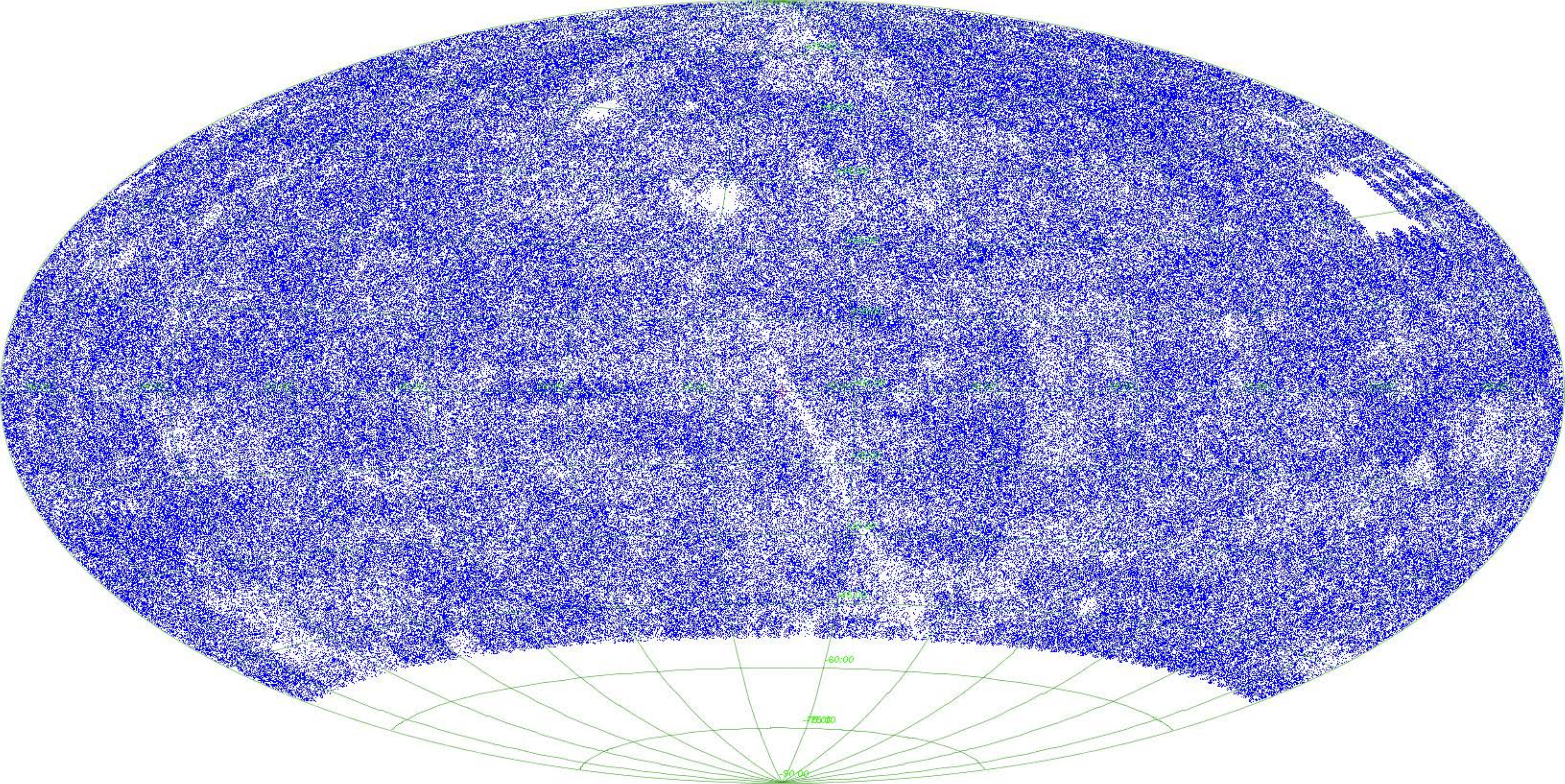}}
\caption{As Figure~\ref{fig:tgss_map}, but now marking the spatial distribution of the \changeone{623,604}~extracted sources in this data release. The extracted source density is correlated with the background RMS noise distribution as depicted in Figure~\ref{fig:rms_map}.}
\label{fig:source_map}
\end{center}
\end{figure*}

\end{appendix}


\end{document}